\documentclass[fleqn,usenatbib]{mnras}

\usepackage{newtxtext,newtxmath}

\usepackage[T1]{fontenc}

\DeclareRobustCommand{\VAN}[3]{#2}
\let\VANthebibliography\thebibliography
\def\thebibliography{\DeclareRobustCommand{\VAN}[3]{##3}\VANthebibliography}

\usepackage{graphicx}	
\usepackage{amsmath}	
\usepackage{enumitem}

\usepackage{hyperref,siunitx}
\usepackage{orcidlink}
\usepackage{multirow}
\newcommand{\Msun}{{\rm M_{\odot}}}

\newcommand{\um}{\SI{}{\micro\meter}}
\newcommand{\eg}{$\textnormal{e.g.,}$}
\newcommand{\ie}{$\textnormal{i.e.}$}

\newcommand{\jwst}{\textit{JWST}}

\newcommand{\colibre}{\mbox{\sc{COLIBRE}}}

\newcommand{\dtm}{\mbox{$\mathcal{DTM}$}}
\newcommand{\dtg}{\mbox{$\mathcal{DTG}$}}
\newcommand{\stl}{\mbox{$\mathcal{STL}$}}

\title[Dust in COLIBRE]{The evolution of galaxy dust scaling relations in the COLIBRE simulations}

\author[Aswin P. Vijayan et al.]{Aswin P. Vijayan\orcidlink{0000-0002-1905-4194}$^{1}$\thanks{E-mail: aswinpvijayan@gmail.com (APV)},
James W. Trayford$^{2}$,
Joop Schaye$^{3}$,
Sylvia Ploeckinger$^{4}$,
Andrea Gebek$^{5}$,
\newauthor
Nick Andreadis$^{5}$,
Maarten Baes$^{5}$,
Alejandro Benítez-Llambay$^{6}$,
Evgenii Chaikin$^{7,3}$,
Carlos S. Frenk$^{7}$,
\newauthor
Filip Hu\v{s}ko$^{3}$,
Robert J. McGibbon$^{3}$,
Alexander J. Richings$^{8,9}$,
Matthieu Schaller$^{3,10}$
\\
$^{1}$Astronomy Centre, University of Sussex, Falmer, Brighton BN1 9QH, UK\\
$^{2}$Centre for Astrophysics Research, Department of Physics, Astronomy and Mathematics, University of Hertfordshire, College Lane, Hatfield, AL10 9AB, UK\\
$^{3}$Leiden Observatory, Leiden University, PO Box 9513, 2300 RA Leiden, the Netherlands\\
$^{4}$Department of Astrophysics, University of Vienna, Türkenschanzstrasse 17, A-1180 Vienna, Austria\\
$^{5}$Department of Physics and Astronomy, Universiteit Gent, Proeftuinstraat 86 N3, B-9000 Ghent, Belgium\\
$^{6}$Dipartimento di Fisica "Giuseppe Occhialini", Università degli Studi di Milano-Bicocca\\
$^{7}$Institute for Computational Cosmology, Department of Physics,
University of Durham, South Road, Durham, DH1 3LE, UK\\
$^{8}$Centre for Data Science, Artificial Intelligence and Modelling, University of Hull, Cottingham Road, Hull, HU6 7RX, UK\\
$^{9}$E. A. Milne Centre for Astrophysics, University of Hull, Cottingham Road, Hull, HU6 7RX, UK\\
$^{10}$Lorentz Institute for Theoretical Physics, Leiden University, PO Box 9506, 2300 RA Leiden, the Netherlands
}
\date{Accepted XXX. Received YYY; in original form ZZZ}

\pubyear{\the\year{}}

\begin{document}
\label{firstpage}
\pagerange{\pageref{firstpage}--\pageref{lastpage}}
\maketitle

\begin{abstract}
We present dust scaling relations across cosmic time ($0 \le z \le 15$) for galaxies in the \textsc{COLIBRE} cosmological simulations.
\textsc{COLIBRE} self-consistently tracks dust production, growth, destruction, and grain size evolution within a multiphase interstellar medium.
Using volumes up to $(400\, {\rm cMpc})^3$ at three mass resolutions ($10^{5}\text{--}10^7\, \Msun$), we predict the dust mass function, cosmic dust mass density, and key dust scaling relations (dust-to-gas ratio, dust-to-metal ratio, grain species fractions, and grain sizes) as functions of galaxy metallicity, stellar mass, and dust mass. 
The model broadly reproduces most observed relations across cosmic time, matching closest at the highest resolution. 
We find that silicates dominate the dust mass ($\gtrsim 70\%$) at all epochs, and while large grains dominate in the early Universe ($z \ge 5$), their mass fraction declines to become comparable to small grains by $z=0$. 
At $z < 1$, the simulated dust mass functions align well with observations, but the cosmic dust mass density is systematically high by $\lesssim 0.3$~dex, while in agreement with observations at higher redshifts. 
Additionally, the simulations underpredict the extreme dust masses of bright sub-millimeter galaxies at $z \ge 2$. 
We demonstrate that scaling relations are sensitive to numerical resolution only in the low-redshift, low-mass regime; while their normalisation is influenced by gas-phase selection.
These findings highlight both the predictive power and resolution-dependent limits of cosmological dust models, providing essential insights to refine ISM physics.

\end{abstract}




\section{Introduction}\label{sec:intro}
Dust is one of the fundamental components of the interstellar medium (ISM) of galaxies. 
Cosmic dust consists of small particles that can range from a few molecules to large complexes of molecules, usually smaller than $1\, \um$.
Dust constitutes less than 1\% of the total baryonic content in galaxies \cite[\eg][]{RemyRuyer2014,Chiang2025}, however it has a significant effect on galaxy formation and evolution \cite[\eg][]{Dorschner1995} as well as on the inference of galaxy properties from astrophysical observations \cite[\eg][]{Calzetti2001,Conroy2013}.

Dust can act as a catalyst for star formation by providing sites for molecular hydrogen formation and gas cooling \cite[\eg][]{Hirashita2002,Wakelam2017,Vogelsberger2019}. 
Dust depletes metals, thus changing the observed gas-phase abundances of galaxies \cite[\eg][]{Jenkins2009,DeCia2018} and affecting the thermal balance of the ISM \cite[\eg][]{Richings2022}. 
It obscures the UV-optical light of galaxies, re-emitting it in the far-infrared (FIR). 
Thus, detecting the dust-continuum in the FIR acts as a good tracer of the star formation activity in the galaxy \cite[\eg][]{Calzetti2010}. 
It is estimated that approximately $30\%$ of all photons in the Universe have been reprocessed by dust during their lifetime \cite[\eg][]{Dwek1998,Bernstein2002,Driver2016}. Therefore, it is important that this component is modelled in galaxy formation and evolution simulations in order to understand and quantify its role in the evolution of galaxies and their observed light.

Multi-wavelength observations have been instrumental in improving our understanding of cosmic dust properties.
The wavelength dependence of interstellar extinction, driven by absorption and scattering in the ultraviolet (UV) and optical regimes, constrains the dust grain size distribution \cite[\eg][]{Mathis1977,Draine2003}. 
Furthermore, specific spectral signatures (such as the 2175\AA\ UV bump, Polycyclic Aromatic Hydrocarbon (PAH) emission and silicate absorption in the IR) and polarisation measurements trace the chemical composition \cite[\eg][]{DraineLi2007,Siebenmorgen2014} and grain shape \cite[\eg][]{Hildebrand1988,Hensley2023,Ysard2024}.
Although laboratory experiments have been used to create and test models based on these properties \cite[\eg][]{Dorschner1995,Gavilan2016}, theoretical models that incorporate dust grain composition, size, and shape only partially reproduce the full range of observable data \cite[\eg][]{Zubko2004,Hensley2017}.

The last decade has seen a surge of multi-wavelength observations of galaxies.
Facilities such as the Atacama Large Millimeter/submillimeter Array (ALMA), \textit{Hubble Space Telescope} (\textit{HST}),
\textit{JWST} and \textit{Herschel Space Observatory} provide broadband photometry or spectroscopy from the observed-frame UV to FIR/sub-mm, enabling the placement of constraints on the gas, metal and dust content of galaxies through the detection of line absorption/emission or dust continuum.
Consequently, these datasets have paved the way to probe key dust scaling relations of galaxies,
such as the dust mass \cite[\eg][]{Santini2014,daCunha2015,Davies2017}, the dust-to-gas ratio \cite[\dtg, \eg][]{RemyRuyer2014,Peroux&Howk2020,Konstantopoulou2024}, and the dust-to-metal ratio \cite[\dtm, \eg][]{DeCia2016,Wiseman2017,Algera2025}, with galaxy stellar mass and metallicity across cosmic time ($z \approx 15-0$). 
Furthermore, the total dust content has been used to construct the galaxy dust mass function \cite[\eg][]{Vlahakis2005,Eales2009,Berta2025} and trace the evolution of the cosmic dust mass density \cite[\eg][]{Pozzi2020,Traina2024}. 
They provide a critical test bed for theoretical models of galaxy formation and evolution that incorporate detailed dust physics.

Similarly, theoretical simulations have made substantial advances, driven by advancements in code and hardware as well as the integration of sophisticated subgrid models for unresolved physical processes into galaxy simulations.
However, the self-consistent inclusion of dust production and destruction has only recently entered the mainstream in both semi-analytical models \cite[SAMs, \eg][]{Popping_dust2017,Vijayan_dust2019,Triani_dust2020,Dayal2022,Yates2024,Osman2025,Parente2026} and hydrodynamical simulations, in both representative volume \cite[\eg][]{Bekki2015,Hirashita2015,McKinnon2017,Aoyama2018,Li2019,Graziani2020,Granato2021,Thesan2022,Parente2022,Lewis2023,Trayford_colibre_dust2025} and zoom simulations of isolated galaxies or clusters \cite[\eg][]{Gjergo2018,Choban2024,Dubois2024,Han2025,RodrguezMontero2026}. 
These simulations typically model dust formation via supernovae (SNe) and asymptotic giant branch (AGB) stars, subsequent grain growth in the dense ISM and destruction via shock heating and sputtering. 
A subset of these works also track dust grain size evolution from processes such as coagulation and shattering \cite[\eg][]{Hirashita2015,McKinnon2018,Li2021,Parente2026_dust,Trayford_colibre_dust2025}, which is important for building galaxy extinction curves. 
However, in the majority of these works \cite[with notable exceptions such as][]{Bekki2015,Thesan2022,Dubois2024,Han2025,RodrguezMontero2026,Trayford_colibre_dust2025} dust remains a passive tracer, and does not actively regulate galaxy evolution through effects such as self-shielding or the catalysis of molecular hydrogen formation.

Simulations that model dust evolution have achieved moderate success in reproducing various observed dust scaling relations. 
Large volume ($\ge (50\, {\rm cMpc})^{3}$) cosmological simulations run to $z=0$ generally recover the scaling relation between galaxy stellar mass and dust mass in the local Universe \cite[\eg][]{McKinnon2017,Popping_dust2017,Aoyama2018,Li2019,Vijayan_dust2019,Triani_dust2020,Osman2025}.
However, these models often struggle to reproduce the extreme dust masses observed in sub-millimeter (sub-mm) galaxies (or dusty star forming galaxies, DSFGs) at high-redshift ($z \ge 2$). 
A similar tension exists for the evolution of the dust mass function, where the models match the normalisation at $z=0$, while they frequently underpredict it at high-redshifts ($z \ge 2$).
Many models match the observed \dtg\ or \dtm\ ratios only at specific redshifts, failing to reproduce the trends consistently across cosmic time \cite[\eg][]{Popping2022}. Furthermore, there is little consensus, with different simulation suites succeeding at different epochs for different observables.
These mixed results highlight the complex dependence of various dust scaling relations on the underlying physics implementations in the different simulations.
It is also important to note that theoretical metal and dust yields from stellar processes remain a significant source of uncertainty \cite[\eg][]{Wiersma2009b,Kobayashi2020}.
Concurrently, the assumptions required to derive galaxy dust masses and metallicities from observational data may introduce artificial trends or biases \cite[\eg][]{Sommovigo2025,Vijayan2025}.

In this paper we present predictions for the evolution of dust scaling relations in the \colibre\ suite of cosmological hydrodynamical simulations \cite[]{schaye_colibre2025,Chaikin_colibre2025}, and connect them to the physics of galaxy formation and evolution implemented in the model.
In \S~\ref{sec:sim_model} we describe the \colibre\ simulation model and in particular the implemented model for dust production and evolution (\S~\ref{sec:dustmodel}). 
\S~\ref{sec:obs_data} briefly describes the observational data used in this work, and the caveats of observationally inferred physical quantities such as the galaxy dust, gas, and metal mass.
In \S~\ref{sec:dustscaling}, we show how dust grows in the simulation by following the different dust scaling relations such as the dust-to-gas ratio (\S~\ref{sec:dustscaling.dtg}, \ref{sec:dustscaling.dtg_dtm_mstar}),  dust-to-metal ratio (\S~\ref{sec:dustscaling.dtm}, \ref{sec:dustscaling.dtg_dtm_mstar}), dust mass--stellar mass relation (\S~\ref{sec:dustscaling.dmass_smass}), and the dust grain species fraction and size distribution (\S~\ref{sec:dustscaling.grain_evo}).
In \S~\ref{sec:dmf} we examine the evolution of the dust mass function and the evolution of cosmic dust mass density in \S~\ref{sec:cdmd}.
\S~\ref{sec:gas_phase} demonstrates the influence of different gas-phase selections on galaxy dust scaling relations.
Finally, we present our conclusions in \S~\ref{sec:conclusions}.

\section{The Colibre Simulation Model}\label{sec:sim_model}
The \colibre\ (\textbf{COL}d \textbf{I}sm and \textbf{B}etter \textbf{RE}solution) simulations are fully described in \cite{schaye_colibre2025} and \cite{Chaikin_colibre2025}.
Here we describe briefly the physics prescriptions incorporated into the model, along with the various box sizes and mass resolutions used in this work.

The \colibre\ simulations are performed using \texttt{SWIFT}, an open-source, highly parallel gravity and hydrodynamics code \cite[]{Schaller2024_swift}.
The hydrodynamics are solved using the \textsc{Sphenix} Smoothed Particle Hydrodynamics (SPH) scheme described in \cite{borrow2022_sphenix}.
The \colibre\ simulations use a sophisticated and comprehensive subgrid physics model designed to explicitly capture the multiphase ISM and the complex interplay between gas, stars, dust, and feedback processes \cite[see Section~3 in][]{schaye_colibre2025}. We only summarise the subgrid prescriptions here, with the dust implementation described in more detail in \S~\ref{sec:dustmodel}.

\colibre\ adopts a flat $\Lambda$CDM cosmology from the Dark Energy Survey year three results \cite[`3$\times$2pt + All Ext.' in][]{DESY32022}.
The simulation runs span three distinct resolution levels, denoted as m5, m6, and m7, corresponding to initial mean baryon particle masses of $2.30 \times 10^5 \, \text{M}_\odot$, $1.84 \times 10^6$, and $1.47 \times 10^7$, respectively. 
The simulations use 4 times more Cold Dark Matter particles, resulting in comparable particle masses as used for the baryonic component, which suppresses spurious energy transfer from CDM to stellar particles \cite[]{Ludlow2019,Ludlow2021,Ludlow2023}.
The volumes range from cubic boxes of 25 comoving Mpc (cMpc) on a side up to 400 cMpc.

Table~\ref{tbl:simulations} lists the simulation volumes and corresponding resolutions used in this paper.
We use the L400m7 and L200m6 simulations for m7 and m6 resolutions, respectively. 
Due to the larger computational expense of runs using m5 resolution, L050m5 and L100m5 have not currently reached $z=0$. Hence, we switch from L100m5 to L050m5 at $z=2$ and from L050m5 to L025m5 at $z=1$. We base our analysis on these runs, unless specified otherwise.
\begin{table}
\centering
\caption{\colibre\ hydrodynamical simulations used in this work. 
The columns specify the simulation identifier, where `L' denotes the side length of the cubic simulation volume in cMpc, `m' gives the rounded logarithm (base-10) of the mean initial particle mass (in solar masses) for both baryons and dark matter. The table also lists 
the comoving box side length $L$; 
the number of baryonic particles $N_\text{b}$; 
the initial mean baryonic particle mass $m_\text{g}$; 
the coagulation factor $f_{\rm co}$;
and the lowest redshift up to which the simulation is used $z$.
The simulations run with the `Hybrid' AGN feedback model is also used in this work for investigating the effect of the AGN feedback mode on dust, labelled with a `h' suffix (\eg\ L100m6h).
For the complete list of simulations that have been performed and the values of subgrid parameters, please refer to Table~1 and 2 in \protect\cite{schaye_colibre2025}.
}
\label{tbl:simulations}
    \begin{tabular}{lcccccc}
	\hline
	Identifier & $L$ & $N_\text{b}$ & $m_\text{g}$ & $f_{\rm co}$ & $z$\\ 
	& (cMpc) & & ($\Msun$)  &&\\
\hline
	    L100m5   & 100 & $3008^3$ & $2.30\times 10^5$ & $1.0$       & $2$\\
        L050m5   & 50  & $1504^3$ & $2.30\times 10^5$ & $1.0$       & $1$\\
        L025m5   & 25  & $752^3$  & $2.30\times 10^5$ & $1.0$       & $0$\\
        L025m5h  & 25  & $752^3$  & $2.30\times 10^5$ & $1.0$       & $0$\\
\hline
        L200m6   & 200 & $3008^3$ & $1.84\times 10^6$ & $1.0$        & $0$\\
        L100m6   & 100 & $1504^3$ & $1.84\times 10^6$ & $1.0$        & $0$\\
        L100m6h  & 100 & $1504^3$ & $1.84\times 10^6$ & $1.0$        & $0$\\
        L050m6   & 50  & $752^3$  & $1.84\times 10^6$ & $1.0$        & $0$\\
        L025m6   & 25  & $376^3$  & $1.84\times 10^6$ & $1.0$        & $0$\\
\hline
        L400m7   & 400 & $3008^3$ & $1.47\times 10^7$ & $10^{-0.5}$  & $0$\\
        L200m7   & 200 & $1504^3$ & $1.47\times 10^7$ & $10^{-0.5}$  & $0$\\
        L200m7h  & 200 & $1504^3$ & $1.47\times 10^7$ & $10^{-0.5}$  & $0$\\
        L050m7   & 50  & $376^3$  & $1.47\times 10^7$ & $10^{-0.5}$  & $0$\\
        L025m7   & 25  & $188^3$  & $1.47\times 10^7$ & $10^{-0.5}$  & $0$\\
\hline
    \end{tabular}
\end{table}
We demonstrate the consistency of our results across different resolutions in the same volume in Appendix~\ref{sec:app_resolution}, where we compare a subset of the dust scaling relations presented in the main text.

\subsection{Calibration}
The subgrid parameters controlling the SNe and active galactic nuclei (AGN) feedback were calibrated to reproduce the observed galaxy stellar mass function at $z=0$ from \cite{Driver2022} and the $z < 0.05$ galaxy size – stellar mass relation (SMR) from \cite{Hardwick2022}.
This was achieved using Gaussian process emulators, trained on $\sim 200$ L050m7 simulations, to fit observational data for stellar masses between $10^9$ and $10^{11.3}\, \Msun$.
Additional manual adjustments were performed on smaller volumes (L025m6 and L10.5m5) to ensure convergence in higher-resolution runs (m5 and m6). See \cite{Chaikin_colibre2025} for a detailed description of the calibration process.

\subsection{Subgrid model}
\subsubsection{Radiative heating and cooling, and chemistry}
Radiative heating and cooling rates are computed using the \textsc{Hybrid-Chimes} model \cite[]{Ploeckinger2025_hybridChimes}. This model combines the non-equilibrium heating and cooling rates for H and He with tabulated rates for metals, the latter of which are corrected using the non-equilibrium free electron density.
This hybrid approach captures the thermal history of gas, while maintaining computational efficiency.
The chemical network \textsc{Chimes} \cite[]{Richings2014_chimes_a,Richings2014_chimes_b} is used for both the non-equilibrium and the tabulated equilibrium species fractions.
Gas and dust cooling rates, dust photoelectric heating, dust shielding and chemical reactions involving dust grains (such as molecular hydrogen formation) use the dust species abundances, depletion of metals and grain sizes tracked within the simulation (see \S~\ref{sec:dustmodel}).
These calculations also account for the presence of the CMB, the UV/X-ray background from galaxies and quasars (modified version of the \citealt{UV_XrayBG2020} spectrum following \citealt{Ploeckinger2020}), an interstellar radiation field \cite[the `weak ISRF' model in][]{Ploeckinger2025_hybridChimes}, and interstellar cosmic rays.
Unlike previous large-volume cosmological simulations that imposed a pressure floor, gas is allowed to cool down to T $\approx 10$ K.

\subsubsection{Star formation, evolution and enrichment}\label{sec:sims.subgrid.stell_evo}
The prescription for star formation is described in \cite{Nobels2024_SF_recipe}. 
Gas particles are stochastically converted into stars if the gas is locally gravitationally unstable.
Eligible gas forms stars following a \cite{Schmidt1959} law with an efficiency per free-fall time of $\epsilon = 0.01$.

Each stellar particle represents a Simple Stellar Population (SSP) with a \cite{Chabrier2003} Initial Mass Function (IMF) spanning $0.1-100 \, \Msun$. 
Star particles at birth inherit the chemical composition of their parent gas. 
Mass loss is calculated by interpolating yield tables based on Zero Age Main Sequence (ZAMS) mass and metallicity.

The prescription for chemical enrichment and stellar mass loss is based on the work of \cite{Wiersma2009b}, tracking the 11 elements that are important for radiative cooling (H, He, C, N, O, Ne, Mg, Si, S, Ca, Fe).
The updates to the model are described in \cite{Correa25}, which include updated nucleosynthetic yields, an updated SNIa delay time distribution, the addition of r- and s-process elements (Sr, Ba, and Eu), and the inclusion of a new model for unresolved small-scale mixing by turbulent diffusion.




\subsubsection{Stellar feedback}\label{sec:sims.subgrid.stell_feedback}
Starting before the explosion of Core Collapse Supernovae (CCSNe), young stars exert feedback via stellar winds, radiation pressure, and growth of H\textsc{ii} regions \cite[]{Llambay2025_preSNe}.
Momentum flux from radiation and winds is derived from \texttt{BPASS} \cite[]{BPASS2.2.1} stellar population synthesis models ($0.1-100\, \Msun$) based on the age and metallicity of the SSP, 
which is implemented kinetically using stochastic kicks of 50 km s$^{-1}$. 
Galactic winds are primarily powered by CCSNe using a stochastic thermal feedback model to overcome numerical overcooling \cite[]{DallaVecchia2012}.
$10\%$ of the SNe energy is injected as low-velocity isotropic kicks ($\Delta v = 50$ km s$^{-1}$) to increase the turbulent velocity dispersion in the ISM \cite[]{Chaikin2023_thermal_kinetic_SNe_feed}.
To improve sampling while suppressing numerical overcooling, the heating temperature increment, $\Delta T_{\rm SN}$, increases with the local gas density \cite[equation 18 in][]{schaye_colibre2025}.
The CCSNe energy budget increases with the stellar birth pressure, $P_{\rm birth}$ \cite[equation 20 in][]{schaye_colibre2025} to account for residual overcooling in high-pressure environments. 

\subsubsection{Supermassive black holes and feedback}\label{sec:sims.subgrid.smbh}
Supermassive black holes (SMBHs) are seeded into haloes that exceed a threshold mass and lack an existing black hole \cite[for the exact values, see Table~1 in][]{schaye_colibre2025}.
BHs grow via gas accretion and mergers. The accretion rate follows the Bondi-Hoyle-Lyttleton (BHL) formula, modified to account for turbulence and vorticity \cite[]{Krumholz2006}.
Accretion is not capped at the Eddington limit ($\dot{m}_{\rm Edd}$), allowing for super-Eddington growth (up to $100\, \dot{m}_{\rm Edd}$) which has a significant impact on high-redshift evolution \cite[\eg][]{Husko2025,Chaikin2026_superedd_highz}.

\textbf{Thermal AGN feedback model}:
This is the fiducial model for AGN feedback in this work.
In this model, BHs accumulate energy to heat the nearest gas neighbour by a temperature increment $\Delta T_{\rm AGN}$ \cite[]{Booth2009}. 
$\Delta T_{\rm AGN}$ scales with the BH mass to improve time sampling across different masses, while suppressing numerical overcooling \cite[equation 38 in][]{schaye_colibre2025}.

\textbf{Hybrid AGN feedback model}:
The hybrid model introduces BH spin evolution and adds kinetic jets, as described in \cite{Husko2025_hybridAGN}. 
The accretion disc is modelled in three states based on the Eddington ratio ($f_{\rm Edd}$).
Feedback includes thermal energy (from winds/radiation) and kinetic jets, with jets implemented by kicking particles with a velocity that scales with the square root of the BH mass.

In Appendix~\ref{sec:app_hybrid} we explore briefly the difference between the thermal and hybrid AGN feedback model on dust properties. Generally, we find the differences in dust properties between the two AGN feedback models to be minimal.

\subsection{Dust}\label{sec:dustmodel}
\colibre\ tracks dust on a per particle basis, including its creation by stellar sources, its evolution through grain growth and destruction, grain size evolution and transport within the ISM. 
The dust model is described in detail in \cite{Trayford_colibre_dust2025}.
Here we summarise its key aspects.

\subsubsection{Dust grain types}
The model tracks six dust grain types: three dust species consisting of graphite (carbonaceous) and two silicate species of the olivine type having Mg (Magnesium, forsterite - Mg$_2$SiO$_4$) and Fe (Iron, fayalite - Fe$_2$SiO$_4$) as their end-member. Each species is represented in two size bins of radius centred at $0.01\,\um$ ($a_{\rm S}$, small grains) and $0.1\,\um$ ($a_{\rm L}$, large grains).
We assume spherical grains, which simplifies dust-evolution equations. 
Although spherical grains minimise the surface-to-volume ratio affecting gas–grain interactions, this effect is typically degenerate with empirically calibrated growth and destruction efficiencies. 
We do not explicitly model PAHs. 
PAH abundances may instead be inferred empirically relative to carbon depletion in small grains \cite[\eg][]{Hirashita2023}.

\subsubsection{Dust grain seeding}\label{sec:sims.grain_seeding}
\begin{table*}
 \caption{The composition and nucleation properties of the grain species used in the \colibre\ model. Graphite is modelled as homonuclear and silicates as heteronuclear (comprising fayalite, Fe$_2$SiO$_4$, and forsterite, Mg$_2$SiO$_4$). For the seeding calculations equipartition of the silicate sub-species is assumed; implying an effective molecule of FeMgSiO$_4$  (\S~\ref{sec:sims.grain_seeding}) and corresponding to a value of $A_{\rm G} = 171.853$ in equation~\ref{eq:nucleation}. This is the same as Table~1 in \protect\cite{Trayford_colibre_dust2025}.}
  \begin{tabular}{lllllllllll}
  \hline
  Species & Sub-species & Seed Composition  &  $A_{\rm G}$ & $\eta_{\rm CCSN}$ & $j$ & $A_j$ & $\tau_{\rm G}$/Myr\\ 
  \hline
  Carbonaceous & Graphite & C & 12.01 & 0.15 & C & 12.01 & 180 \\
  Silicate (Olivine) & Fayalite, Forsterite & FeMgSiO\textsubscript{4} & 171.85
  &  3.5$\times10^{-4}$ &  Si | Fe | Mg &  28.09 | 55.85 | 24.31 & 99.3\\
  \hline
 \end{tabular}
 \label{table:grains}
\end{table*}
Dust-phase metals are tracked by budgeting from the total metal yields and stellar mass return rates.
The dust yields are consistent with the \cite{Correa25} stellar nucleosynthetic yields (\S~\ref{sec:sims.subgrid.stell_evo}) so that elemental abundances are conserved.
For stars with initial mass in the range $1\,\Msun < {\rm M}_{\rm init} < 8\,\Msun$, the model uses AGB/SAGB dust yields from \cite{DellAgli2017_AGB_yields}.
Dust production is assigned only within this mass range. Carbon grains are associated with AGB stars of ${\rm M}_{\rm init} \leq 3.5\, \Msun$, where carbon-rich atmospheres form, while more massive AGBs produce silicates due to carbon being destroyed from hot-bottom burning.

CCSNe provide an additional dust formation channel, although yields remain uncertain because reverse shocks may destroy newly formed grains \cite[\eg][]{Nozawa2007,Priestley2022}. 
CCSN dust yields follow \cite{Zhukovska2008} nucleation efficiencies, applied to the model elemental yields. The mass of each grain species, $M_{\rm G}$ is calculated as
\begin{equation}
\label{eq:nucleation}
    M_{\rm G} = \eta_{\rm G} M_j\frac{A_{\rm G}}{A_j},
\end{equation}
where $M_j$ is the ejecta mass of the ``bottleneck element" $j$ (element that limits the number of grain molecules that can be produced), $\eta_{\rm G}$ the species condensation efficiency, $A_{\rm G}$ the grain molecular mass, and $A_j$ the atomic mass of element $j$. 
A fixed $1:1$ ratio of Fe- and Mg-endmember grains is assumed, yielding an effective seed molecule FeMgSiO$_4$. 
This choice has minimal impact because grain growth in the ISM dominates final dust compositions.

The dust injection model assumes an initial grain size distribution dominated by large grains (radius $a=0.1\, \um$), which is consistent with models of AGB winds \cite[\eg][]{Yasuda2012,Asano2013} and CCSNe \cite[\eg][]{Nozawa2007}.
Specifically, $90\%$ of the seeded dust mass is allocated to large grains and $10\%$ to small grains. 
The precise seed distribution has limited influence because grain growth, destruction, and mass transfer between grain sizes in the ISM regulate the final dust content. However, in the early Universe and pristine environments, where these ISM processes have not had sufficient time to dominate, the assumed stellar seeding dictates the dust grain sizes.
Understanding this primordial dust production is currently a subject of intense research, driven in part by recent \jwst\ discoveries of abundant, UV-bright galaxies at high redshifts \cite[also see][]{Narayanan2025}.

The model excludes Type Ia SNe as dust sources due to a lack of observed dust formation, and neglects AGN dust formation for simplicity.

\subsubsection{Grain growth}\label{sec:sims.grain_growth}
Once grains have nucleated in stellar ejecta, their subsequent evolution includes growth by accretion in the ISM. 
The grain growth model adopts the accretion timescale equations of \cite{Hirashita2014} with some modifications:
\begin{equation}
\label{eq:grain_growth}
\tau_{\rm acc} = \tau_{\rm G}
\left(\frac{r_g}{0.1 \text{\textmu m}}\right)
\left(\frac{\epsilon_{j,\odot}}{\epsilon_j}\right)
\left(\frac{10 \;{\rm cm}^{-3}}{n^{\prime}_{\rm H}}\right)
\left(\frac{10 \;{\rm K}}{T}\right)^{0.5}
\left(\frac{0.3}{S_{\rm acc}}\right),
\end{equation}
where $r_{\rm g}$ is the grain radius, $n^{\prime}_{\rm H}$ is the effective local number hydrogen density, $n^{\prime}_{\rm H} = \mathcal{C} n_{\rm H}$, where $\mathcal{C}$ is a clumping factor (defined later in this section), $T$ the gas temperature, 
S$_{\rm acc}=0.3$ the sticking probability, $\tau_{\rm G}$ the accretion timescale normalisation for a given grain species (see Table~\ref{table:grains}), $\epsilon_j$ and $\epsilon_{j,\odot}$ the local gas-phase and solar abundance (by number, relative to H) of the bottleneck element. 
Notably, grain growth is an order of magnitude faster for small grains compared to large grains, serving as the dominant pathway for growth of small grains.


For carbon grains, the bottleneck element is always C ($\epsilon_j = \epsilon_{\rm C}$), with the maximum depletion capped to 2/3 to account for carbon locked as CO in molecular clouds \cite[\eg][]{Fuente2019} similar to other dust modelling works \cite[\eg][]{Zhukovska2008,Vijayan_dust2019}. 
In contrast, the limiting element for silicates (composed of O, Mg, Si, and Fe) is determined by the maximum of $\epsilon_j / \epsilon_{j,\odot}$ across its constituent elements \cite[see equation~5 and 6 in][]{Trayford_colibre_dust2025}. 
After computing the total silicate accretion, the accreted mass is divided between Mg- and Fe-endmember species in proportion to their diffuse abundances.

Accretion is expected to dominate in dense gas environments \cite[\eg][]{Hirashita2000}. Therefore, properly capturing this process requires resolving molecular clouds.
The \colibre\ model has a multi-phase ISM and can thus attempt to track dust in both the molecular and the diffuse ISM phase. 
However, the \colibre\ simulations we analyse do not resolve the dense clouds where accretion is most efficient. Therefore, a clumping factor, $\mathcal{C}$, is introduced that boosts the density used in the accretion rate, $n^{\prime}_{\rm H} = \mathcal{C} n_{\rm H}$, where 
\begin{equation}\label{eq:clumping_factor}
  \mathcal{C}(n_{\rm H})=
\begin{cases}
1, & \text{if} \; n_{\rm H} \leq n_{\rm H,\, min} \\
\left(\frac{n_{\rm H}}{n_{\rm H,\, min}} \right)^m , & \text{if} \; n_{\rm H,\, min} < n_{\rm H} \leq n_{\rm H,\, max} \\
C_{\rm max}, & n_{\rm H} > n_{\rm H,\, max}
\end{cases}
\end{equation}
where $n_{\rm H, min}=0.1$ cm$^{-3}$ and $n_{\rm H, max}=100$ cm$^{-3}$ are the minimum and maximum densities of the clumping transition, $\mathcal{C}_{\rm max}=100$ is the maximum clumping value and $m = {\log_{10}(\mathcal{C}_{\rm max})/\log_{10}(n_{\rm H,\, max}/n_{\rm H,\, min})}=2/3$. 
This prescription reproduces observed ISM dust-to-metal ratios for $n_{\rm H} \gtrsim 0.1$ cm$^{-3}$ \cite[]{Trayford_colibre_dust2025}. Our results are insensitive to the value of $\mathcal{C}_{\rm max}$ for high density gas ($n_{\rm H} \gg 10$ cm$^{-3}$), provided it is $\gtrsim 10$ \cite[]{Trayford_colibre_dust2025}.

\subsubsection{Grain destruction}
Dust grains in the ISM can shrink or be destroyed through interactions with their surrounding media, which can be through the sputtering of grains, astration or SNe shocks. 

\textbf{Sputtering}:
Sputtering describes grain erosion through high-velocity collisions with gas particles in hot environments.
The model employs the \cite{Tsai1995} prescription, which gives the grain shrinkage rate  
\begin{equation}
    \frac{{\rm d} r_{\rm g}}{{\rm d} t} = -3.2\times 10^{-18} \; {\rm cm \; s^{-1}} \left(\frac{n_{\rm H}}{1\;{\rm cm^{-3}}} \right)
    \left[1 + \left(\frac{T}{T_0}\right)^{-2.5}\right]^{-1},
\end{equation}
where $T_0 = 2\times 10^6$~K.
The corresponding sputtering timescale is
\begin{equation}\label{eq:sputtering}
    \tau_{\rm sp} = 0.85 \; {\rm Myr} \left(\frac{r_{\rm g}}{0.1\;\text{\textmu m}} \right)\left(\frac{n_{\rm H}}{1\;{\rm cm^{-3}}} \right)^{-1}
    \left[1 + \left(\frac{T}{T_0}\right)^{-2.5}\right].
\end{equation}
Because of the steep temperature dependence, grains in gas with $T > T_0$ are rapidly destroyed. 
Because it is a diffuse-gas process, sputtering uses the unmodulated density $n_{\rm H}$ (as opposed to $n^{\prime}_{\rm H} = \mathcal{C} n_{\rm H}$).
Due to the linear dependence of the timescale on the grain-size, sputtering timescales are longer for large grains, in our case an order magnitude higher.

\textbf{Astration}:
Astration refers to grain destruction when dust-bearing gas is converted into stars. In \colibre, when a gas particle is converted into a stellar particle, its dust mass is reassigned to the particle’s diffuse elemental mass fractions based on its grain composition. These elemental fractions then define the chemical composition of the resulting star particle.

\textbf{Supernova shocks}:
Although SN shocks can be viewed as an extension of sputtering, shocks of individual SNe are unresolved in galaxy-scale simulations. 
In \colibre, feedback is stochastic; therefore, dust is assumed to be completely destroyed in any gas particle selected to receive SN thermal feedback. This is also the case for AGN feedback.
As with astration, the dust mass is redistributed into the diffuse elemental components and dust abundances in the affected particle are set to zero.

\subsubsection{Grain size evolution}\label{sec:dustmodel.grain_size_evo}
Dust mass can be conserved while the grain size distribution evolves through shattering and coagulation, both of which alter accretion and sputtering rates because of their dependence on grain size.

\textbf{Shattering}:
Energetic grain–grain collisions fragment large grains into many smaller ones, transferring mass from the large to the small size component. This process is implemented  following the \cite{Aoyama2017} formulation, modified by \cite{Granato2021}:
\begin{equation}\label{eq:shattering}
\tau_{\rm sh} = \tau_{\rm sh, 0} \; \left(\frac{a_{\rm L}}{0.1\;\text{\textmu m}} \right) \left(\frac{\mathcal{DTG}^{\prime}}{0.01}\right)^{-1} \times \begin{cases}
\left(\frac{n_{\rm H}}{{\rm cm^{-3}}}\right)^{-1} & \text{for } \frac{n_{\rm H}}{{\rm cm^{-3}}} < 1, \\ 
\left(\frac{n_{\rm H}}{{\rm cm^{-3}}}\right)^{-\frac{1}{3}} & \text{for } \frac{n_{\rm H}}{{\rm cm^{-3}}} \geq 1,
\end{cases}
\end{equation}
where $\tau_{\rm sh, 0} = 54.1\, {\rm Myr}$, and 
\dtg$^{\prime}$\ is the ratio of the dust mass to the total gas mass.
Because shattering is a diffuse-gas process, the calculation uses the unmodulated density $n_{\rm H}$.

\textbf{Coagulation}:
Low-velocity collisions allow small grains to merge into larger ones, transferring mass from small to large grain sizes. 
Following the prescription from \cite{Aoyama2017}, the coagulation timescale is 
\begin{equation}
\tau_{\rm co} = \tau_{\rm co,0} \; f_{\rm co} \left(\frac{a_S}{0.01\;\text{\textmu m}}\right) \left(\frac{\mathcal{DTG}^{\prime}}{0.01}\right)^{-1} \left(\frac{v_{\rm co}}{0.2 \; {\rm km \; s^{-1}}}\right)^{-1} \left(\frac{n^\prime_{\rm H}}{{\rm cm}^{-3}}\right)^{-1},
\label{eq:coag}
\end{equation}
with $\tau_{\rm co,0}=54.2$ Myr, and we assume a fixed coagulation velocity $v_{\rm co}=0.2$ km/s.
$f_{\rm co}$ is a resolution-dependent factor (see Table~\ref{tbl:simulations}) to improve convergence.
Coagulation depends on the boosted density, $n^{\prime}_{\rm H}$, which is modulated by the clumping factor (equation~\ref{eq:clumping_factor}).
The normalisation $\tau_{\rm co,0}$ remains uncertain, especially since non-spherical grains could enhance collision rates and accelerate growth.
This could in principle be calibrated using the observed distribution of small-to-large grains, which is however hard to constrain using current SED fitting techniques.
Comparing equations~\ref{eq:shattering} and \ref{eq:coag}, it is evident that coagulation will always dominate over shattering, while the rates will be similar only for $n_{\rm H} < 0.1\, {\rm cm}^{-3}$. 

\subsection{Galaxy identification}
The simulation snapshots are post-processed using three sequential algorithms to identify galaxies and clusters within each snapshot. 
First, a Friends-of-Friends (FoF) halo finder algorithm runs on dark matter  particles, linking particles separated by less than $0.2$ times the mean inter-particle distance. 
This resulting FoF catalogue is then processed by the subhalo finder \textsc{hbt-herons} \cite[][a modified version of the \textsc{hbt+} code, \citealt{Han2012,Han2018}]{HBT_Herons2025} to assign particles to self-bound substructures.
Finally, the \textsc{soap} tool \cite[]{SOAP2025} runs on the outputs from \textsc{hbt-herons} to calculate  (sub)halo properties across various apertures.

\subsection{Measurement of physical quantities}
In this work, galaxy properties are measured using bound particles within a 3D aperture of 50 proper kpc (pkpc) centred on the subhalo.
This aperture is 
consistent with previous \colibre\ works. We show the variation of dust mass as a function of stellar mass for different aperture sizes in Appendix~\ref{sec:app.aperture}, finding minimal difference in the dust mass for apertures $>3$~pkpc, well within the limits of typical observational apertures. 
For the analyses in the main text, we only select galaxies with stellar mass, M$_{\star} > 10$ times the mean initial baryonic particle mass at each resolution \ie\ corresponding to $> 10$ star particles for a galaxy. This also mitigates sampling noise and prevents single or rare enrichment events from disproportionately skewing the dust and metallicity measurements. 
In Appendix~\ref{sec:app_resolution}, where we test convergence across resolutions, we do not enforce a particle cut, and demonstrate its effect.
For the simulated dust masses, we consider the total dust mass summed over all gas phases. However, we show in \S~\ref{sec:gas_phase} that the choice of gas-phase selection significantly influences the measured dust mass.

Gas masses are measured by summing up the gas mass in atomic and molecular hydrogen (neutral gas), multiplied by 1.36 to account for helium contribution. 
Although dust can exist outside the atomic and molecular gas phases, the majority of the dust mass resides in the neutral phase across cosmic time (\eg\ $>60\%$ at $z=0$ and $>80\%$ for $z>2$).
We adopt this definition because observations estimate gas masses from inferred atomic and molecular gas measurements (also see \S~\ref{sec:obs_data}).
We compute the metal mass fraction in the galaxy as the mass fraction of metals (including dust) in the cool and dense gas, defined as gas with temperatures below $10^{4.5}$ K and densities n$_{\rm H}>0.1\, {\rm cm}^{-3}$. 
For metallicity, we use the mass-weighted number density of oxygen (does not include the value depleted onto dust) to hydrogen in cool, dense gas:
\begin{equation}
    {\rm O/H} = \frac{1}{\Sigma_i m_i} \sum_i m_i \frac{n_{\mathrm{O},i}}{n_{\mathrm{H},i}}.
\end{equation}
Here $m_i$, $n_{\mathrm{O},i}$, and $n_{\mathrm{H},i}$ are the particle mass, number density of oxygen, and number density of hydrogen, respectively for particle $i$.
This definition is well suited for comparison with observational metallicities, which are derived from both absorption lines along Gamma Ray Burst (GRB) or quasar sight lines and emission lines from H\textsc{ii} regions. 

There are multiple ways to define these quantities that may not directly align with the heterogeneous set of observations (see \S~\ref{sec:obs_data}) we compare to, where the gas mass or metallicity (or metal mass) are measured in different ways. In \S~\ref{sec:gas_phase}, we examine the impact of different gas-phase selection criteria on the dust scaling relations presented in this work. We find that specific choices significantly affect the normalisation of these relations.

\section{Observational data used in this work}\label{sec:obs_data}
In Table~\ref{table:obs_data} we describe the observational data used to compare the dust scaling relations explored in this work. For those wishing to skip the data-specific details, the main results begin in \S~\ref{sec:dustscaling}. 
However, we recommend referring back to this section for necessary context and caveats for the observational comparisons in subsequent sections.
\begin{table*}
 \caption{Observational data used in this work. Columns provide the reference for the data, redshift (z), derived quantity, and relevant survey/technical details, respectively.}
  \begin{tabular}{llll}
  \hline
   Reference & Redshift & Derived quantity & Notes \\
  \hline
  \cite{Algera2025} & $6 < z < 8$ & \multirow{3}{3cm}{Dust mass, \dtg, \dtm, metallicity, stellar mass} & \multirow{4}{9cm}{Gas masses from [C\textsc{ii}]158\um\, (via metallicity-dependent scaling), dust masses from dust continuum (MBB$^{1}$ fit), metallicities from strong-line calibration, stellar masses from \textsc{Bagpipes}$^{2}$ SED fit, galaxy data from REBELS \cite[]{Bouwens2022} survey}\\
  & & & \\
  & & & \\
  & & & \\

  \cite{Beeston2024} & $0 \le z \le 0.45$ & \multirow{2}{3cm}{Dust mass function, cosmic dust mass density} & \textsc{Magphys}$^{3}$ SED fit to H-ATLAS \cite[]{Ward2022} galaxies\\
  & & & \\

  \cite{Berta2025} & $0.6 < z \le 7.2$ & \multirow{2}{3cm}{Dust mass function, cosmic dust mass density} & \textsc{Magphys} and \textsc{Sed3fit}$^{4}$ SED fit to N2CLS \cite[]{Bing2023} galaxies\\
  & & & \\

  \cite{Chiang2025} & $0.1 \le z \le 3.85$ & Cosmic dust mass density & \multirow{2}{9cm}{Cross correlates Cosmic Infrared Background (intensity maps from \textit{Planck}, \textit{IRAS}, \textit{Herschel}) anisotropies with spectroscopic galaxies and quasars}\\
  & & & \\

  \cite{Clemens2013} & $z=0$ & Dust mass function & \textsc{Magphys} SED fit to WISE, \textit{Spitzer}, \textit{IRAS} and \textit{Herschel} data \\

  \cite{Davies2017} & $z \approx 0$ & Dust mass, stellar mass & \textsc{Cigale}$^{5}$ SED fit to DustPedia \cite[]{Davies2017} galaxies\\

  \cite{Driver2018} & $0.02 \le z \le 1.75$ & Cosmic dust mass density & \multirow{3}{9cm}{\textsc{Magphys} SED fit to GAMA \cite[]{Driver2011,Liske2015}, G10-COSMOS \cite[]{Davies2015,Andrews2017}, 3D-HST \cite[]{Momcheva2016} data}  \\
  & & & \\
  & & & \\

  \cite{DSilva2026} & $0 < z \lesssim 3$ & Cosmic dust mass density & \textsc{ProSpect}$^{6}$ SED fit to GAMA, DEVILS \cite[]{Davies2025} data\\

  \cite{Dunne2011} & $0.05 \le z \le 2.5$ & \multirow{2}{3cm}{Dust mass function, cosmic dust mass density} & MBB fit to H-ATLAS \cite[]{HATLAS2010} galaxies\\
  & & & \\

  \cite{daCunha2015} & $2 \le z \le 6$ & Dust mass, stellar mass & \textsc{Magphys} SED fit to ALESS \cite[]{Hodge2013} sub-mm galaxies\\

  \cite{DeCia2016} & $2 < z < 3$ & \dtm, metallicity & \multirow{2}{9cm}{Quasar sightlines to derive neutral hydrogen weighted dust depletion and absorption-line metallicities}\\
  & & & \\

  \cite{Eales2009} & $z\approx1$ & Dust mass function & MBB fit to BLAST survey \cite[]{BLAST2009} data\\

  \cite{Ealse2024} & $0.28 \le z \le 5.13$ & Cosmic dust mass density & \multirow{2}{9cm}{MBB fit to stacked samples from COSMOS \cite[]{Davidzon2017,Simpson2017}}\\
  & & & \\

  \cite{Fudamoto2024} & $z \approx 11$ & Dust mass, stellar mass & \multirow{2}{9cm}{Stellar mass from \textsc{Prospector}$^{7}$ SED fit to \jwst\ data, dust mass upper limit from MBB fit to NOEMA data}\\
  & & & \\

  \cite{Gillman2024} & $0.5 \le z \le 4$ & Dust mass, stellar mass & \multirow{3}{9cm}{SMGs: AS2UDS \cite[]{AS2UDS2019} covered by \jwst/NIRCam, field galaxies: K-band selected (matched redshift distribution and similar mass to the SMGs) from \cite{Dudzeviciute2020}, both fit with \textsc{Magphys}}\\ 
  & & & \\
  & & & \\

  \cite{Heintz2023} & $1.7 < z < 6.3$ & \dtg, \dtm, metallicity & \multirow{2}{9cm}{GRB$^{8}$ sightlines to derive neutral hydrogen weighted dust depletion and absorption-line metallicity}\\
  & & & \\

  \cite{Heintz2025} & $z \approx 7$ & \multirow{2}{3.1cm}{Dust mass, \dtg, \dtm, metallicity, stellar mass} & \multirow{2}{9cm}{ALMA \& \jwst/NIRSpec observation of single galaxy, similar method as \cite{Algera2025}}\\ 
  & & & \\

  \multirow{2}{2cm}{\cite{Konstantopoulou2024}} & $0.6 < z < 6.3$ & \dtg, \dtm, metallicity & Same as \cite{Heintz2023}\\
  & & & \\

  \cite{Laporte2017} & $z \approx 8.3$ & Dust mass, stellar mass & \textsc{Magphys} SED fit to a Lyman-break galaxy\\
  
  \cite{Lee2024} & $0.5 \le z \le 4$ & Dust mass, stellar mass & \multirow{2}{9cm}{Compilation of passive galaxies from their work and previous works (see their Table B1) fit with MBB}\\
  & & & \\

  \cite{Mancini2015} & $7 \le z \le 8$ & Dust mass, stellar mass & MBB fit to ALMA and PdBI observations, mostly upper limits on dust mass\\

  \cite{Manning2025} & $2 \le z \le 6$ & Dust mass, stellar mass & \textsc{Cigale} SED fit to SCUBADive \cite[]{McKinney2025} sub-mm galaxies\\

  \cite{Magnelli2020} & $0.3 \le z \le 5.5$ & Cosmic dust mass density & MBB fit to ASPECS \cite[]{Walter2016} data\\

  \multirow{2}{2cm}{\cite{Mitsuhashi2025}} & $z \approx 12, 14$ & Dust mass, stellar mass & Dust mass (upper limits) from MBB fit, stellar mass from \textsc{Bagpipes} SED fit\\
  & & & \\

  \multirow{2}{2cm}{\cite{Peroux&Howk2020}} & $0 \le z \le 5$ & \multirow{2}{3.1cm}{\dtg, \dtm, cosmic dust mass density} & Quasar and GRB sightlines, same as \cite{DeCia2016,Heintz2023}\\
  & & & \\

  \cite{Pozzi2020} & $0.1 \le z \le 2.5$ & Cosmic dust mass density & MBB fit to \textit{Herschel} PEP survey \cite[]{Lutz2011} data\\

  \cite{Relano2022} & $z=0$ & \multirow{2}{3cm}{Dust mass, \stl\ grain mass ratio, stellar mass} & \multirow{4}{9cm}{SED fitting with dust SED templates for different grain species and sizes, galaxy data from DGS \cite[]{Madden_2013}, HiGH \cite[][]{DeVis2017}, HRS \cite[]{HRS2010}, JINGLE \cite[]{JINGLE2018}, and KINGFISH \cite[]{Kennicutt_2011} surveys} \\
  & & & \\
  & & & \\
  & & & \\

  \multirow{2}{2cm}{\cite{RemyRuyer2014}} & $z=0$ & \multirow{2}{3.1cm}{Dust mass, \dtg, \dtm, metallicity, stellar mass} & \multirow{4}{9cm}{Gas masses from summing the H$\textsc{i}$ and H$_{2}$ (via a CO-to-H$_{2}$ conversion factor) gas masses (multiplied by 1.36 to account for helium), metallicities from strong-line calibrations, and dust mass and stellar mass from SED fitting, galaxy data from DGS and KINGFISH surveys}\\
   & & & \\
   & & & \\
   & & & \\

   \cite{Santini2014} & $1 \le z \le 3$ 7 & Dust mass, stellar mass & \multirow{2}{9cm}{SED fitting using \cite{DraineLi2007} templates on COSMOS, GOODS-S, GOODS-N fields with \textit{Herschel} data}\\
   & & & \\

   \cite{Traina2024} & $0.5 < z \le 6$ & \multirow{2}{3.1cm}{Dust mass function, cosmic dust mass density} & \multirow{2}{9cm}{\textsc{Cigale} SED fit using \cite{DraineLi2007} templates to A$^3$COSMOS \cite[]{A3COSOS2019} data}\\
   & & & \\

   \cite{watson2015} & $z \approx 7.5$ & Dust mass, stellar mass & Dust mass from MBB fit, and custom SED fit for stellar mass for a single galaxy\\

   \cite{Wiseman2017} & $2 < z < 5$ & \dtm, metallicity & Similar method as \cite{Heintz2023}\\
   \hline
   \\
 \end{tabular}
 \label{table:obs_data}
  \raggedright
  \textbf{Notes:} $^{1}$Modified blackbody, $^{2}$\cite{Bagpipes2018,Bagpipes2019}, $^{3}$\cite{magphys}, $^{4}$\cite{SED3FIT2013}, $^{5}$\cite{cigale2019}, $^{6}$\cite{Prospect2020}, $^{7}$\cite{Prospector2021}, $^{8}$Gamma-ray bursts
\end{table*}

\subsection{Caveats and uncertainties on the derived physical properties}\label{sec:obs_data.caveats}
Comparing simulated and observational data requires caution due to several systematic uncertainties inherent in converting raw fluxes into physical quantities (dust, gas, and metal masses). These methods rely heavily on various empirical relations and simplistic assumptions that can introduce biases.

\subsubsection{Assumptions on dust temperature and opacity coefficient}
Majority of the high-redshift observational datasets used in this work rely on modified blackbody (MBB) fitting to derive dust masses. Usually, the dust mass is derived assuming an optically-thin dust and single dust temperature for the FIR SED using:
\begin{equation}\label{eq:mdust}
    M_{\rm dust} = L_{\nu} / (4\pi\, B_{\nu}(T_{\rm dust})\, \kappa_{\nu}),
\end{equation}
where $M_{\rm dust}$ is the dust mass, $L_{\nu}$ is the monochromatic luminosity at frequency $\nu$, $B_{\nu}$ is the Planck function, $T_{\rm dust}$ is the dust temperature, and $\kappa_{\nu}$ is the dust mass opacity coefficient. 
At FIR wavelengths ($\lambda \gtrsim 50\, \um$), $\kappa_{\nu}$ is parametrised as, $\kappa_{\nu} = \kappa_{0}(\nu/\nu_{0})^{\beta}$,
where $\kappa_{0}$ and $\beta$ (dust emissivity index) are dependent on the underlying dust properties such as grain composition and size distribution \cite[\eg][]{Draine2003}.
These properties may vary significantly among different galaxy populations as well as spatially within a single galaxy.
Hence, calculating the dust mass from equation~\ref{eq:mdust} requires explicit assumptions regarding $\kappa_{0}$, $T_{\rm dust}$ and $\beta$. 
Most studies fix $\kappa_{\nu}$ based on known grain compositions \cite[\eg][]{Dunne2011,Clemens2013,watson2015}. 
Studies such as \cite{Clemens2013} have also shown that $T_{\rm d}$ and $\beta$ are highly degenerate when performing a modified blackbody fit; consequently, estimating these parameters through fitting is non-trivial.

The assumptions on optical depth also play a critical role: invoking an optically-thick dust model generally yields higher inferred dust temperatures compared to optically-thin models \cite[\eg][]{Liang2019,Vijayan2022,Sommovigo2025}, subsequently leading to lower derived dust masses.
Consequently, applying local Milky Way-like dust properties \cite[\eg][]{DraineLi2007} and dust opacity assumptions to heterogeneous galaxy populations introduces unconstrained systematic shifts in the measured dust mass, resulting in typical uncertainties of $0.5 - 1$ dex \cite[\eg][]{Sommovigo2025}.

\subsubsection{Assumptions in SED fitting}
Multi-wavelength SED fitting codes like \textsc{Cigale}, \textsc{Magphys}, or \textsc{ProSpect} employ dust energy balance technique \cite[\eg][]{Walcher2011}, where the energy emitted in the IR corresponds to the energy absorbed by dust in the UV-optical. 
While powerful, the assumption of a global energy balance may break down in galaxies \cite[\eg][]{Saftly2015,Lower2022,Qin2022,Faucher2023}. 
To model the dust SED, these codes must assume both an attenuation curve and empirical dust emission templates \cite[\eg][]{DraineLi2007}. 
Crucially, these assumptions are inherently dependent on the underlying galaxy dust properties, leading to systematic uncertainties in dust mass estimates of $0.5-0.7$ dex \cite[\eg][]{Decleir2026}.

\subsubsection{Difficulty in measuring gas and metal mass}
Local observations measure the total gas mass by detecting molecular gas using low-J CO transitions and atomic gas using 21cm H\textsc{i} observations. 
At higher redshifts, these specific transitions are largely inaccessible with current observational facilities due to severe surface brightness dimming, atmospheric/ionospheric constraints, and extreme sensitivity limitations.
Therefore, high-redshift studies \cite[\eg][]{Heintz2025,Algera2025} rely on detections of [C\textsc{ii}]158\um\ to estimate total atomic and molecular gas mass.
However, estimating gas masses via either [C\textsc{ii}]158\um\ or CO observations remains inherently limited by significant uncertainties in their respective conversion factors—namely $\alpha_{\text{C II}}$ and $\alpha_{\text{CO}}$—both of which depend strongly on local metallicity and environmental conditions.
These factors are known to vary by nearly an order of magnitude between galaxy environments and for different metallicities \cite[\eg][]{Bolatto2013,Madden2020}.

Indirect metal mass derivations propagate errors directly from absorption/emission line gas-phase metallicity calibrations, which are then compounded by uncertainties in the gas mass estimates following the relation M$_{Z} = {\rm M}_{\rm gas} \times Z$. 
The use of different calibrations can introduce systematic offsets up to $\sim 0.7$ dex in the oxygen abundance \cite[\eg][]{Kewley2019}, fundamentally altering the inferred gas-phase metal budget. 
Furthermore, converting a numerical oxygen abundance to a mass metallicity fraction ($Z$) inherently depends on the assumed abundance pattern, which can vary across galaxies and cosmic time.
Studies such as \cite{Schady2024} have also shown that emission line metallicities correlate with those inferred from absorption systems, but not in a one-to-one manner.

Standard quasar sightline samples are heavily biased toward low-density, diffuse environments, often tracing the metal-poor outskirts or circumgalactic medium (CGM) of galaxies \cite[\eg][]{Arabsalmani2023}. This may lead to systemically lower inferred \dtg\ and \dtm\ ratios compared to emission-based measurements. 
Conversely, GRB sightlines sample the dense, inner regions of galaxies where dust growth via accretion is highly efficient. This may result in systematically higher dust depletion factors and heightened \dtm\ ratios that maybe unrepresentative of the global average of the host galaxy. Furthermore, the intense UV radiation field of the GRB itself can destroy dust grains and ionise gas within the immediate vicinity of the explosion site, altering the localised \dtg\ and \dtm\ properties before they can be reliably measured \cite[\eg][]{Ledoux2009}.

In short, the observationally derived dust properties are highly uncertain, with expected systematic errors of up to $1$~dex.

We also investigate the effect of the chosen gas phase on these dust properties in \S~\ref{sec:gas_phase}, where we show that the normalisation of the scaling relations is sensitive to the selected gas phase.

\section{Dust Scaling Relations}\label{sec:dustscaling}
In \S~\ref{sec:dustscaling.dtg}-\ref{sec:dustscaling.dtg_dtm_mstar}, we quantify the dust content of galaxies using two fundamental dust scaling relations, the 
dust-to-gas (\dtg\footnote{Note that this is different from dust-to-gas ratio usage in \S~\ref{sec:dustmodel.grain_size_evo}, where the ratio is measured with respect to the total gas mass of the particle. Here the denominator is the sum of the atomic and molecular gas mass.}) 
ratio and the dust-to-metal (\dtm) ratio. 
While the \dtg\ ratio tracks the total dust budget that influences the ISM chemistry, the \dtm\ ratio isolates the efficiency of dust grain production, growth, and destruction relative to metal production.
In \S~\ref{sec:dustscaling.dmass_smass}, we explore the evolution of dust mass as a function of stellar mass, and quantify dust grain evolution in \S~\ref{sec:dustscaling.grain_evo}, focussing on dust grain species and sizes.

\subsection{Dust-to-gas ratio}\label{sec:dustscaling.dtg}
\begin{figure*}
    \centering
    \includegraphics[width=\textwidth]{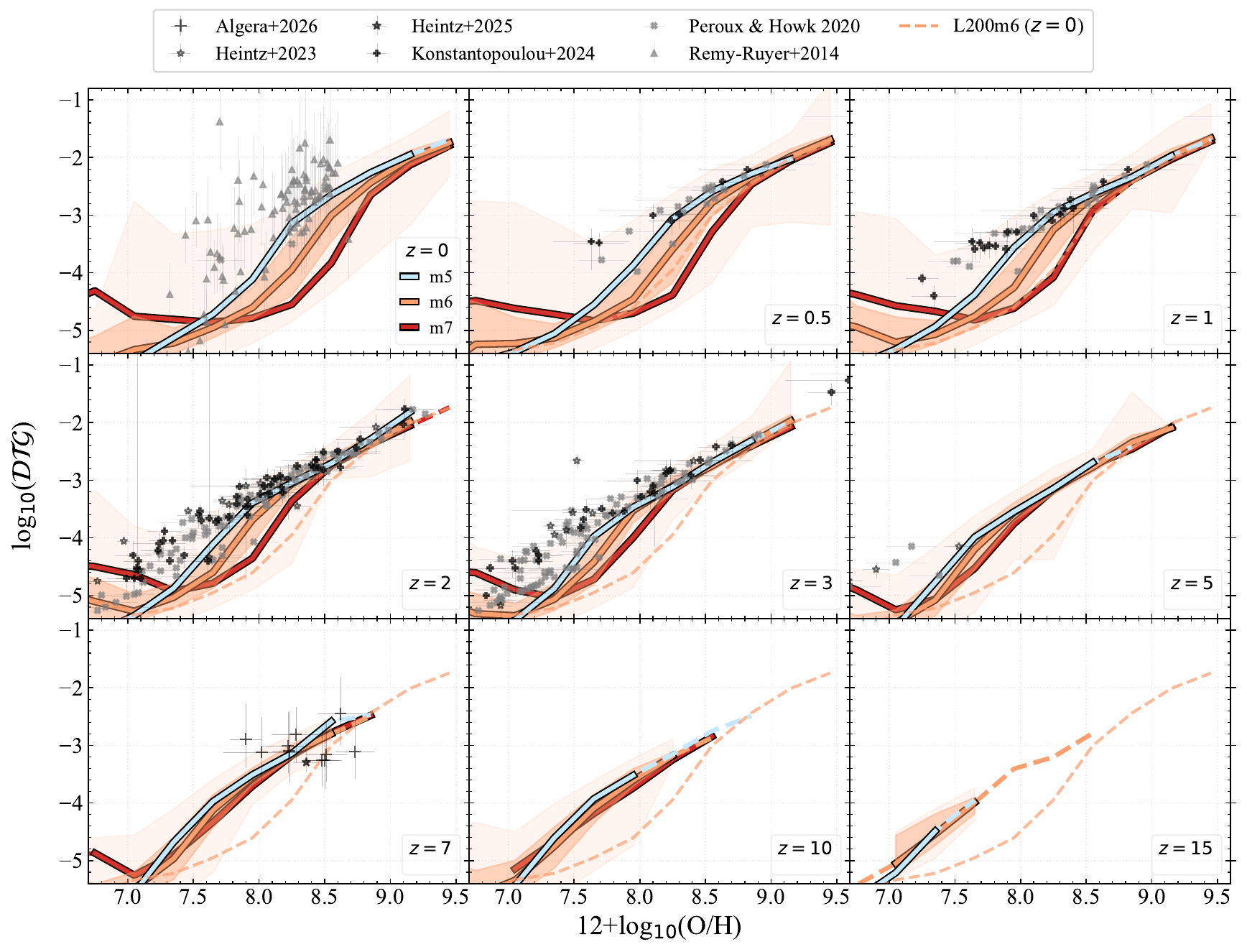}
    \caption{The dust-to-gas (DTG) ratio of galaxies in the \colibre\ m5 (L025m5 or L050m5 or L100m5 in light-blue), m6 (L200m6 in orange), and m7 (L400m7 in red) simulations as a function of their gas phase metallicity for $z \in [0,15]$. The solid line shows the median result, with the darker and lighter shaded regions
    denoting the $16^{\rm th} - 84^{\rm th}$ ($1\sigma$) and $99.865^{\rm th}$ - $0.135^{\rm th}$ ($3\sigma$) percentile spread. The median relation for bins with less than 5 galaxies are plotted as dashed lines.
    The data points with errorbar show observations from \protect\cite{RemyRuyer2014}, \protect\cite{Peroux&Howk2020}, \protect\cite{Heintz2023,Heintz2025}, \protect\cite{Konstantopoulou2024}, and \protect\cite{Algera2025}. In $z>0$ panels, we overplot the $z=0$ median relation from the L200m6 box (lighter dashed line). The different resolutions show excellent convergence across the metallicity range for $z \gtrsim 5$, with convergence degrading towards lower redshifts and achieved at progressively higher metallicities with decreasing resolution. The higher resolution m5 simulations show better agreement with data compared to lower resolutions. Note that for the m5 resolution, we switch from L100m5 simulation to L50m5 for $z<2$, and to L25m5 for $z<1$.
    }
    \label{fig:dgr_Z_z0_9}
\end{figure*}
The \dtg\ ratio is widely used to estimate gas masses from FIR dust continuum measurements, often assuming a constant value ($\approx 0.01$ as estimated for the Milky Way) or a metallicity-dependent scaling \cite[\eg][]{Magdis2011,RemyRuyer2014}. Its widespread use arises from the fact that dust-continuum emission is often detected even when traditional gas tracers such as CO or [C\textsc{ii}]$158$\um\ are undetected or unavailable.

Figure~\ref{fig:dgr_Z_z0_9} presents the evolution of the galaxy-integrated \dtg\ ratio as a function of the gas-phase metallicity for galaxies in \colibre\ over the redshift range, $z\in[0,15]$. 
We plot the median relations derived for the different resolutions in bins with a width of $0.3$~dex. 
The shaded region indicates the percentile spread for the L200m6 volume: the darker region shows the $16^{\rm th} - 84^{\rm th}$ ($1\sigma$) percentile, while the lighter region covers the $0.135^{\rm th} - 99.865^{\rm th}$ ($3\sigma$) percentile. 
To highlight evolutionary trends, we overlay the median relation corresponding to $z = 0$ from the L200m6 box at all higher redshifts. 
Figure~\ref{fig:dtg_met_median} in Appendix~\ref{sec:app.median_scaling} shows the median \dtg-metallicity evolution from $z=0$ to $z=15$ for all resolutions in a single panel, facilitating a direct comparison of the redshift evolution.

Overall, our simulations reveal an approximately linear correlation between the \dtg\ ratio and metallicity, with the ratio increasing as galaxies become more metal-rich. 
While this trend holds across all resolutions, the precise quantitative value of this relation is resolution-dependent. 
Specifically, convergence among resolutions is achieved at progressively higher metallicities with decreasing resolution, which is reached at metallicities corresponding to the depletion of available bottleneck elements onto dust grains.
Conversely, the lack of convergence at low- to intermediate- metallicities highlights the increasing efficiency of dust grain growth at higher resolutions which can resolve higher densities in the ISM. 
Since progressively higher resolutions better resolve clumping, the clumping factor (equation~\ref{eq:clumping_factor}) may require re-calibration at each resolution to ensure convergence across the entire metallicity range.
With increasing redshift, convergence shifts toward progressively lower metallicities. 
At $z \ge 4$, for metallicities $12+{\rm log}_{10}({\rm O/H})\gtrsim 7$, all the resolutions exhibit excellent convergence. We attribute this to the higher fraction of dense-gas present in the early universe, enabling earlier saturation of dust growth.

At $z \le 2$, the m6 and m7 \colibre\ simulations produce a median relation with a characteristic sigmoid or "S-shape" (as referred to in literature), similar to that seen in the \dtm-metallicity relation (Figure~\ref{fig:DTM_vs_Z}). This is also seen in dust models implemented in other cosmological simulations like those in \cite{Popping_dust2017,Vijayan_dust2019,Triani_dust2020} (SAMs) and \cite{Hou2019,Li2019,Graziani2020} (hydrodynamical simulations). The presence of such a shape at these metallicities is resolution-dependent, since the m5 resolution does not exhibit this shape.
However, we note that this characteristic shape naturally arises from distinct metallicity regimes characterised by dust mass maintained through stellar production, grain growth, and regions of saturated dust growth (also see \S~\ref{sec:dustscaling.dtm}).
This also results in a steeper slope at low-metallicities ($12+{\rm log}_{10}({\rm [O/H]})\lesssim 8$ for m5, and progressively higher metallicity limits for lower resolutions) compared to the high-metallicity end.

The median \dtg-metallicity relation exhibits a mild redshift evolution from $z=0$ to $2$, showing a gradual increase in the ratio at intermediate metallicities ($7.5 \lesssim 12+{\rm log}_{10}({\rm [O/H]})\lesssim 8.5$ for m5), with the increase more pronounced for lower resolutions (see Figure~\ref{fig:dtg_met_median}).
This trend is the result of two key physical factors: the dominance of dust grain growth within this metallicity range, coupled with a strong evolution of the gas mass that is in dense gas in this redshift range.
Beyond $z=2$, the dense gas fraction barely evolves for this metallicity range (not shown), leading to a lack of evolution in the \dtg\ ratio.

We compare the simulated relations to observational datasets (see Table~\ref{table:obs_data} for details) from 
\citet[][$z=0$]{RemyRuyer2014}, \citet[][$0 \le z < 5$]{Peroux&Howk2020}, \citet[][$1.7 < z < 6.3$]{Heintz2023}, \citet[][$0.6 < z < 6.3$]{Konstantopoulou2024}, \citet[][$z \approx 7$]{Heintz2025}, and \citet[][$6 < z < 8$]{Algera2025}. There are no observational constraints on the \dtg\ ratio beyond $z>7$.

Overall, the \colibre\ model reproduces observed trends across all redshifts, with the higher-resolution m5 simulations providing improved agreement at every epoch. 
At $z=0$, most of the \cite{RemyRuyer2014} data agree with the median m5 results within large observational uncertainties ($\sim 1$ dex). However, the observed median shows a higher normalisation compared to the \colibre\ median for $12+{\rm log}_{10}({\rm [O/H]}) > 7.5$.  
For $z>0$ (excluding the \citealt{Algera2025} sample at $z=7$), the observational data follow an approximately linear relationship with a slope similar to that predicted by \colibre. 
Specifically, in the redshift range $0.5 \le z \le 3$, the m5 simulations provide excellent agreement with the observations.
At $z=2$ and $z=3$, where the observations cover the low-metallicity end, the observed median is higher compared to the \colibre\ median ($\sim 0.5$ dex with respect to m5), although this deviation is still within the $3\sigma$ spread.
Furthermore, while the observations in this regime show a steepening of the relation, this is less pronounced than in the simulations.
Finally, at $z=7$, the \cite{Algera2025} data exhibit constant \dtg\ with metallicity; however, the \colibre\ median remains within the observational uncertainties.
As discussed in \S~\ref{sec:obs_data.caveats}, systematic uncertainties on derived quantities can be up to one dex.

\subsection{Dust-to-metal ratio}\label{sec:dustscaling.dtm}
\begin{figure*}
    \centering
    \includegraphics[width=\textwidth]{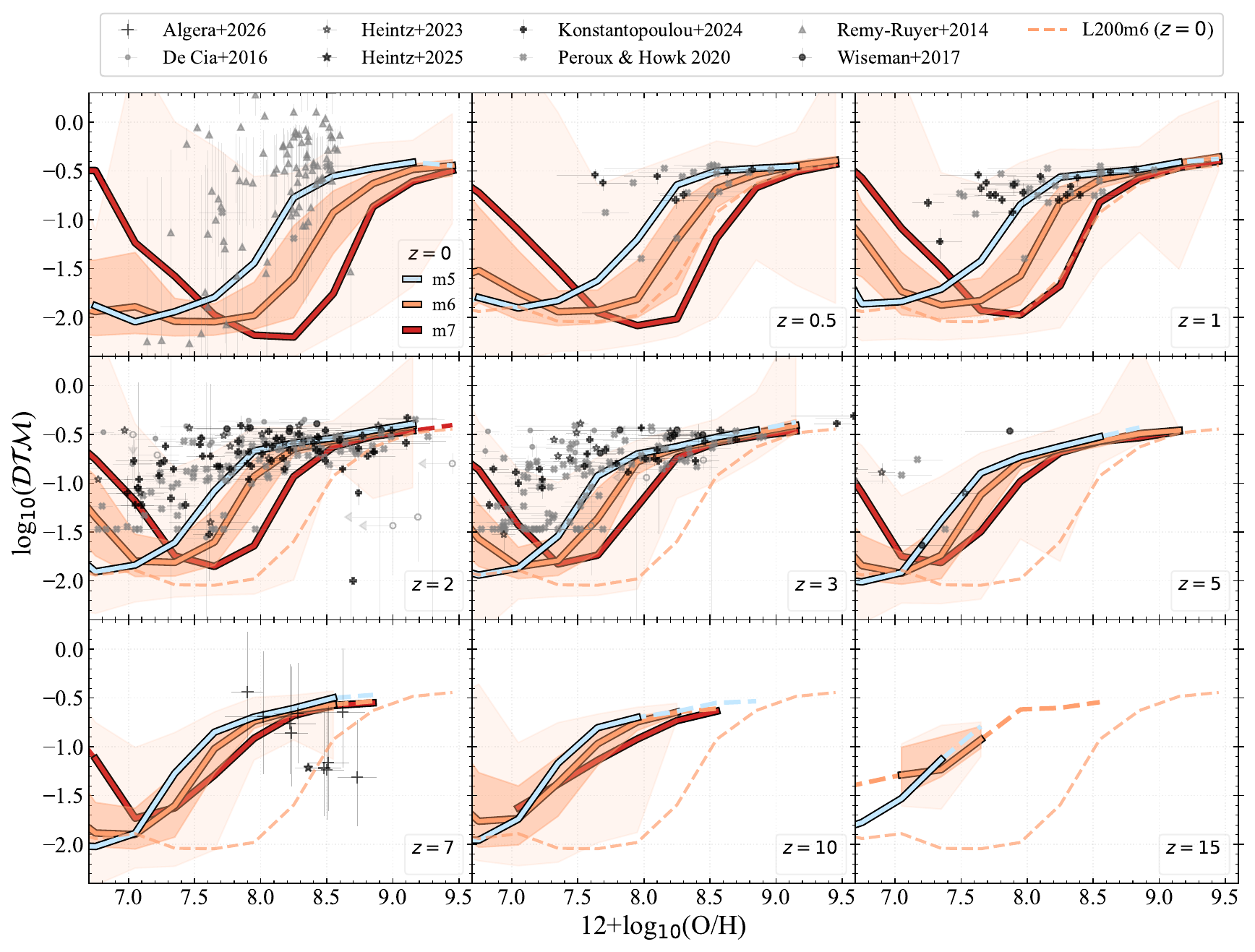}
    \caption{The dust-to-metal (\dtm) ratio of galaxies in the \colibre\ m5 (L025m5 or L050m5 or L100m5), m6 (L200m6), and m7 (L400m7) simulations (solid light-blue, orange, and red curves representing each resolution, respectively) as a function of their gas phase metallicity for $z \in [0,15]$. The solid line shows the median result (dashed line for bins with fewer than 5 galaxies), with the darker and lighter shaded regions
    denoting the $1\sigma$ and $3\sigma$ spread. 
    We show observations from \protect\cite{RemyRuyer2015A}, \protect\cite{DeCia2016}, \protect\cite{Wiseman2017}, \protect\cite{Peroux&Howk2020}, \protect\cite{Heintz2023,Heintz2025}, \protect\cite{Konstantopoulou2024} and \protect\cite{Algera2025}, with upper limits on the metallicity denoted by arrows.
    In $z>0$ panels, we overplot the $z=0$ median relation from the L200m6 box (lighter dashed line). The different resolutions show convergence at the high-metallicity end, with convergence achieved at progressively higher metallicities with decreasing resolution. The m5 simulations show better agreement with data compared to lower resolutions. Note that for the m5 resolution, we switch from L100m5 simulation to L50m5 for $z<2$, and to L25m5 for $z<1$.}\label{fig:DTM_vs_Z}
\end{figure*}
The dust-to-metal (\dtm) ratio serves as a direct tracer of the balance between dust seeding, grain growth efficiency, and destruction. 
We define the \dtm\ ratio here as the total dust mass divided by the total metal mass in atomic and molecular gas. Here, the metal mass also includes the fraction already locked in dust.

Figure~\ref{fig:DTM_vs_Z} shows the \dtm\ ratio as a function of the gas-phase metallicity. 
The median relation (using bins with a width of $0.3$~dex) for the simulation is shown by the solid line, with the shaded regions indicating the $1\sigma$ and $3\sigma$ spread for the L200m6 simulation.
Also shown is the median relation from L200m6 at $z = 0$ at all higher redshifts to highlight the evolution. 
The median \dtm-metallicity evolution from $z=0$ to $z=15$ is shown for individual resolutions in the same panel in Figure~\ref{fig:dtm_met_median} to allow for easier comparison of redshift trends.

The median \dtm-metallicity relation traces a characteristic sigmoid or "S-shape" curve at all redshifts (though this deviates at low metallicities for the m7 simulation, as discussed later). This feature is common in models where grain growth timescales are inversely proportional to metallicity \cite[\eg][]{Popping_dust2017,Li2019,Triani_dust2020}. Physically, the shape reflects a transition in dust production mechanisms:
\begin{itemize}
    \item Low metallicity: Dust production is dominated by stellar injection, with the resulting \dtm\ ratio reflecting the initial input abundance.
    \item Intermediate metallicity: Grain growth in the dense ISM dominates, causing a sharp rise in the \dtm\ ratio.
    \item High metallicity: The \dtm\ ratio plateaus at a value of $\approx 0.3$ (i.e. log$_{10}$(\dtm) $ \approx -0.5$). This saturation reflects an equilibrium between dust grain growth and destruction, alongside total depletion of key refractory elements onto dust, limiting further growth.
\end{itemize}
The precise location of these transition regimes is sensitive to resolution, with higher resolutions shifting the onset toward lower metallicities. 
This occurs because grain growth timescales also depend inversely on gas density, and higher resolution simulations better resolve higher densities in the ISM.

The \dtm\ saturation value is primarily governed by the depletion of carbon and oxygen onto dust, since they dominate over Si, Mg or Fe within the total metal mass fraction \cite[][]{Asplund2009}. The maximum depletion of carbon is capped to $2/3$ in the model to account for carbon locked in CO. Oxygen depletion is stoichiometrically limited by the availability of Mg and Fe. Hence the saturation fraction in the model is crucially determined by the cap we introduced for carbon. 
Section~3.8 in \cite{Trayford_colibre_dust2025} shows the depletion of O, Fe, Mg and Si to be $\approx 25\%$, $\approx 3\%$, $\approx 3\%$, and $\approx 10\%$, respectively. Assuming solar metal mass fractions from \cite{Asplund2009}, this corresponds to a \dtm\ ratio of $\approx 0.3$, similar to the median saturation value reached in the model.
Several other simulations that model dust evolution also show a saturation in the \dtm\ ratio around $0.3-0.4$ at high-metallicity \cite[\eg][]{Vijayan_dust2019,Osman2025}.
This value is also similar to what is typically assumed in simulations that do not model dust and assume a constant \dtm\ ratio \cite[\eg][]{Trayford2017,Ma2019}.

In the metallicity regime where the \dtm\ ratio has not converged, it increases significantly at fixed metallicity from $z = 0$ to $2$, with weak evolution for $z>2$ (also see Figure~\ref{fig:dtm_met_median}), both of which are also seen in the \dtg\ ratio. The evolution from $z = 0$ to $2$ is also stronger for lower resolution runs in this metallicity regime. 

The median relation for different resolutions show good convergence in the high metallicity regime where the \dtm\ ratio has saturated. The metallicity at which convergence occurs shifts to progressively higher metallicity with decreasing resolution. The convergence between resolutions also improve with increasing redshift, similar to the \dtg-metallicity relation. 
At lower metallicities where the simulations have not fully converged, the higher resolution runs exhibit higher \dtm\ values at fixed metallicity. This behaviour is expected, as this metallicity range corresponds to the transition regime where grain growth in the ISM is the dominant dust production channel.
Since the grain growth timescale depends inversely on gas density, higher-resolution simulations which can resolve higher densities facilitate more efficient growth. 
Therefore, the characteristic upturn in the \dtm\ ratio shifts to lower metallicities with increasing resolution.

The m6 and m7 simulations display a rise in the \dtm\ ratio at the low-metallicity end. 
The exact metallicity is dependent on both resolution and redshift: it tends to be higher for lower resolutions, and decreases with increasing redshift ($12+{\rm log}_{10}({\rm O/H})\lesssim 7.25$ for m6 at $z=0$, but higher for m7).
This behaviour can be attributed to our choice to use the dust mass calculated across all gas phases, while the metal mass is restricted only to the neutral gas phase. 
Thus, at low metallicities, where the neutral gas mass fraction is inherently small, much of the dust produced via stellar injection exists outside the measurable neutral phase, thereby elevating the calculated \dtm\ ratio.

The \dtm-metallicity relation exhibits a $1\sigma$ scatter of approximately $0.5$ dex. However, the $3 \sigma$ scatter is considerably larger ($\gtrsim 1$ dex), primarily driven by passive galaxies. 
While some of these passive galaxies still possess significant dust reservoirs, their neutral gas content is negligible, leading to a low calculated metal mass according to our definition. Galaxies falling below the relation's median demonstrate a relatively low dust content compared to their measured metal mass.

We also overplot observational data (see Table~\ref{table:obs_data} for details) from 
\citet[][$z=0$]{RemyRuyer2014}, \citet[][$0 \le z \le 5$]{Peroux&Howk2020}, \citet[][$0.6 < z < 6.3$]{Konstantopoulou2024}, \citet[][$2<z<3$]{DeCia2016}, \citet[][$2<z<5$]{Wiseman2017}, \citet[][$1.7 < z < 6.3$]{Heintz2023}, \citet[][$z \approx 7$]{Heintz2025} and \citet[][$6 < z < 8$]{Algera2025}.
For metallicities above $12+{\rm log}_{10}({\rm [O/H]})\gtrsim 8$, the \dtm\ ratio is nearly constant ($\dtm \approx 0.3$).
At lower metallicities, the ratio shows a gradual decline with decreasing metallicity. 
The simulations generally mirror this trend, though the low-metallicity slope is steeper in the models, and the transition metallicity remains resolution-dependent. 
Similar to the \dtg\ ratio, the m5 resolution provides the best agreement with observations across all redshifts.
While the general simulation data overlaps well with observation at all epochs, detailed examination reveals a few differences.
The $1 \sigma$ scatter in the simulated data is smaller than that of the scatter in the observational data. However, we caution against over-interpreting the difference in scatter, as the observational errors are significant ($\sim 1$ dex for the \dtm\ ratio) and encompass diverse measurement techniques, which may enhance the model differences.
Additionally the observations are flux-limited, while the simulations are volume-limited.
Hence the two scatters cannot be directly compared.
At $z=7$, and for $12+{\rm log}_{10}({\rm O/H})\gtrsim 8$, the observed \dtm\ ratio decreases with increasing metallicity, while the median relation in \colibre\ remains almost constant, though these \dtm\ observations carry considerable observational ($\sim 0.5$ dex) and systematic (upto $1$~dex, see \S~\ref{sec:obs_data.caveats}) uncertainties.
Currently, there are no observational constraints on the \dtm\ ratio for $z>7$.

To our knowledge, \colibre\ is the first simulation to model dust evolution within a multi-phase ISM and coupled to the ISM chemistry in a cosmological volume run to $z=0$.
Hence, quantitative comparison to previous cosmological hydrodynamical simulations \cite[e.g.][]{Hou2019,Li2019} or semi-analytical models \cite[e.g.][]{Popping_dust2017,Vijayan_dust2019,Triani_dust2020} is difficult, due to the quantities compared not being exactly the same. 
Generally, several models \cite[\eg][]{Popping_dust2017,Li2019,Triani_dust2020} predict a `critical metallicity' across all redshifts above which the \dtm\ ratio increases rapidly with metallicity, indicating efficient dust grain growth in the ISM. 
This results in little evolution of the \dtm-metallicity relation with redshift. 
\cite{Vijayan_dust2019} adopt a grain growth timescale that is inversely dependent on the dust mass in dense gas, showing gradual \dtm\ evolution that only saturates at low redshifts.
\cite{Hou2019} restrict grain growth to dense gas and assume a fixed dense gas fraction of $0.1$, thus restricting the efficiency of grain growth at high redshift, while showing strong evolution towards higher values at lower redshift.
\citet[][\textsc{SIMBA} simulations]{Li2019} do not restrict grain growth to the dense gas, similar to the one implemented here, predicting little redshift evolution for the median \dtm\ ratio.
In contrast, \colibre\ shows higher \dtm\ ratios in the intermediate-metallicity range ($7.5 \lesssim 12+{\rm log}_{10}({\rm O/H}) \lesssim 8.5$) at high redshift, and demonstrate a strong redshift evolution of this transition regime from $z=0$ to $2$. 
This behaviour is driven by the multi-phase ISM model, which naturally predicts denser gas at higher redshifts, effectively pushing the transition threshold toward lower metallicities. 
Beyond $z=2$, the median \dtm\ ratio barely evolves with redshift, similar to the findings from the SIMBA simulations.

\subsection{Trends of \dtg\ and \dtm\ with stellar mass} \label{sec:dustscaling.dtg_dtm_mstar}
\begin{figure*}
    \includegraphics[width=\textwidth]{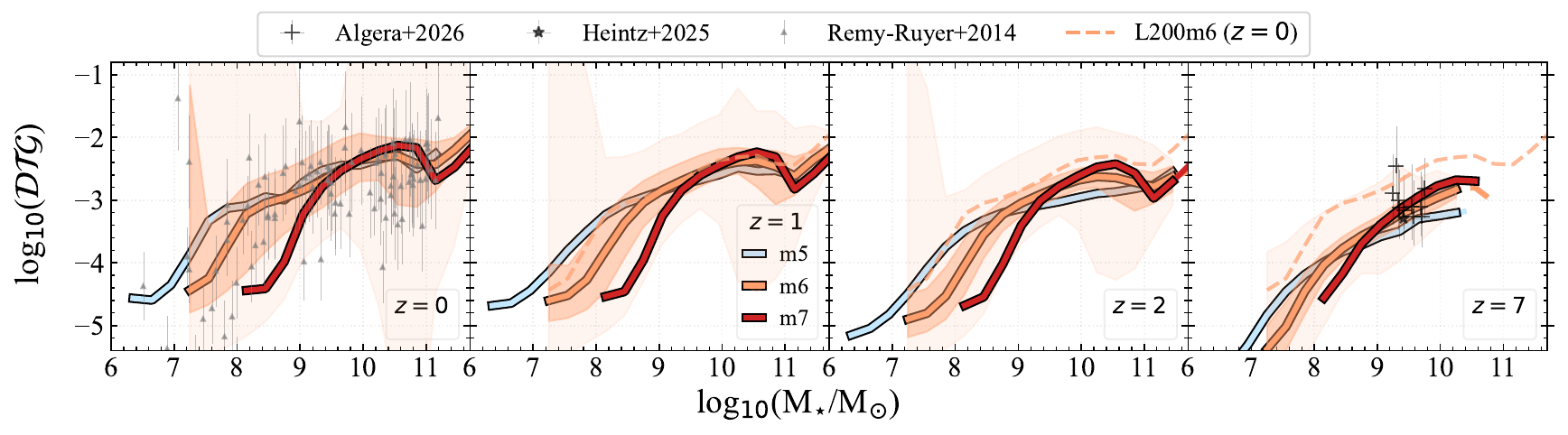}
    \includegraphics[width=\textwidth]{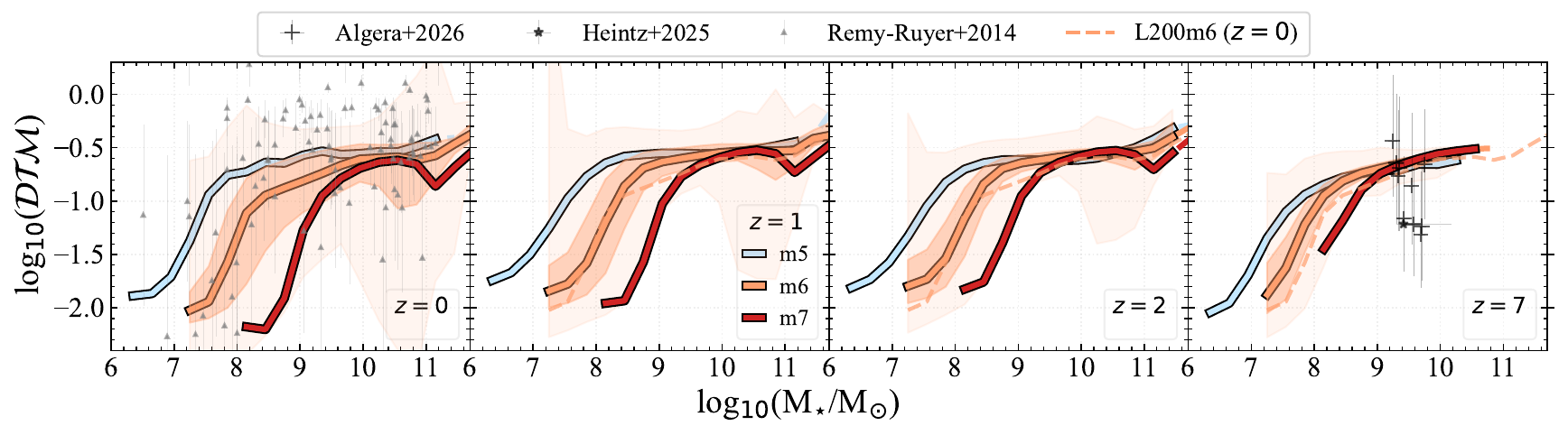}
    \caption{\dtg\ (top panel) and \dtm\ (bottom panel) ratio as a function of stellar mass for $z=0, 1, 2, {\rm and}\, 7$ of galaxies in the COLIBRE m5 (L025m5 or L050m5 or L100m5), m6 (L200m6), and m7 (L400m7) simulations. The solid lines show the median relation (solid light-blue, orange, and red curves representing m5, m6, and m7, respectively), with the darker and lighter shaded region showing the $1 \sigma$ and $3 \sigma$ spread. The median relation for bins with less than 5 galaxies are plotted as dashed lines. 
    We plot observational data from \protect\cite{RemyRuyer2014,Heintz2025} and \protect\cite{Algera2025}.
    In $z>0$ panels, we overplot the $z=0$ median relation from the L200m6 box (lighter dashed line). The different resolutions show convergence at the high stellar mass end (M$_{\star} \gtrsim 10^{9}\, \Msun$ for m5 and m6), with convergence achieved at progressively higher stellar mass with decreasing resolution. Note that for the m5 resolution, we switch from L100m5 simulation to L50m5 for $z=1$, and to L25m5 for $z=0$.
    }
    \label{fig:dtg_dtm_mstar_z0_7}
\end{figure*}
Figure~\ref{fig:dtg_dtm_mstar_z0_7} illustrates the median \dtg\ (top panel) and \dtm\ (bottom panel) as a function of stellar mass (using bins with a width of $0.3$~dex). 
The median \dtg-stellar mass relation shows weak evolution from $z = 0\, {\rm to}\, 2$, beyond which it hardly evolves. 
This contrasts with the \dtm-stellar mass relation, which barely evolves with redshift.
Both relations feature a low $1\sigma$ scatter ($\lesssim 0.5$ dex), while the $3\sigma$ scatter is large ($\gtrsim 1$ dex) at both the high- and low-mass ends, primarily driven by the presence of quiescent galaxies, as discussed earlier. 
A secondary driver of this scatter is our definition of the \dtg\ (and \dtm) ratio: the total dust mass (which includes dust residing outside the neutral phase) against only neutral gas mass (metal mass in neutral gas). 

Low-mass (M$_{\star} < 10^9\, \Msun$ for m6) quiescent galaxies exhibit elevated \dtg\ and \dtm\ ratios due to their depleted gas reservoirs, whereas actively star-forming low-mass galaxies lie below the median (not shown) because they contain a significant reservoir of neutral gas, while still building up their dust reservoir, similar to the trend seen for the gas mass -- metallicity relation \cite[\eg][]{Yates2012,DeRossi2017,Torrey2019}.
At $z \ge 2$ the median \dtg\ ratio falls well below the local relation, indicating that high-redshift galaxies in the model host gas reservoirs that are dilute in dust in comparison to their $z=0$ counterparts. 
The \dtg\ ratio generally increases with increasing stellar mass, while the \dtm-M$_{\star}$ relation is approximately flat for M$_{\star} = 10^{8}\, \Msun$ to $10^{11}\, \Msun$, showing negligible variation, even across redshift. 
This demonstrates that the stellar mass exerts a minimal role in anchoring the values of the \dtm\ ratio in the model, with the gas-phase metallicity acting as the dominant driver. 
This behaviour is consistent with results from other cosmological simulations that implement dust evolution, such as the semi-analytical models of \cite{Popping_dust2017,Triani_dust2020}, though we note that in \cite{Vijayan_dust2019}, the median \dtm\ ratio at fixed stellar mass decreases with increasing redshift.

The \dtg\ and \dtm\ ratios show a rise at low stellar mass that is resolution dependent, $\sim 10^7\, \Msun, \sim 10^8\, \Msun,\, {\rm and}\, \sim 10^{9}\, \Msun\,$ for m5, m6, and m7 respectively. 
The \dtg\ ratio is approximately converged at high stellar masses (except at the highest stellar masses, M$_{\star}\sim 10^{11}\, \Msun$, due to different passive galaxy fractions), with the convergence shifting to higher stellar masses for progressively lower resolutions (m5 and m6 converged for $\gtrsim 10^{9}\, \Msun$, while M$_{\star} \gtrsim 10^{10}\, \Msun$ for m6 and m7, at $z=0$). This is also seen for the \dtm-stellar mass relationship, which shows a tighter convergence across resolutions (bottom panel). In Figure~\ref{fig:dgr_Z_z0_9}, we saw that for high-metallicities ($12+{\rm log}_{10}([{\rm O/H}]) \gtrsim 8.5$) the \dtg\ and \dtm\ ratios are converged. \cite{schaye_colibre2025} showed that higher resolution \colibre\ simulations predict systematically lower metallicities for fixed stellar mass (their Figure 20). This leads to lower resolution simulations having slightly higher \dtg\ ratios ($\gtrsim 0.2$ dex compared to the immediate higher resolution) at the highest masses ($\ge 10^{10}\, \Msun$) where the \dtg\- and \dtm -metallicity relations are converged.

We compare our predicted relations to observations from  \cite[][$z=0$]{RemyRuyer2014}, \citet[][$z=7$]{Heintz2025}; and \cite[][$z=7$]{Algera2025}. 
The bulk of the observations for the data shown in Figure~\ref{fig:dgr_Z_z0_9} and \ref{fig:DTM_vs_Z} at $z=1\, {\rm and}\, z=2$ do not quote stellar masses, hence observational comparisons are only plotted for $z=0\, {\rm and}\, 7$.
At $z=0$, the model underpredicts the \cite{RemyRuyer2014} observations for M$_{\star} \lesssim 10^8\, \Msun$ ($\sim 0.5$ dex for \dtg, $\sim 1$ dex for \dtm, in case of m5).
The scatter exhibited by the \cite{RemyRuyer2014} \dtg\ and \dtm\ data is large, $\ge 1$ dex, with similar uncertainties on the measurement. 
At $z=7$, the \dtm\ (\dtg) ratio is within the uncertainties of the observations from \cite{Heintz2025,Algera2025}. 
Additional observations across the full redshift range will be necessary to fully evaluate the model’s performance in this parameter space.

\subsection{Dust mass - stellar mass relation}\label{sec:dustscaling.dmass_smass}
\begin{figure*}
    \centering
    \includegraphics[width=\textwidth]{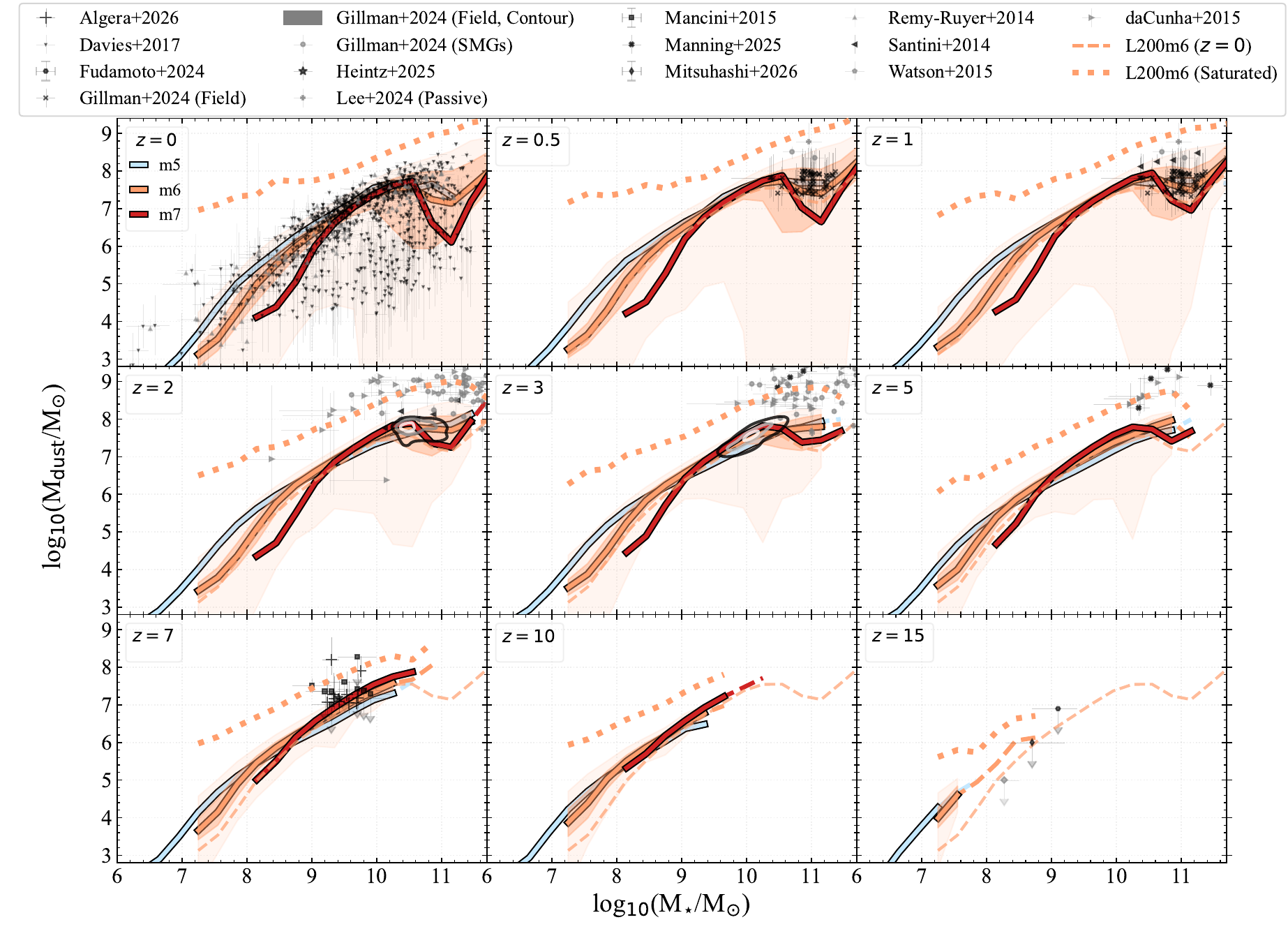}
    \caption{The stellar mass-dust mass relation for galaxies in the m5 (L025m5 or L050m5 or L100m5), m6 (L200m6), and m7 (L400m7) \colibre\ simulations with the solid light-blue, orange, and red curves representing the median each resolution for redshifts in the range $z \in [0,15]$. The darker and lighter shaded region shows the $1\sigma$ and $3\sigma$ spread. The observational constraints from \protect\cite{RemyRuyer2014,Santini2014,daCunha2015,Mancini2015,watson2015,Davies2017,Laporte2017}; \protect\citet[][gray and black for field galaxies and SMGs respectively]{Gillman2024}; \protect\cite{Lee2024,Fudamoto2024,Heintz2025,Manning2025,Mitsuhashi2025,Schouws2025,Algera2025} are shown as scatter points with errorbars. At $z \ge 2$, we plot the \protect\cite{Gillman2024} field galaxy data as contours ($16, 50\, \rm{and}\, 84$ percentiles), due to the large number of data points in a small region (mean errors on the stellar mass and dust mass are $\sim 1$ dex and $0.5$ dex respectively, similar to the lower redshift).
    In $z>0$ panels, the lighter orange dashed line shows the $z=0$ median relation from the L200m6 simulation. The thick orange dotted line shows the maximum mass of metals in the  atomic and molecular gas (also the maximum expected dust mass at saturation) in the L200m6 box in different stellar mass bins.
    The simulated median barely evolves with redshift. The simulations agree with the bulk of the observational data, however, it does not capture the extreme dust masses observed in bright sub-mm galaxies.
    Note that for the m5 resolution, we switch from L100m5 simulation to L50m5 for $z<2$, and to L25m5 for $z<1$.
    }
    \label{fig:dmsm_z0_10}
\end{figure*}
We plot the dust mass - stellar mass relation in Figure~\ref{fig:dmsm_z0_10} for $z \in [0,15]$, along with observational data from various studies. 
The median relation (using bins with a width of $0.3$~dex) for the m5, m6, and m7 simulations is shown by coloured solid lines. The shaded regions indicate the $1\sigma$ and $3\sigma$ spread for L200m6 simulation.
Also shown is the median relation from L200m6 at $z = 0$ at all higher redshifts to highlight the evolution. The dotted line shows the total mass of metals in the atomic and molecular gas within the L200m6 box for different stellar mass bins. 
Figure~\ref{fig:mdust_mstar_median} shows the median dust mass - stellar mass evolution from $z=0$ to $15$ for all resolutions in a single panel, facilitating a direct comparison of the redshift evolution.

The different resolutions converge at the high stellar mass end, corresponding to a minimum of about $100$ star particles in the lower-resolution simulations (\eg\ $10^8\, \Msun$ for m5 and m6).
The convergence slightly worsens at $z>1$ in this stellar mass regime, with the lower resolutions exhibiting higher dust masses,  due to the resolution dependence of the stellar mass - metallicity relation in \colibre.
At the highest stellar masses (M$_{\star} \gtrsim 10^{11}\, \Msun$) for $z < 6$, lower resolution simulations show a stronger dip in the median dust mass compared to higher resolutions. This is driven by a higher quiescent fraction in the lower resolution run \cite[see][]{Chandro-Gomez2025}.

We overplot observational data (see Table~\ref{table:obs_data} for details) from \citet[][$z=0$]{RemyRuyer2014}, \citet[][$1 \le z \le 3$]{Santini2014}, \citet[][$2 \le z \le 6$]{daCunha2015}, \citet[][$7 \le z \le 8$]{Mancini2015}, \citet[][$z \sim 7.5$]{watson2015}, \citet[][$z \sim 0$]{Davies2017}, \citet[][$z \sim 8.3$]{Laporte2017}, \citet[][$0.5 \le z \le 4$]{Gillman2024}, \citet[][$0.5 \le z \le 4$]{Lee2024}, \citet[][$z \approx 11$]{Fudamoto2024}, \citet[][$z \approx 7$]{Heintz2025}, \citet[][$2 \le z \le 6$]{Manning2025}, \citet[][$6 < z < 8$]{Algera2025}, and \citet[][$z \approx 12, 14$]{Mitsuhashi2025}.

Notably, at $z=0$, the median \colibre\ relation matches the observational data quite well \cite[also see Section~7.2.5 in][]{schaye_colibre2025}. However, in detail, there are galaxy populations in \cite{Davies2017} that are significantly dust-poor compared to the median value ($<2-3$ dex) in the simulations. This spread to lower values at high stellar masses is due to the presence of elliptical early-type galaxies (ETGs) with lower dust and gas content in the \cite{Davies2017} sample. 
The simulations reproduce these galaxies within the $3\sigma$ region. 
We also note a lack of high-dust-mass galaxies at ${\rm M}_{\star} < 10^{8}\, \Msun$ in our simulations compared to some observational samples. This suggests that dust destruction may be over-efficient in the simulation for these lower-mass systems.
This could also be attributed to observational selection biases favouring  IR-bright galaxies, and hence high dust masses.
In \S~\ref{sec:dustscaling.dmass_smass.ms_md_sSFR}, we explore the median dust mass - stellar mass relation in sSFR bins.

At $0.5 \leq z \leq 4$, the \colibre\ galaxies are representative of the bulk of the main-sequence population, reproducing the galaxy population from \citet[][their field galaxy population]{Gillman2024} and \cite{Santini2014}. However, \colibre\ struggles to reproduce the large dust masses estimated from the heavily dust-obscured population of high-redshift ($z \in [2, 4]$) sub-millimeter galaxies (SMGs, or dusty star-forming galaxies) studied in \cite{daCunha2015,Gillman2024,Manning2025}, also seen in several other cosmological simulations that model dust \cite[\eg][]{Li2019,Vijayan_dust2019,Triani_dust2020}.
These objects represent some of the most extreme starbursts with significantly high dust reservoirs at their respective epochs. They are typically selected as bright sources in single-dish (\eg\ JCMT/SCUBA) pointings and subsequently followed up with ALMA or NOEMA. Thus, they are inherently biased toward the most dust-rich systems. 
Also high-resolution imaging often resolves single-dish sources into multiple distinct galaxies \cite[\eg][]{Hodge2013}. Furthermore, as these are some of the rarest and most FIR-luminous objects in the Universe, it is possible that the $(200\, {\rm cMpc})^3$ and $(400\, {\rm cMpc})^3$ volumes of \colibre\ are too small to sample the extreme peaks of the density field where such rare SMGs reside. 
For instance, \cite{Kumar2025} demonstrated that while the smaller-volume $(300\text{ cMpc})^3$ TNG simulations \cite[]{TNG2018} underpredict SMG number counts, the $(1\text{ cGpc})^3$ FLAMINGO simulations \cite[]{Flamingo2023} successfully reproduce them, demonstrating the effect of cosmic variance.

\cite{Gillman2024} also found that applying a standard dust surface density (using dust masses from SED fitting and half-light radii) to  A$_{\rm V}$ (attenuation in the V-band) relation, one obtains A$_{\rm V}>100$, which would imply fully obscured stars in the V-band. 
The detection of light at shorter wavelength than the infrared in these objects therefore implies a patchy star-dust geometry rather than the well-mixed geometry assumed in standard SED fitting. This could lower the dust mass estimates and thus bring the values closer to the simulated ones. 
This can be tested using \colibre, which will be explored in future work by forward modelling the galaxies.

Another factor that can overestimate the dust mass of high-redshift far-IR luminous objects is errors in the estimated photometric redshifts of dusty galaxies \cite[\eg][]{Jin2024,McKay2026}, due to the negative k-correction in the FIR combined with degeneracies between the photo-z solution and the FIR emission line detection. Galaxies that have been incorrectly placed at higher redshifts, will result in a dramatic over-estimation of their dust masses. Also see \S~\ref{sec:obs_data.caveats} for a brief discussion on dust mass measurement uncertainties.

We overplot the maximum achievable dust content in the L200m6 run for $z \in [0, 15]$ by assuming that $100\%$ of the metals in the neutral gas phase have condensed into dust. While this scenario is unphysical, as galaxies always maintain a significant gas-phase metal fraction, it serves as an upper bound. Yet, even under this extreme assumption, the simulated maximum dust masses remain roughly $0.5$ dex below those of the most extreme observed SMGs at $2 \le z \le 6$. Reproducing such extreme systems in the simulation would require much higher dense gas fractions and elevated metallicities, driven by alternative, highly efficient early enrichment channels \cite[]{Gall2018}.

These trends have thus motivated various theoretical \cite[\eg][]{Jones2023,Sommovigo2025} and observational \cite[\eg][]{Schreiber2018,Viero2022} studies to investigate whether dust masses could be systematically overestimated. 
Many studies using simulations \cite[\eg][]{Liang2019,Parente2026} and observations \cite[\eg][]{Schreiber2018,Sommovigo2022} have shown that the dust temperature rises with increasing redshift. 
For instance, increasing the dust temperature by $10$ K, say from $20$ K $\to$ $30$ K reduces the dust mass by a factor of $\sim 10$, for the same infrared luminosity, $L_{\rm IR}$, given\footnote{$L_{\rm IR}$ can be obtained by integrating equation~\ref{eq:mdust} for $\int_{0}^{\infty} L_{\nu} {\rm d}\nu$.} that $M_{\rm dust} \propto L_{\rm IR} T_{\rm dust}^{-(4+\beta)}$ for optically thin dust (assuming $\beta=2$).

At $5 \le z \le 7$, dust mass measurements remain scarce. This scarcity is due to the observational challenges of detecting the dust continuum, which requires prohibitively long integration times with interferometers like ALMA, while single-dish telescopes are inherently biased toward bright, rare sub-mm galaxies.
As expected, the sub-mm galaxies in \cite{daCunha2015,Manning2025} exhibit high dust masses ($\ge 1$ dex above the median), while the bulk of the star forming galaxies in \cite{Mancini2015,Heintz2025,Algera2025} are in good agreement with the \colibre\ median.
Interestingly, the two galaxies that are most discrepant from the median relation ($\gtrsim 0.5$ dex) in the \cite{Algera2025} sample are those with multi-band ALMA measurements. For these two galaxies, the dust temperature is measured to be $T_{\rm d} = 30-35$ K, whereas a higher temperature of $T_{\rm d} = 40-45$ K was assumed for the rest of the sample. This lower dust temperature subsequently increases the estimated dust mass by $\sim 0.4$ dex.

Spectroscopic observations have detected galaxies at redshifts as high as $z \sim 14$ \cite[\eg][]{Naidu2025}. Their spectra reveal metal emission lines in the UV, indicating that their ISM has already undergone metal enrichment. It is therefore likely that these galaxies also contain dust. Significant efforts have been undertaken by the community to follow up these extremely high-redshift, spectroscopically confirmed sources using FIR facilities. These campaigns have produced mixed results: while FIR metal emission lines have been detected, no dust-continuum emission has yet been observed. Even in the absence of dust-continuum detections, these observations provide upper limits on the dust masses of these galaxies under the assumption of a dust temperature and slope of the Rayleigh-Jeans tail \cite[\eg][]{Fudamoto2024,Mitsuhashi2025}. 
These upper limits are only marginally consistent with our median dust mass - stellar mass relation, with the upper limit at $z = 12$ is offset by $\approx 0.5$ dex to a lower dust mass, indicating that dust grain growth in \colibre\ galaxies might be overly efficient \cite[see also the discussion on the \colibre\ high-redshift UV luminosity function in][]{Lu2026_highzuvlf}. 

Another notable feature is that the simulated median relation shows very little redshift evolution, albeit with a slight, monotonic increase toward higher redshifts (less than $0.5$~dex from $z=0$ to $15$, also see Figure~\ref{fig:mdust_mstar_median}).
This trend aligns with the results of several semi-analytical models \cite[\eg][]{Popping_dust2017,Triani_dust2020}.
However, observational constraints show competing scenarios. 
For instance, \cite{Jolly2025}, which stacks galaxies from the ALMA Lensing Cluster Survey \cite[ALCS,][]{ALCS2023}, find that at fixed stellar mass, dust mass generally decreases with increasing redshift, a trend that directly contrasts with our \colibre\ predictions.
Conversely, \cite{Casey2026}, stacks galaxies from the COSMOS-Web Survey \cite[]{Cosmosweb}, combining \jwst\ imaging with FIR data, finds that at fixed stellar mass, the dust mass increases by nearly an order of magnitude from $z=0$ to $z=3$, with shallow evolution at higher redshifts. This increasing trend is similar to the general behaviour in \colibre, while suggesting a much stronger redshift evolution.

\colibre\ has very few dust-rich, low- to intermediate-mass ($10^{8} - 10^{10.5}\, \Msun$) passive galaxies at $z \in [5,7]$. Their abundance increases dramatically over cosmic time, resulting in a significant number of dust-poor quiescent galaxies at lower redshifts. 
Compared to the observational sample from \cite{Lee2024} (also see \S~\ref{sec:dustscaling.dmass_smass.ms_md_sSFR}), the quiescent galaxies in \colibre\ possess a lower dust content on average; instead their sample is representative of our main-sequence galaxies. 
It remains unclear what fraction of quiescent galaxies are truly dust-rich, since most individual systems at $z \ge 2$ remain undetected in the dust continuum. Consequently, dust mass estimates are largely restricted to stacked samples \cite[\eg][]{Magdis2021}, which mask the underlying variations of the population.

We also present the redshift evolution of the dust mass-to-stellar mass ratio as a function of stellar mass in Appendix~\ref{sec:app.dustscaling.dmass_smass_ratio}.
The dust mass-to-stellar mass ratio in \colibre\ follows an inverted-parabolic shape at all redshifts driven by the balance between dust production and destruction, with higher values at higher redshifts.

\subsubsection{Dust mass - stellar mass dependence on sSFR}\label{sec:dustscaling.dmass_smass.ms_md_sSFR}
\begin{figure*}
    \centering
    \includegraphics[width=\textwidth]{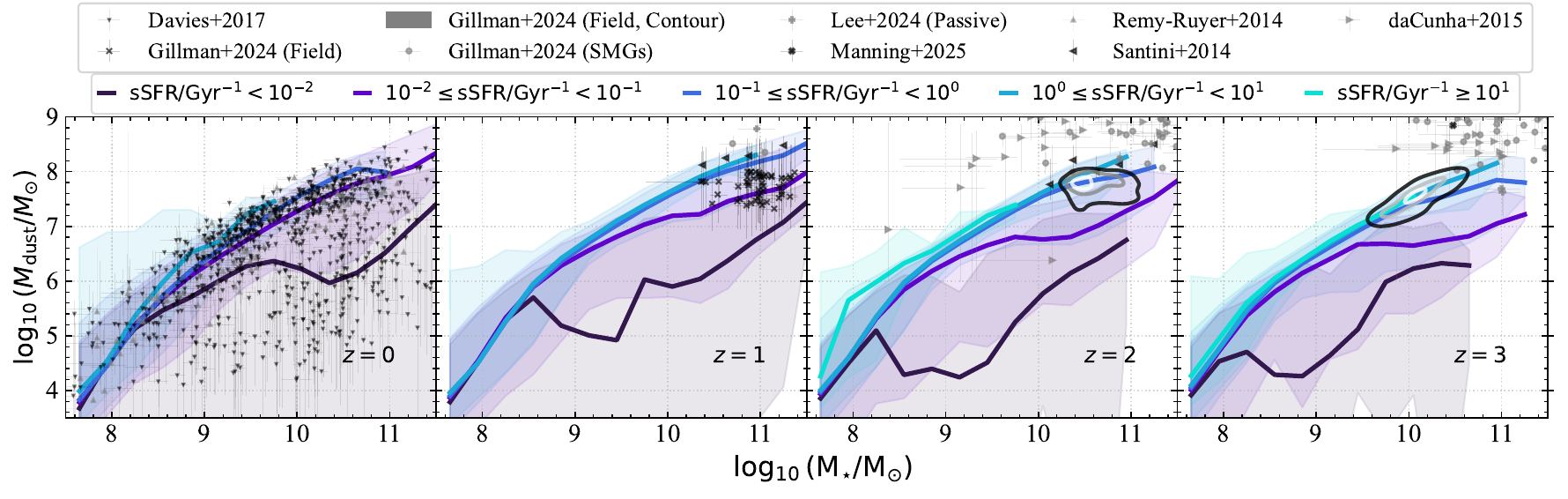}
    \caption{Galaxy dust mass as a function of the stellar mass, split into different specific star formation rate (sSFR) bins for the L200m6 simulation for $z=0,1,2$ and $3$. We overplot observational data from \protect\cite{RemyRuyer2014,Santini2014,Davies2017,Gillman2024,Lee2024,Manning2025}. At $z=2\, {\rm and}\, 3$, we show the \protect\cite{Gillman2024} field galaxy data as contours.
    }
    \label{fig:ms_md_sSFR}
\end{figure*}
Figure~\ref{fig:ms_md_sSFR} shows the median dust mass as a function of the galaxy stellar mass, split into different specific star formation rate (sSFR) bins for the L200m6 simulation. This provides a relation for the dust mass analogous to the Fundamental Metallicity Relation \cite[FMR, ][]{FMR_Garnett2002,FMR_Tremonti2004} observed for galaxy metallicity.

It is evident that galaxies with low (high) dust mass at constant stellar mass exhibit low (high) sSFR.
The lowest bin (sSFR/Gyr$^{-1}< 10^{-2}$) can be considered to be a strict selection criterion for passive galaxies. 
Under this criterion, the low dust mass galaxies are consistent with the early-type galaxy population from \cite{Davies2017} within the $3 \sigma$ scatter at $z=0$.
The \cite{Lee2024} sample for $1 \le z \le 3$ have sSFR $< 10^{-2}$ Gyr$^{-1}$, while having dust masses comparable to the star forming galaxies in the simulation. 
This passive galaxies are at M$_{\rm dust} > 10^{7.6}\, \Msun$ and M$_{\star} > 10^{10}\, \Msun$, while there are no passive galaxies that are similarly dust enriched at $z \ge 1$ in the simulation. 
As discussed, dust mass estimates carry notable uncertainty. \cite{Lee2024} assume $T_{\rm dust} \sim 25\text{ K}$; if the true dust temperature in these passive systems is higher, these masses would be systematically overestimated.

Furthermore, preferentially selecting galaxies with higher sSFR in our simulations fail to reproduce the estimated dust masses of sub-mm galaxies \cite{daCunha2015,Gillman2024,Manning2025}. Further analysis of this population with forward modelled observables will be presented in future works.

\subsection{Dust grain evolution}\label{sec:dustscaling.grain_evo}
\begin{figure*}
    \centering
    \includegraphics[width=\textwidth]{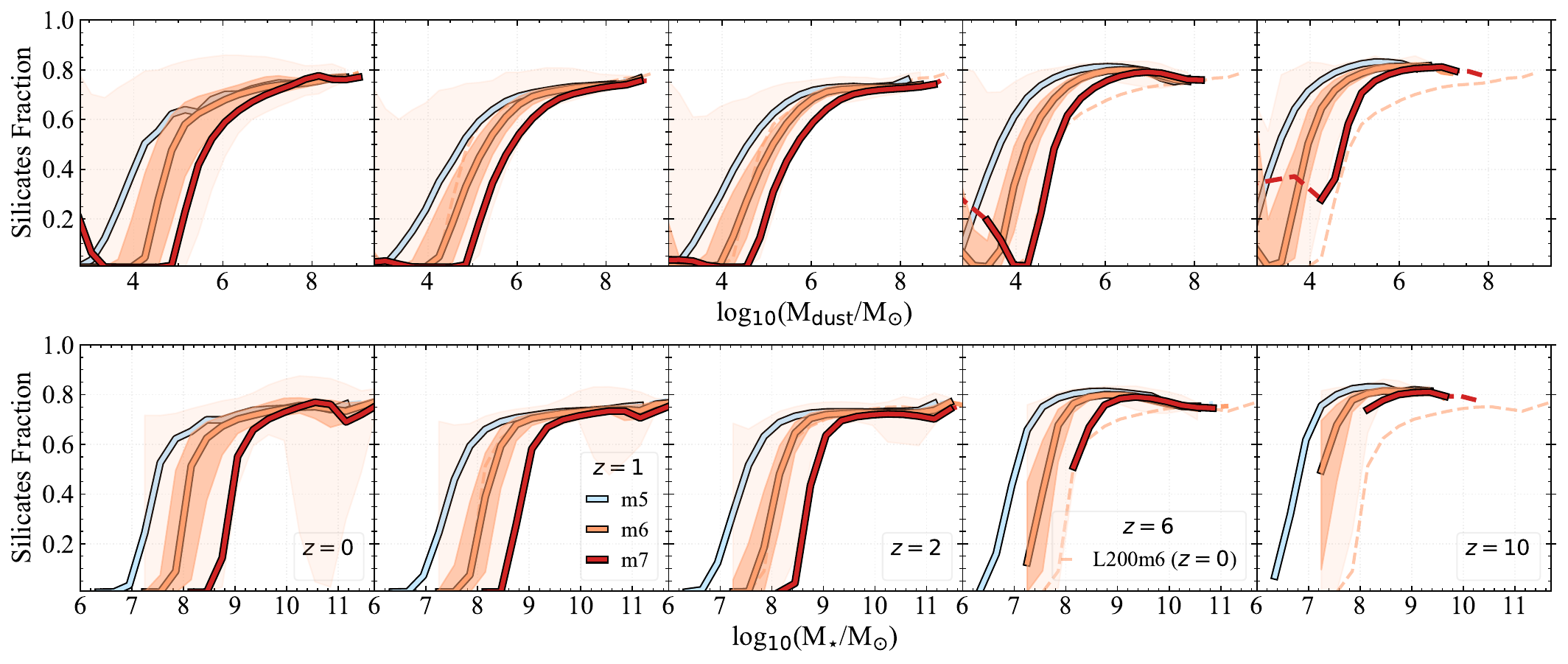}
    \caption{The silicates mass fraction as a function of galaxy stellar mass (bottom panel) and galaxy dust mass (top panel) for the m5 (L025m5 or L050m5 or L100m5), m6 (L200m6), and m7 (L400m7) \colibre\ simulations in the redshifts $z = 0, 1, 2, 6, {\rm and}\, 10$. The solid coloured line (light-blue, orange, and red curves for m5, m6, and m7 respectively) shows the median, with the darker and lighter shaded region denoting the $1 \sigma$ and $3 \sigma$ spreads. The fainter dashed line at $z>0$ shows the $z=0$ median relation from the L200m6 simulation.
    Note that for the m5 resolution, we switch from L100m5 simulation to L50m5 for $z<2$, and to L25m5 for $z<1$.
    }
    \label{fig:silicates_fraction}
\end{figure*}
The \colibre\ simulations track the evolution of graphite and silicate grains in two grain sizes, each as described in \S\ref{sec:dustmodel}. 
Given that carbonaceous and silicate grains possess different extinction coefficients, their relative abundance and size distribution are critical to both determining the galaxy's extinction curve and accurately inferring the total dust mass.

\subsubsection{Grain species}\label{sec:dustscaling.grain_evo.species}
In Figure~\ref{fig:silicates_fraction} we plot the fraction of dust mass that is in silicates (sum of Fe and Mg end-member species) as a function of the galaxy stellar mass (bottom panel, in bins of $0.3$~dex) and the dust mass (top panel, in bins of $0.3$~dex) for redshifts $z = 0, 1, 2, 6, {\rm and}\, 10$ for the m5, m6, and m7 \colibre\ simulations. We show the different resolutions, and also show the $1 \sigma$ (darker shaded region) and $3 \sigma$ (lighter shaded region) spread for the L200m6 simulation. 
Currently, no observational constraints exist on the mass fraction of silicates or graphite grains in galaxies.

In the early Universe, stellar dust seeding is dominated by CCSNe. 
In our model, the dust condensation from these sources is dominated by graphite \cite[$\eta_{\rm CCSN}$ in Table~\ref{table:grains}, see also Figure~11 in][]{Zhukovska2008}.
Notably, significant uncertainty remains regarding the amount of dust destroyed in the SN reverse shock, with estimates ranging from 20\% to 100\% \cite[\eg][]{Nozawa2007,Priestley2022}.
AGB stars also produce predominantly carbonaceous dust.
Consequently, the mass fraction of silicates remains low at low stellar masses.

In \colibre, the grain growth timescale for silicate grains is half of that of graphite grains for the same ISM conditions, albeit not accounting for the bottleneck element in the process. 
The maximum amount of carbon that can condense in dust is restricted to $2/3$, with the rest assumed to be unavailable due to being locked up as CO. 
In contrast, the key bottleneck elements for silicates--Si, Mg, and Fe--deplete heavily onto dust. This shorter growth timescale drives the rapid increase in the silicate mass fraction once galaxies enter the grain growth dominated regime.

Another important factor is that for the same grain number abundance in graphite and silicate grains, the silicate grains in the model (olivine type, see Table~\ref{table:grains}) are about $14$ times more massive.
Meanwhile, the mass yield per SN of the bottleneck elements C, Fe, Mg or Si, is similar \cite[see Figure~1 of][]{Correa25}.
Therefore, while the silicate fraction starts low as initial abundances are dictated by stellar seeding, it ramps up quickly as silicate production dominates subsequent grain growth.
At high dust ($> 10^{4}\, \Msun$ for m5, higher for m6 and m7) and stellar masses ($>10^{7}\, \Msun$ for m5, higher for m6 and m7), the median silicates mass fraction in dust is high ($\approx 80\%$) across all redshifts. 
The median value in the model is close to what was assumed by \cite{WD01,DraineLi2007} for the fraction of dust mass in silicates, $\approx 73\%$. 

In low-mass, low-redshift galaxies, dust production is dominated by stellar seeding. These galaxies evolve their dust content more slowly due to their lower dense-gas content.
Dust grain growth, destruction and size transfer depend on gas densities, hence these galaxies in general evolve at a much slower pace.
Therefore, they are carbon-rich throughout their lifetime as grain growth plays a minor role. 
Dust destruction by SNe and astration are also reduced in these galaxies due to their low specific SFRs.

Across all redshifts, the silicate fractions from different resolutions converge at high dust and high stellar masses. This convergence point shifts to higher mass values for progressively lower resolution (m5 and m6 converge for M$_{\rm dust}\gtrsim 10^{5}\, \Msun$ and M$_{\star}\gtrsim 10^{8}\, \Msun$, while for m6 and m7 it is M$_{\rm dust}\gtrsim 10^{6}\, \Msun$ and M$_{\star}\gtrsim 10^{9}\, \Msun$).
As previously noted, convergence occurs in the high-mass regime where dust growth has saturated, resulting in a near-fixed silicate grain mass fraction ($\approx 80\%$).
At lower masses, the median silicate fraction is highly sensitive to the initial seeding and the exact onset of the grain-growth transition regime. As previously discussed, higher-resolution simulations resolve higher gas densities, allowing this transition to occur earlier. Hence, at a fixed dust or stellar mass below convergence, the higher-resolution simulations yield a higher silicates fraction.

\subsubsection{Grain size}\label{sec:dustscaling.grain_evo.sizes}
\begin{figure*}
    \centering
    \includegraphics[width=\textwidth]{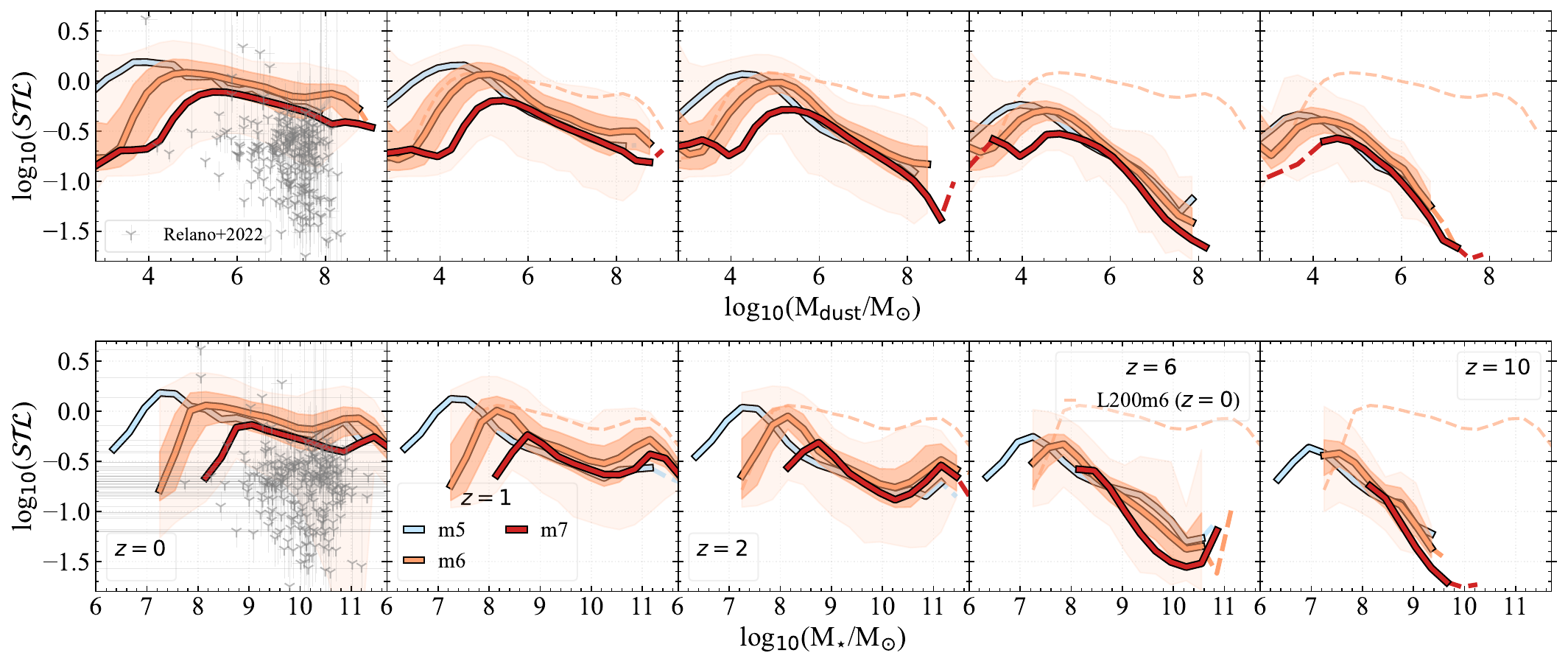}
    \caption{The small-to-large grain mass as a function of galaxy stellar mass (bottom panel) and dust mass (top panel) for $z = 0, 1, 2, 6, {\rm and}\, 10$ for the m5 (L025m5 or L050m5 or L100m5), m6 (L200m6), and m7 (L400m7) \colibre\ simulations. The solid line (light-blue, orange, and red for m5, m6, and m7 respectively) shows the median, with the darker and lighter shaded region denoting the $1 \sigma$ and $3 \sigma$ spread. In $z>0$ panels, the fainter orange dashed line shows the $z=0$ median relation from the L200m6 simulation.
    We plot observational data from \protect\cite{Relano2022} for the small-to-large grain mass as a function of the galaxy dust mass and stellar mass.
    Note that for the m5 resolution, we switch from L100m5 simulation to L50m5 for $z<2$, and to L25m5 for $z<1$.
    }
    \label{fig:Ds_Dl_ratio}
\end{figure*}
In our model, stellar dust injection predominantly produces large grains, which are initialised with a small-to-large grain mass (\stl) ratio of $1:9$ (log$_{10}(\stl) \approx -0.9$).
The mass fraction of small grains subsequently grows rapidly because the grain growth timescale is an order of magnitude shorter for small grains than for large grains.

Figure~\ref{fig:Ds_Dl_ratio} shows the \stl\ ratio as a function of dust mass (top panel) and stellar mass (bottom panel)—both binned in intervals of $0.3$~dex—at $z = 0, 1, 2, 6,$ and $10$ for the m5, m6, and m7 \colibre\ simulations.
The figure also displays the $1\sigma$ (darker shaded region) and the $3\sigma$ (lighter shaded region) scatter for the L200m6 simulation, with the $z=0$ median plotted as the lighter orange dashed line for comparison  at $z>0$.

At the lowest dust masses (M$_{\rm dust} \sim 10^3\, \Msun$), specifically in the m6 and m7 simulations, the ratio approaches the value at seeding. 
As the dust mass increases, the \stl\ ratio rises quickly, driven by the shorter grain growth timescale on the seed small grains. 
The \stl\ ratio peaks around M$_{\rm dust} / \Msun \sim 10^{4}, \sim 10^{5}, {\rm and}\, \sim 10^{6}$ for m5, m6, and m7 respectively, regardless of redshift, before subsequently decreasing. 
This peak corresponds to the saturation value, the point where dust grain growth in small grains is balanced by coagulation.  
The subsequent decrease can be attributed to grain coagulation, which effectively transfers dust mass from small to large grains.
Since coagulation always dominates over shattering in our model, the fraction of dust in large grains increases steadily with increasing dust mass once dust production has saturated.

The \stl\ ratio shows strong redshift evolution of the median value at and beyond the peak. Generally, the ratio is higher at higher redshifts, reflecting the higher gas densities during these early epochs, resulting in shorter coagulation times, and more efficient conversion of small grains into large grains.

The different resolutions converge at high dust mass, with the exact value shifting to higher masses as resolution decreases (m5 and m6 approximately converge for M$_{\rm dust} \gtrsim 10^{5}\, \Msun$; m6 and m7 for M$_{\rm dust} \gtrsim 10^{6}\, \Msun$). This illustrates the impact of resolution on both coagulation and grain growth in dense gas. 
In this high-dust- and high-stellar-mass regime, coagulation regulates the grain size distribution. 
Notably, the peak of the \stl\ ratio shifts to lower dust masses with increasing resolution, since higher-resolution simulations can resolve higher gas densities, lowering the relevant evolutionary timescales (see equation~\ref{eq:grain_growth} and \ref{eq:coag}).

We overplot the observed \stl\, -- dust mass data from \cite{Relano2022} for galaxies at $z=0$. This dataset is heterogeneous, drawing from multiple surveys (see Table~\ref{table:obs_data}), and is not volume-limited, covering star-forming galaxies along the main sequence.
Despite the systematic differences, specifically that the bulk of observed data favours a lower \stl\ ratio than our simulations, the large observational uncertainty ($\sim 0.4$ dex) means most points fall within our $3\sigma$ scatter. A small subset of their galaxies, predicted to have extremely low \stl\ ratios, aligns with \colibre\ predictions for higher gas density thresholds (see Figure~\ref{fig:gas_phase_STL} for $n_{\rm H} > 10/{\rm cm}^3$). 
However, recovering these values observationally is challenging: SED fits have substantial errors, the sample lacks coverage below $3\, \um$ (lacking features like the UV 2175\AA\ bump needed to systematically constrain small carbonaceous grains), and their templates assume a pure carbon composition for small grains. 
Our model includes both silicates and carbonaceous (graphite) grains. In this context, the observed small-grain mass from \cite{Relano2022} represents a lower limit, as graphite grains in general have higher dust extinction coefficient than silicates \cite[\eg][]{Draine_physics_of_ISGM}.
Despite these caveats, the simulated median \stl\ ratio exhibits a decrease with increasing dust mass (\eg\ for M$_{\rm dust} > 10^4\, \Msun$ in the m5 simulation), which qualitatively aligns with the $z=0$ predictions from \cite{Relano2022} (see the median line in their Figure 11). 

The bottom panel shows the \stl\ grain mass ratio relation with stellar mass (see also Figure~11 in \citealt{Trayford_colibre_dust2025} and Figure~23 in \citealt{schaye_colibre2025}).
The \stl\ ratio shows a dependence on stellar mass similar to its dependency on dust mass.
We see convergence in the \stl\ ratio between resolutions for $\approx 10^2$ star particles. 
The \stl\ ratio decreases until M$_{\star}\sim 10^{10}\, \Msun$, beyond which it increases.
This upturn coincides with the onset of effective AGN feedback; AGN heating decreases gas densities, thereby suppressing coagulation.
Small grains possess a higher surface-to-volume ratio, which accelerates all evolutionary processes, such as growth and sputtering, with a net effect of increased growth on small grains.
This rise in the \stl\ ratio reaches a maximum at M$_{\star}\sim 10^{11}\, \Msun$. In this regime, there is a large fraction of passive galaxies that rely on stellar dust production channels (due to lack of dense gas), injecting large grains, causing the ratio to decline once more.

We compare the median \stl\ ratio relation with stellar mass against the observed data in \cite{Relano2022}. 
As with the dust mass relationship, the large observational uncertainty implies that most points fall within the $3\sigma$ scatter of our median, with a small subset of observed galaxies with extremely low \stl\ ratio are similar to what the model predicts for higher density gas (see Figure~\ref{fig:gas_phase_STL} and also Figure~11 in \citealt{Trayford_colibre_dust2025}).

The relations in Figure~\ref{fig:silicates_fraction} and \ref{fig:Ds_Dl_ratio} provide insight into the extinction properties of \colibre\ galaxies without presenting the extinction curves directly, which would require further assumptions about dust optical properties.
A method to derive the full grain size distribution from the two grain size bins is described in \cite{Gebek_skirt2026}, implemented through the dust radiative transfer code \textsc{skirt} \cite[]{skirt2020}. 
It is important to note that observations of galaxies measure \textit{attenuation} curves, as different lines of sight are mixed by scattering, and by the presence of obscured and unobscured stars \cite[see][for further discussion on the distinction between extinction and attenuation]{Salim&Narayanan2020}.
We defer derivation and analysis of dust attenuation curves to future work using the \textsc{skirt} radiative transfer code (Andreadis et al. in prep).

\section{Dust mass function}\label{sec:dmf}
\begin{figure*}
    \includegraphics[width=\textwidth]{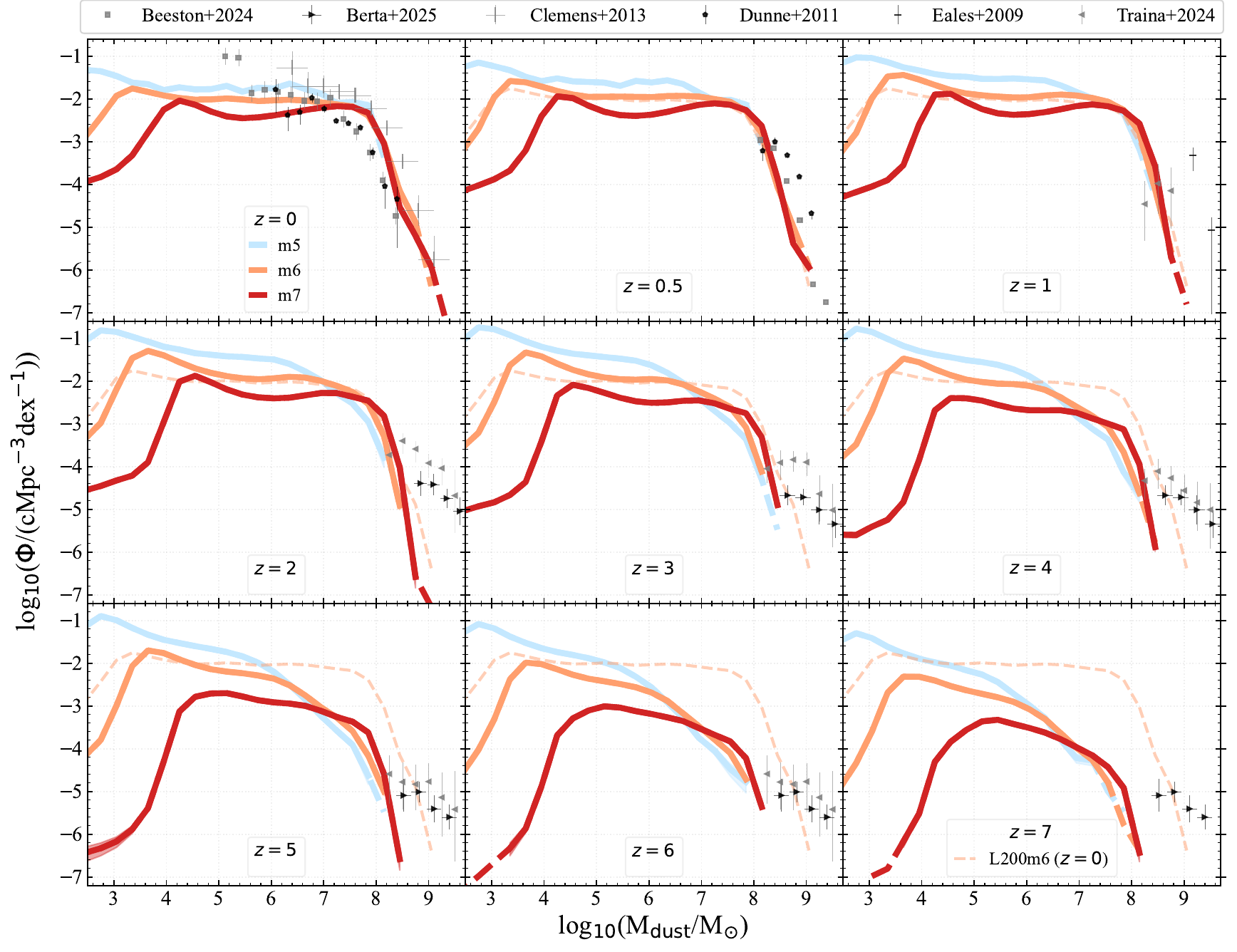}
    \caption{The dust mass function (DMF) for the m5 (L025m5 or L050m5 or L100m5), m6 (L200m6), and m7 (L400m7) \colibre\ simulations for $z\in[0,7]$ shown as the solid light blue, orange, and red lines, respectively. Bins with less than 5 galaxies are denoted by coloured dashed lines. The fainter orange dashed line repeated in the $z>0$ panels, is the L200m6 DMF at $z=0$. The \colibre\ DMF barely evolves from $z=0$ to $2$, beyond which its normalisation is lower compared to the DMF at $z=0$. 
    We show observed DMFs from \protect\cite{Eales2009,Dunne2011,Clemens2013,Beeston2024,Traina2024,Berta2025}.
    The \colibre\ DMF shows reasonable agreement with the $z=0$ and the  $z=0.5$ data from \protect\cite{Beeston2024}. At $z>1$, all the observed data are at dust masses not covered by the simulations.
    Note that for the m5 resolution, we switch from L100m5 simulation to L50m5 for $z<2$, and to L25m5 for $z<1$.
    }
    \label{fig:dmf}
\end{figure*}
\begin{figure*}
    \centering
    \includegraphics[width=\textwidth]{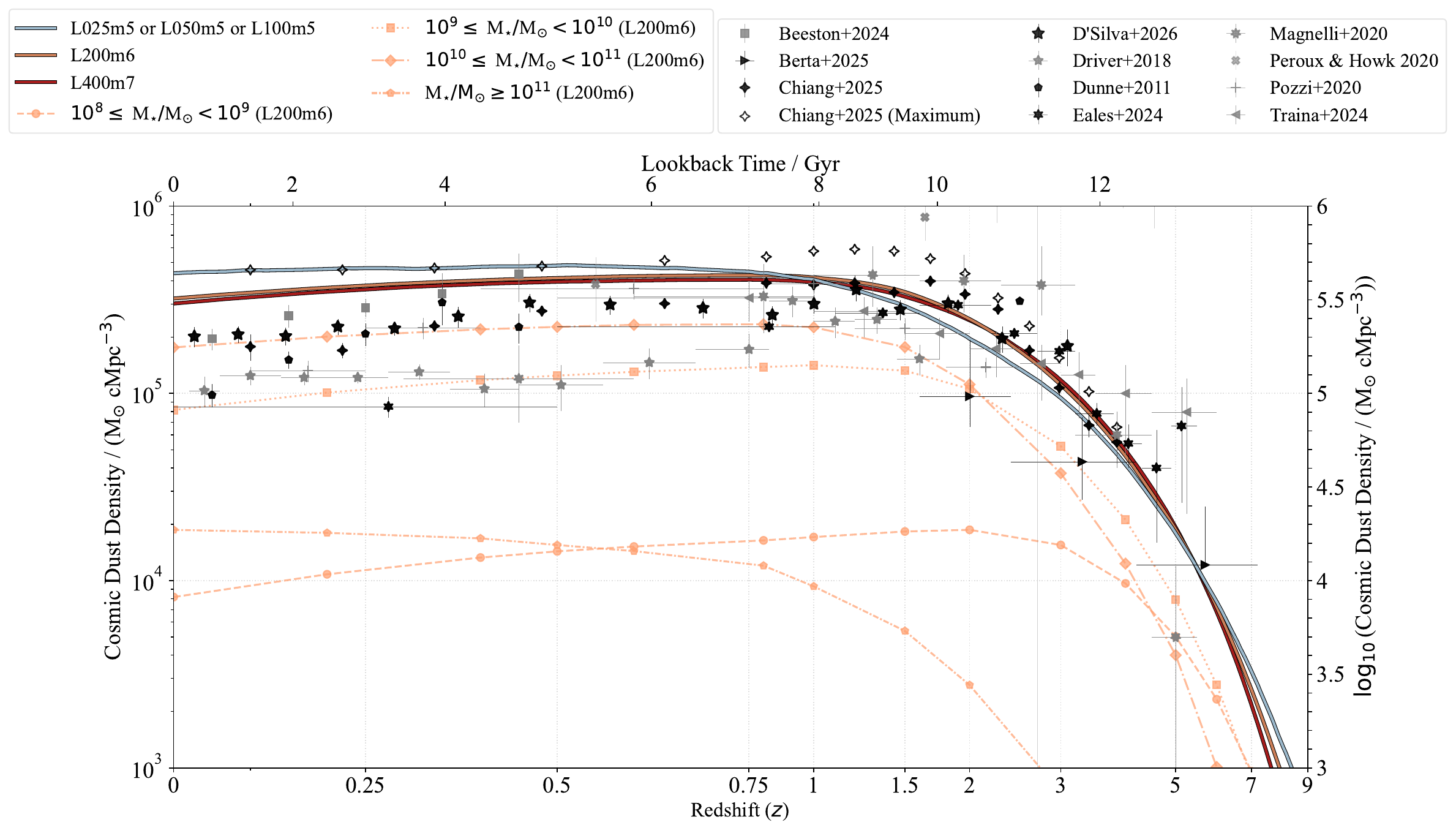}
    \caption{The evolution of the cosmic dust mass density for the m5 (L025m5 or L050m5 or L100m5), m6 (L200m6), and m7 (L400m7) \colibre\ simulations (solid light-blue, orange, and red lines respectively). The dashed (with circle marker), dotted (square marker), dash-dotted (diamond marker) and densely dash-dotted (pentagon marker) lines shows the dust mass density evolution in the L200m6 volume in stellar mass range of of $10^{8} \leq$ M$_{\star}/$M$_{\odot} < 10^{9}$, $10^{9} \leq$ M$_{\star}/$M$_{\odot} < 10^{10}$, $10^{10} \leq$ M$_{\star}/$M$_{\odot} < 10^{11}$, and M$_{\star}/{\rm M}_{\odot} \geq 10^{11}$, respectively.
    We plot observational data from \protect\cite{Dunne2011,Driver2018,Magnelli2020,Peroux&Howk2020,Pozzi2020,Traina2024,Beeston2024,Berta2025,DSilva2026}. We also plot the CDMD derived using the Cosmic Infrared Background (CIB) in \protect\cite{Chiang2025} and the maximum CDMD allowed by the CIB. 
    Note that for the m5 resolution, we switch from L100m5 simulation to L050m5 for $z<2$, and to L025m5 for $z<1$; which is the reason for the discontinuities in the cosmic dust mass density at $z=1\, {\rm and}\, 3$.
    }
    \label{fig:cdmd_evo}
\end{figure*}
The dust mass function (DMF) describes the number density of galaxies as a function of their dust mass. Figure~\ref{fig:dmf} shows the DMF for the m5, m6, and m7 \colibre\ simulations in the redshift range $z\in[0,7]$ using dust mass bins of $0.3$~dex. 
We repeat the $z=0$ L200m6 DMF in all panels to highlight the redshift evolution.

The m6 and m7 simulations show a downturn at M$_{\rm dust} \sim 10^{3.5}\, \Msun$ and $\sim 10^{4.5}\, \Msun$, respectively, due to the imposed stellar mass cut of a minimum of 10 stellar particles, which artificially lowers the number counts at low dust masses (as it is not seen for the m5 simulations). 
Above these limits, the DMF shows little evolution between $z=2$ and $z = 0$, with the normalisation of the knee reaching a maximum at $z=1$. 
The DMF evolves strongly at $z > 2$; specifically, the normalisation at the knee is lower by $\approx 1.5$ dex at $z = 7$ compared to $z = 0$, while the dust mass at the knee is lower by $\approx 0.8$ dex over the same redshift range for m5 (a trend that becomes less pronounced in lower-resolution simulations).
At $z \le 1$ and for dust masses above the resolution-cut, the DMFs of the different resolutions agree within $\sim 0.2$ dex for m5 and m6, and $\sim 0.4$ dex for m6 and m7. 
However, at $z \ge 2$, the convergence between resolutions worsens slightly with increasing redshift, with m5 and m6 agreeing within $\sim 0.4$ dex, while m6 and m7 agree within $\sim 0.6$ dex.

Similar to the previous sections, this lack of convergence is due to the inverse dependence of grain growth timescales on gas density, allowing lower-mass galaxies to enter the grain growth dominated regime earlier in higher resolution simulations. At $z>1$, this effectively shifts the knee of the dust mass function to lower masses for higher resolutions. 
Additionally, higher resolution allows stars to form earlier in the simulation, which leads to earlier chemical enrichment, accelerating the onset of dust growth in the early Universe. This steepens the low-mass slope of DMF at higher redshifts, with the steepest slopes observed in higher-resolution simulations.
At $z>1$ and at the highest stellar masses (above the knee of the DMF), higher resolution simulations show lower normalisation. This is not a result of the smaller volume probed by the higher resolution simulations, as this trend persists for DMFs from the same volume across resolutions (see Figure~\ref{fig:dmf_res_conv}).
As discussed in \S~\ref{sec:dustscaling.dmass_smass} and \ref{sec:dustscaling.dtg_dtm_mstar}, this is because lower-resolution \colibre\ simulations exhibit a higher normalisation of the mass-metallicity relation, consequently producing dustier galaxies at the high-mass end.
We have also checked that at all redshifts, the DMF is dominated by star-forming galaxies, irrespective of resolution (see also Figure~\ref{fig:ms_md_sSFR}).

We overplot observational data (see Table~\ref{table:obs_data} for more details) from 
\citet[][$z=1$]{Eales2009}; \citet[][$z=0\, {\rm and}\, 0.5$]{Dunne2011}; \citet[][$z=0$]{Clemens2013}; \citet[][$z=0$]{Beeston2024}; \citet[][$0.5 < z \le 6$]{Traina2024}; \citet[][$0.6 < z \le 7.2$]{Berta2025}. 

At $z=0$, the predicted DMF agrees reasonably well with the observational data, being within the envelope of the observationally-derived DMFs for all resolutions. 
The \cite{Clemens2013} DMF has a knee at a similar value as the simulations, with the knee of the \cite{Dunne2011,Beeston2024} data being $\approx 0.5$ dex lower. 
At $z=0.5$, the DMFs from \cite{Dunne2011} and \cite{Beeston2024} do not agree with each other (while showing better agreement at $z=0$), with the \cite{Dunne2011} data showing higher normalisation. 
The \colibre\ data agrees better with \cite{Beeston2024}. 
If we compare the DMF of the simulation at $z=0$ and $z=0.5$, there is little evolution, however the normalisation of the \cite{Dunne2011,Beeston2024} DMF 
is $\approx 1$ dex higher at $z=0.5$ than at $z=0$ (at M$_{\rm dust}\sim10^{8.5}\, \Msun$).
In the simulations, the DMF barely evolves from $z=2$ to $0$, with the normalisation at the knee reaching a maximum at $z=1$.

For higher redshifts ($z>1$), observational data \cite[]{Traina2024,Berta2025} is available only at the highest dust masses. 
We also caution against over-interpreting the $z\ge0.5$ observations.
At these redshifts there is a bias towards dust-rich galaxies (M$_{\rm dust}>10^8\, \Msun$), with the sample being incomplete below this mass.
Correspondingly, the knee of the Schechter fits to these observed DMFs shifts to higher dust masses with increasing redshift, a trend that persists even beyond cosmic noon.
Setting aside the caveats regarding dust mass measurements and cosmic variance, the observations at $z>1$ clearly indicate the presence of galaxies that are substantially dustier than those produced in the simulation. This lack of dust-rich galaxies is seen in all current simulations that model dust \cite[\eg][]{McKinnon2017,Popping_dust2017,Vijayan_dust2019,Li2019,Triani_dust2020,Osman2025}.

\section{Cosmic dust mass density}\label{sec:cdmd}
Figure~\ref{fig:cdmd_evo} shows the redshift evolution of the cosmic dust mass density (CDMD) in \colibre\ for different resolutions (solid coloured lines). This quantity is calculated by taking the total dust mass in the simulation (including dust outside our aperture of 50 pkpc radius centred on the galaxy) and dividing by the comoving volume of the simulation box. Observational works typically integrate their Schechter function \cite[][]{Schechter1976} fits to the DMF to obtain the dust mass density.

Overall, the CDMD demonstrates excellent numerical convergence for m6 and m7 simulations across cosmic history. 
Different redshift ranges reveal systematic discrepancies linked to resolution effects and the simpler nature of the calibration employed at the m5 resolution. At $z < 0.75$, m5 yields a noticeably higher dust density ($\approx 0.1$ dex higher compared to m6); which is consistent with the DMF, where the m5 resolution shows higher normalisation at the low-mass end. A similar, albeit weaker, difference can be observed between m6 and m7. 
At $1 \le z \le 6$, the m5 simulation shows a lower normalisation, which is due to systematically lower number density of high dust mass galaxies in higher resolution simulations. Similarly, in this redshift range, the m7 resolution exhibits a negligibly higher dust mass density compared to m6. 
At $z>7$, higher resolutions consistently predict elevated dust densities due to earlier onset of dust enrichment as discussed earlier.

Figure~\ref{fig:cdmd_evo} also shows the contribution to the CDMD in $1$ dex stellar mass bins starting at $10^{8}\, \Msun$. The upper edge of the mass bin is left unbounded, to include all galaxies with M$_{\star} > 10^{11}\, \Msun$. 
The largest contribution to the CDMD is from galaxies in the stellar mass range of $10^{9} \le {\rm M}_{\star} / {\rm M}_{\odot} < 10^{11}$ (spanning two 1 dex bins) for the L200m6 simulation; contributing more than $90\%$ to the CDMD at $z<5$. At $z<0.75$, the bulk of the observations closely match the CDMD contributed by the  $10^{9} \le {\rm M}_{\star} / {\rm M}_{\odot} < 10^{10}$ bin.

At $z>3$, the contribution of the lowest mass bin, $10^{8} \le {\rm M}_{\star} / {\rm M}_{\odot} < 10^9$, to the dust mass density is comparable to that from the $10^{9} \le {\rm M}_{\star} / {\rm M}_{\odot} < 10^{10}$ and $10^{10} \le {\rm M}_{\star} / {\rm M}_{\odot} < 10^{11}$ bins. This highlights the build-up of dust in the early Universe in low-mass galaxies as well as the resolution limit and small volume of the simulations at these redshifts. 
For instance, the contribution from $10^{8} \le {\rm M}_{\star} / {\rm M}_{\odot} < 10^9$ is negligible for the L400m7 simulation, while in m5 resolution, this comparable contribution persists till $z \approx 2$ (not shown).
Below $z=2$, in the L200m6 simulation, the contribution from this bin is less than $3\%$ (while for m5 it is less than $10 \%$). 
The most massive stellar mass bin, ${\rm M}_{\star} / {\rm M}_{\odot} > 10^{11}$, contributes less than $7\%$ for $z<1$ (less than $3 \%$ for L400m7 due to its  higher passive fraction), beyond which its contribution rapidly declines to less than a percent by $z=3$.

We compare our predictions to observational data (see Table~\ref{table:obs_data} for details) from 
\citet[][$0.05 \le z \le 2.5$]{Dunne2011}; \citet[][$0.02 \le z \le 1.75$]{Driver2018}; \citet[][$0.3 \le z \le 5.5$]{Magnelli2020}; \citet[][$0.0 \le z < 5$]{Peroux&Howk2020}; \citet[][$0.1 \le z < 2.5$]{Pozzi2020}; \citet[][$0 \le z \le 0.45$]{Beeston2024}; \citet[][$0.28 \le z \le 5.13$]{Ealse2024}; \citet[][$0.5 \le z < 6$]{Traina2024}; \citet[][$1.6 < z \le 7.2$]{Berta2025}; \citet[][$0.1 \le z < 3.85$]{Chiang2025}; and \citet[][$0 \le z \lesssim 3$]{DSilva2026}. 
The observed cosmic dust mass density exhibits significant scatter, with differences between $\ge 1$ dex observed between some datasets at $z \ge 3$. This scatter decreases with decreasing redshift, becoming $\lesssim 0.3$ dex by $z=0$. The variation among observational datasets highlights both the diverse assumptions used to measure dust masses in these studies and the heterogeneous nature of the datasets themselves.

At $z < 0.7$, the simulated cosmic dust mass density is higher than the bulk of the plotted observations.
However, the simulations are within the uncertainties of the \cite{Driver2018,Pozzi2020,Ealse2024,Traina2024,Berta2025,DSilva2026} datasets for $z > 0.7$. 
The \cite{Pozzi2020,Beeston2024,DSilva2026} data is within $\approx 0.1$ dex (for m6 and m7, $\approx 0.3$ dex for m5) for $z<0.7$. Both \cite{Driver2018} and \cite{DSilva2026} use the GAMA survey, however, they differ in dust mass estimates by a factor of $\approx 3$. \cite{DSilva2026} attribute this to the assumption of a constant dust-to-hydrogen ratio assumed for the dust mass to dust luminosity scaling in \textsc{ProSpect}, while \cite{Driver2018} uses \textsc{Magphys}, which assumes contribution to the warm and cold dust using multiple modified blackbodies.
The \cite{Traina2024,Berta2025} datasets probe more massive dust-rich systems, lying outside the region probed by our simulations (as seen in Figure~\ref{fig:dmf} for the DMFs). Nevertheless, the general agreement in the cosmic dust mass density implies that the extrapolation of their observational Schechter fits to lower masses is consistent with our simulated DMFs. 
The \cite{Chiang2025} data derived from the CIB is the best analogue to our plotted data that includes all the dust mass within the simulations. 
At $z>0.75$, their values are in excellent agreement with \colibre. However, below this redshifts, they are lower by $\approx 0.1$ dex for m6 and m7, and by $0.3$ dex for m5.
\cite{Chiang2025} also derive a maximum dust mass density, the maximum allowed from the CIB. This data is higher than the predictions for all redshift for the m6 and m7 simulations, while it coincides with the m5 simulation for $z<0.5$.

All datasets (that follow the redshift evolution to $z \sim 0$), with the exception of \cite{Driver2018} and \cite{DSilva2026}, display a decline of $\gtrsim 0.3$ dex in the dust mass density for $z = 1$ to $0$. This trend is not reproduced by the simulations. The simulated dust mass density show negligible evolution ($\approx 0.1$ dex for m6 and m7) from $z=1$ to $0$, mirroring the behaviour of the DMF.
However, the peak at $z \approx 1$ is similar to what is seen in the observations

Several theoretical models fail to reproduce this observed evolution of the DMF \cite[\eg][]{Popping_dust2017,Vijayan_dust2019}, instead predicting a minimal decline. 
\cite{Parente2023} investigated this issue using the L-Galaxies semi-analytical model (SAM), incorporating the same dust evolution framework as in \cite{Parente2022}.
They found that reducing the cold gas content using an updated treatment of disc-instability provides a steeper decline in the cosmic dust mass density evolution from $z=1$ to $0$. 
This suggests that the overabundance of dust at late times in \colibre\ may be closely tied to the available cold gas reservoir. The cold gas content of \colibre\ galaxies will be explored in detail in Richings et al. (in prep). However, a global reduction in the dust masses would lead to an underestimation of the stellar mass -- dust mass relation, which matches observations well (Figure~\ref{fig:dmsm_z0_10}).
\section{Influence of gas phase selection}\label{sec:gas_phase}
\begin{figure*}
    \centering
    \includegraphics[width=0.7\textwidth]{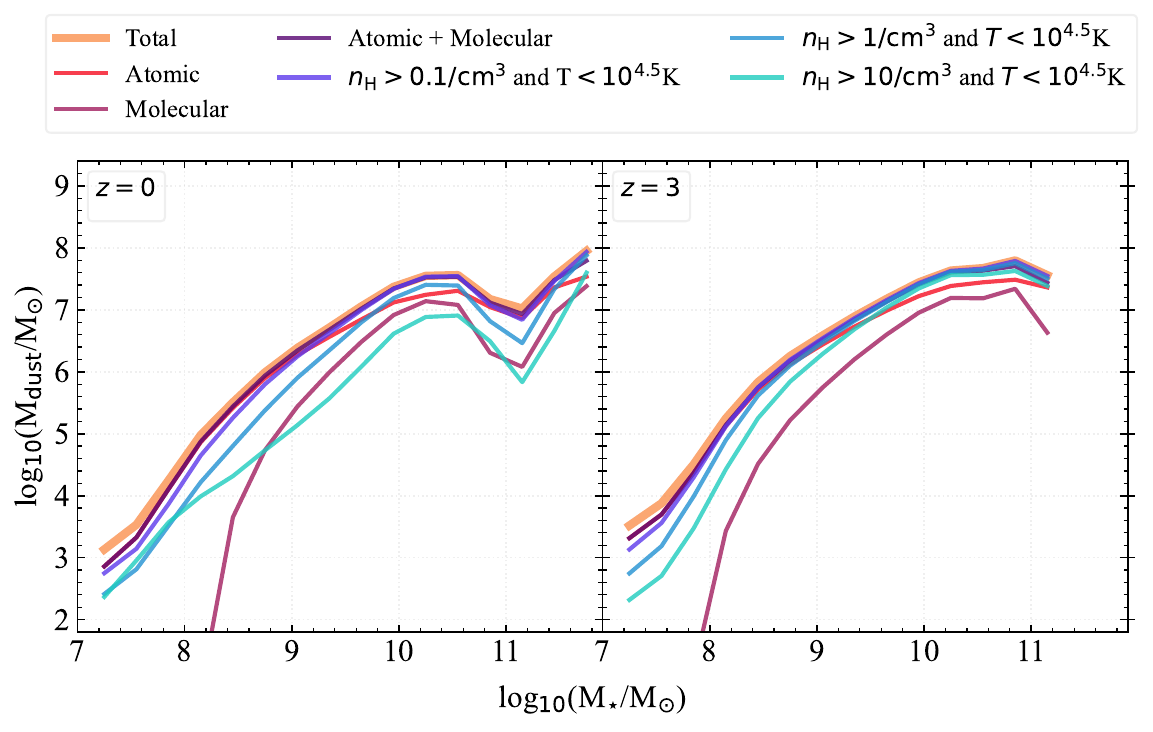}
    \caption{The median dust mass as a function of galaxy stellar mass for the L100m6 \colibre\ simulation at $z=0\, {\rm and}\, 3$. The coloured solid lines show dust mass measured in gas phase corresponding to all gas (Total, fiducial choice), atomic, molecular, both atomic and molecular, and specific density/temperature cuts (T$<10^{4.5}$ K and $n_{\rm H}>0.1, 1, {\rm or }\, 10\, {\rm cm}^{-3}$). }
    \label{fig:gas_phase_mdms}
\end{figure*}

\begin{figure*}
    \centering
    \includegraphics[width=0.7\textwidth]{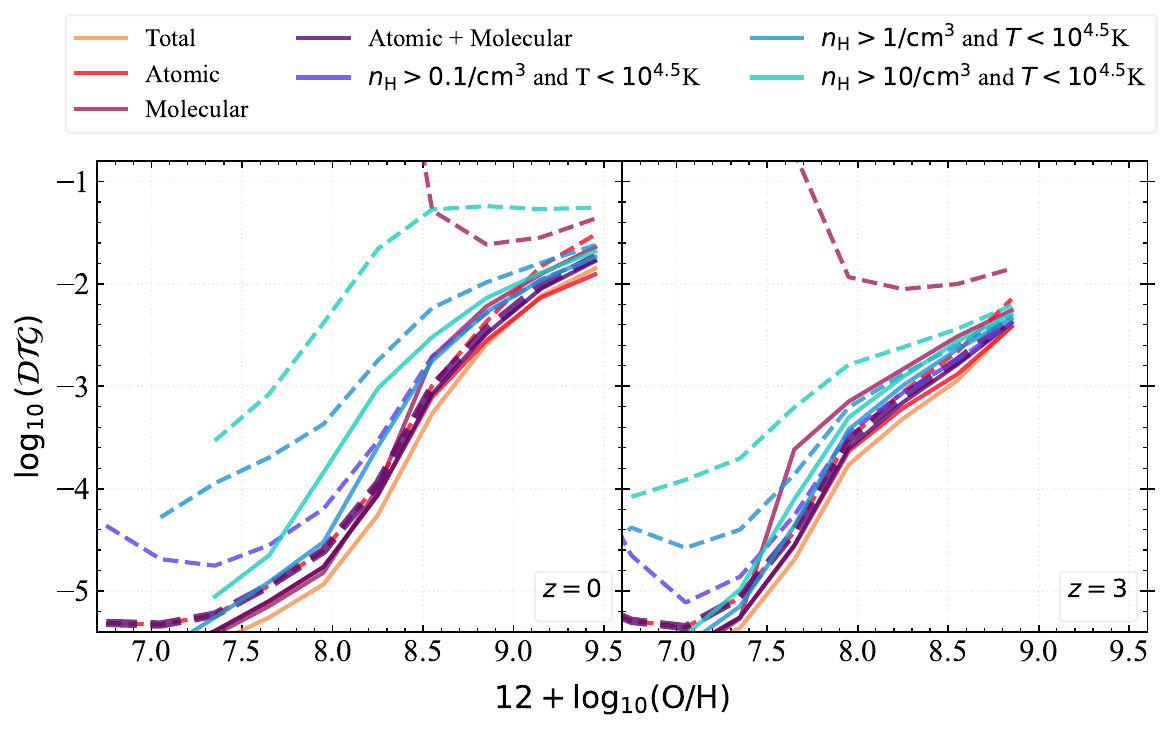}
    \caption{The median dust-to-gas (\dtg) ratio as a function of galaxy gas-phase metallicity for the L100m6 \colibre\ simulation at $z=0\, {\rm and}\, 3$. The coloured solid lines show \dtg\ ratio measured using dust and gas mass corresponding different gas phases: all gas (Total), atomic, molecular, both atomic and molecular, and specific density/temperature cuts (T$<10^{4.5}$ K and $n_{\rm H}>0.1, 1, {\rm or }\, 10\, {\rm cm}^{-3}$). The dashed coloured lines (excluding `Total') show the \dtg\ ratio for the same gas selections, but with the dust mass measured for all dust across all gas phases, giving a higher \dtg\ ratio (the value for molecular gas falls above our y-axis range at low-metallicities). The fiducial model choice is represented by the dashed line corresponding to `Atomic + Molecular'. }
    \label{fig:gas_phase_dgr}
\end{figure*}

\begin{figure*}
    \centering
    \includegraphics[width=0.8\textwidth]{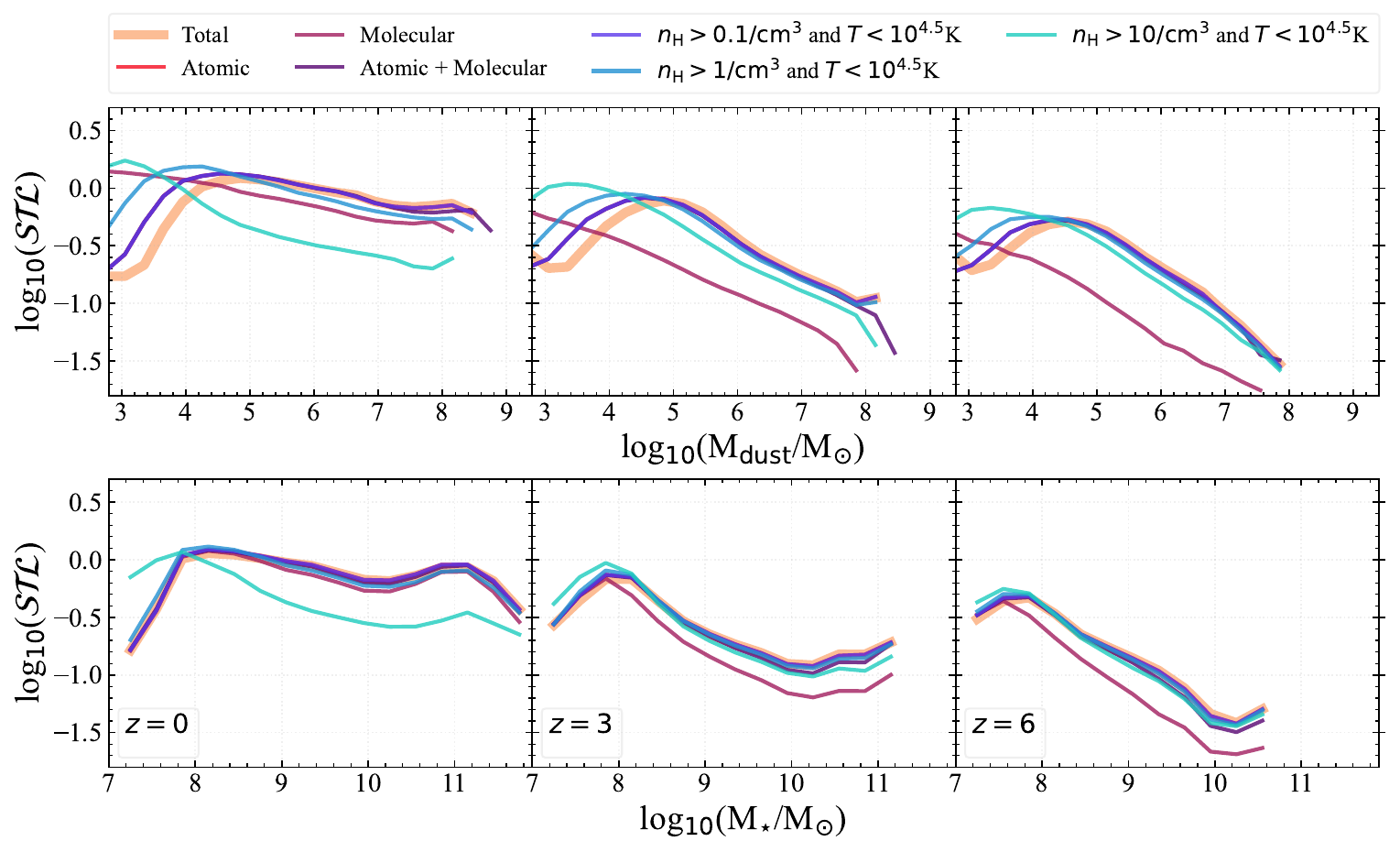}
    \caption{The median small-to-large (\stl) grain mass ratio as a function of galaxy dust mass (top panel) and stellar mass (bottom panel) for the L100m6 \colibre\ simulation at $z=0, 3,\, {\rm and}\, 6$. The coloured solid lines show dust mass measured in gas phase corresponding to all gas (Total), atomic, molecular, both atomic and molecular, and specific density/temperature cuts (T$<10^{4.5}$ K and $n_{\rm H}>0.1, 1, {\rm or }\, 10\, {\rm cm}^{-3}$). The fiducial model choice is represented by the solid line corresponding to `Total'. The lines for Atomic and Atomic+Molecular lie on top of each other.}
    \label{fig:gas_phase_STL}
\end{figure*}

\begin{figure}
    \centering
    \includegraphics[width=\columnwidth]{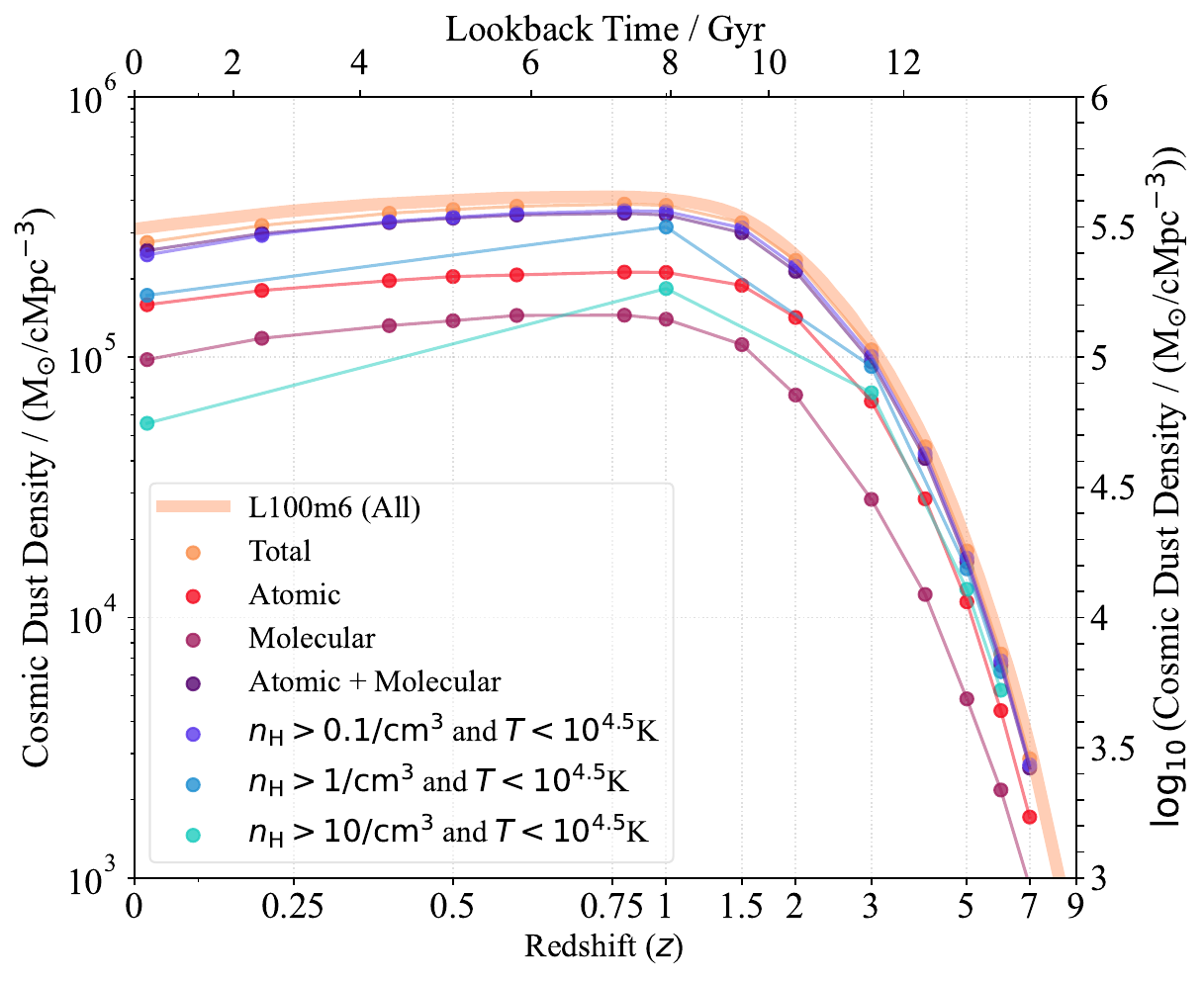}
    \caption{Same as Figure~\ref{fig:cdmd_evo}, now showing the cosmic dust mass density for different gas phases. The coloured solid lines show dust mass measured in gas phase corresponding to all gas (Total), molecular, and specific density/temperature cuts (T$<10^{4.5}$ K and $n_{\rm H}>0.1, 1, {\rm or }\, 10\, {\rm cm}^{-3}$). The fiducial model choice shown in Figure~\ref{fig:cdmd_evo} is the solid line corresponding to `All'.
    We have calculated the dust mass density only at $z=0, 1, 3, 5,$ and $6$ for gas with $T < 10^{4.5}\text{ K}$ and densities of $n_{\rm H} > 1$ and $10\text{ cm}^{-3}$.
    }
    \label{fig:gas_phase_cdmd}
\end{figure}

In the previous sections, we demonstrated the successes and shortcomings of the \colibre\ simulations in reproducing observed galaxy dust scaling relations across cosmic time, primarily using total dust mass and neutral gas mass. We now test the sensitivity of these scaling relations to our measurement definitions, focusing specifically on how different gaseous phase selections affect the results.

The multi-phase nature of gas poses a non-trivial challenge when attempting to isolate dust mass contributions from observational data. While observations can measure galaxy gas masses across different phases using distinct emission-line tracers, isolating the dust mass residing within these individual phases remains difficult. Instead, the multi-phase nature of the gas enters observational dust mass estimates indirectly through the modelling techniques used to derive them. For instance, in modified blackbody fitting, the range of assumed or fitted dust temperatures serves as a proxy for contributions from different phases. Similarly, in techniques employing template dust SED fitting, the distribution of starlight intensities across the galaxy's dust mass \cite[\eg\, equation 23 in][]{DraineLi2007} indirectly maps onto these phases. This challenge is compounded by observational detection limits; dust residing in low surface brightness regions, characteristic of galactic outskirts with low dust column densities, is frequently missed. Moreover, because global dust luminosity is heavily dominated by warmer dust with temperatures of $T_{\rm dust} \approx 30\text{--}40$~K \cite[\eg][]{Singh2021}, reservoirs of the coldest dust remain difficult to detect.

To quantify the influence of these phase definitions, we perform the following analysis using the L100m6 \colibre\ simulations at a few discrete redshifts. 
We analyse the resulting dust scaling relations based on six distinct gas mass choices: the atomic, molecular, sum of the atomic and molecular gas mass (neutral phase), and the following density cuts for a maximum temperature of $T<10^{4.5}$ K for the gas: $n_{\rm H}>0.1\, {\rm cm}^{-3}$, $1\, {\rm cm}^{-3}$, and $10\, {\rm cm}^{-3}$. 
Consistent with \S~\ref{sec:dustscaling}, we measure metallicity within the cool, dense gas ($n_{\rm H}>0.1\, {\rm cm}^{-3}$ and $T<10^{4.5}$ K). However, to determine the corresponding phase-specific metal mass, we multiply the metal fraction of this cool, dense gas by the mass of the specific gas phase being considered. All these quantities are measured within a 50 pkpc aperture.

Figure~\ref{fig:gas_phase_mdms} shows the median relation between dust mass and stellar mass for the L100m6 \colibre\ simulation at $z=0$ and $z=3$. The coloured solid lines track the dust mass contained within the specific gas phases defined above, with the total dust mass included for comparison. Across all stellar masses and redshifts, the bulk of the dust resides in the neutral phase, constituting over $60\%$ of the total dust mass (the fraction is not shown). At $z=0$, the dust mass in the densest gas phases—namely the molecular phase and the $n_{\rm H} > 10\, {\rm cm}^{-3}$ cut—is $\sim 0.5\text{--}1$~dex lower than the total dust mass. All other gas selections yield dust masses within $0.4$~dex of the total across both redshifts. Notably, the fraction of dust residing in the molecular gas remains very similar between $z=0$ and $z=3$.

Figure~\ref{fig:gas_phase_dgr} shows the median relation between the \dtg\ ratio and gas-phase metallicity for galaxies in the L100m6 simulation at $z=0$ and $z=3$. The solid coloured lines show the \dtg\ ratio calculated using the dust mass within that specific phase. For comparison, the dashed lines show the \dtg\ ratio calculated using the \textit{total} dust mass divided by that specific phase's gas mass. This latter approach yields systematically higher \dtg\ ratios, occasionally resulting in unphysically high values for the molecular and $n_{\rm H} > 10\, {\rm cm}^{-3}$ phases. At both redshifts, the atomic, neutral, and total gas selections exhibit similar \dtg\ trends, with the `Total' gas selection yielding the lowest overall values. Conversely, the densest gas phases (molecular, and $n_{\rm H} > 1$ or $10\, {\rm cm}^{-3}$) show significantly elevated \dtg\ ratios. This offset is substantial at $z=0$, reaching $\approx 0.4$~dex at high metallicity ($12+{\rm log}_{10}({\rm O/H}) > 8.5$) and up to $\approx 1$~dex at lower metallicities, with the largest divergence seen in the $n_{\rm H} > 10\, {\rm cm}^{-3}$ phase. This disparity largely diminishes by $z=3$, where all phases maintain \dtg\ ratios within $\approx 0.4$~dex of one another (with the molecular phase showing the largest deviation). 
Using the total dust mass for the molecular, and $n_{\rm H}>10\, {\rm cm}^{-3}$ and $T<10^{4.5}$~K gas phases exacerbates these offsets to $\ge 1$~dex at the low-metallicity end ($12+{\rm log}_{10}({\rm O/H}) < 8$).
We also find that the ``S-shape" of the \dtg-metallicity relation (as discussed in \S~\ref{sec:dustscaling.dtg}) persists in all gas-phase selections, except when using the total dust mass for the molecular gas phase.

We find analogous trends for the \dtm\ ratio as a function of metallicity, which are presented in Figure~\ref{fig:app_gas_phase_dtm} of Appendix~\ref{sec:app_gas_phase}. We also find that the relation between silicate fraction and stellar mass is not affected by the gas-phase selection, while the relation with dust mass is affected at low-dust masses ($M_{\rm dust}<10^5\, \Msun$), with the denser gas phases showing higher silicate fraction (Figure~\ref{fig:app_gas_phase_silicates}).

Figure~\ref{fig:gas_phase_STL} shows the small-to-large grain mass (\stl) ratio as a function of dust mass (top panel) and stellar mass (bottom panel) at $z=0, 3,$ and $6$. The coloured lines show the \stl\ values for the same gas phases as in Figure~\ref{fig:gas_phase_dgr}. 
Since both the grain growth and coagulation timescales depend inversely on local density, the \stl\ ratio shows differences depending on the selected gas phase. The onset of accelerated grain growth occurs earlier in denser gas. Consequently, dense gas phases exhibit a higher fraction of small grains at low dust and stellar masses, reaching the equilibrium between grain growth and coagulation earlier, which corresponds to the peak in the \stl\ ratio. In the coagulation-dominated regime (at higher dust and stellar masses), because denser gas is more efficient at transferring mass from small to large grains, the denser gas shows lower \stl\ ratios.

Figure~\ref{fig:gas_phase_cdmd} illustrates the evolution of the cosmic dust mass density for the various gas-phase selections compared against the fiducial total dust density (`L100m6 (All)', using all the dust mass in the simulation). 
When using the dust mass content within the total gas, neutral gas, or gas with $n_{\rm H} > 0.1\, {\rm cm}^{-3}$, the resulting cosmic dust density evolution is quantitatively similar to the fiducial choice. 
In contrast, selections using the molecular gas, or gas with $n_{\rm H} > 1$ and $10\, {\rm cm}^{-3}$, reveal a significant decline in the normalisation of the dust mass density by $>0.2$~dex between $z=1$ and $z=0$, mirroring the trend seen in the observational data. For $z > 1$, the dust mass densities obtained using the $n_{\rm H} > 1\, {\rm cm}^{-3}$ cut track the fiducial evolution closely, whereas the molecular and $n_{\rm H} > 10\, {\rm cm}^{-3}$ selections yield systematically lower values. 
As noted previously, this suggests that observations may be biased toward the high-density regions of galaxies while potentially underestimating the dust contribution from more diffuse phases, particularly at lower redshifts. This disparity emphasises the critical need to mimic observational selection effects and techniques directly through the forward modelling of simulations.

\section{Conclusions}\label{sec:conclusions}
We have presented an analysis of the dust scaling relations and dust evolution within the \colibre\ cosmological simulation suite \cite[]{schaye_colibre2025,Chaikin_colibre2025}. 
\colibre\ implements a comprehensive model for the production, growth, destruction, and size evolution of dust grains within a multi-phase ISM, coupled self-consistently to the radiative cooling and chemistry networks. 
We analysed the \colibre\ simulations across a range of cosmological volumes and resolutions: m5 resolution with $m_{\rm gas} = 2.30 \times 10^{5}\, \Msun$ in $(100\, {\rm cMpc})^3$, $(50\, {\rm cMpc})^3$, and $(25\, {\rm cMpc})^3$ volumes; m6 resolution with $m_{\rm gas} = 1.84 \times 10^{6}\, \Msun$ in $(200\, {\rm cMpc})^3$; and m7 resolution with $m_{\rm gas} = 1.47 \times 10^{7}\, \Msun$ in $(400\, {\rm cMpc})^3$.
We investigated the relationship of the dust-to-gas (\dtg) ratio with metallicity and stellar mass; dust-to-metal (\dtm) ratio with metallicity and stellar mass; dust mass with stellar mass; dust species fraction and small-to-large grain mass (\stl) ratio with dust mass and stellar mass; the dust mass function and the cosmic dust mass density. We compared these predictions against a wide range of observational constraints from $z=0$ to $z \sim 15$. Our main conclusions are as follows:
\begin{itemize}
    \item The \colibre\ model broadly reproduces the observed evolution of important dust scaling relations. 
    This includes the median trend of \dtg\ ratio with metallicity (Figure~\ref{fig:dgr_Z_z0_9}) and stellar mass (Figure~\ref{fig:dtg_dtm_mstar_z0_7}), the \dtm\ ratio with metallicity (Figure~\ref{fig:DTM_vs_Z}) and stellar mass (Figure~\ref{fig:dtg_dtm_mstar_z0_7}), and the dust mass - stellar mass relations (Figure~\ref{fig:dmsm_z0_10}).
    
    \item The simulation produces the characteristic ``S''- or sigmoid-shape of the \dtg-metallicity (Figure~\ref{fig:dgr_Z_z0_9}) and \dtm-metallicity relations (Figure~\ref{fig:DTM_vs_Z}) seen in several other simulations that model dust evolution. 
    This shape arises due to the transition from a regime dominated by stellar dust production to one dominated by rapid and saturated grain growth in the ISM. We find that the metallicity corresponding to this transition—and thus the emergence of the ``S-shape''—is resolution-dependent, with the value shifting to lower metallicities for higher-resolution models. While the normalisation of the relations is sensitive to the gas-phase selection (Figure~\ref{fig:gas_phase_dgr}), the overall shape persists across all phases (except when combining the total dust mass with the molecular gas mass).
    
    \item The model captures the observed evolution of the galaxy dust mass - stellar mass relation well, with the median relation showing little evolution with redshift (higher-redshifts have slightly higher median). However, the extreme dust masses seen in submillimeter galaxies (or dusty star-forming galaxies) at $z \ge 2$ are not reproduced (Figure~\ref{fig:dmsm_z0_10}, \ref{fig:dmf}). The normalisation of the dust mass - stellar mass relation is sensitive to the selected gas-phase (Figure~\ref{fig:gas_phase_mdms}).
    
    \item Exploring the new redshift frontier ($z=10-15$) opened by \jwst, \colibre\ predicts rapid dust enrichment in the early Universe. 
    The high dust masses found at these epochs in \colibre\ support the scenario of efficient early dust production, with the predicted values marginally consistent with recent observational upper limits (Figure~\ref{fig:dmsm_z0_10}).
    
    \item The \colibre\ model predicts that at all redshifts and at high dust ($> 10^{4}\, \Msun$ for m5, higher for m6 and m7) and high stellar ($>10^{7}\, \Msun$ for m5, higher for m6 and m7) masses,   silicate type grains dominate ($\ge 70\%$, Figure~\ref{fig:silicates_fraction}). 
    
    \item The model reveals an overall increase in the \stl\ ratio (Figure~\ref{fig:Ds_Dl_ratio}) with decreasing redshift, driven by the shorter growth timescales of small grains compared to large dust grains.
    The normalisation of the \stl\ ratio is sensitive to the specific gas phase hosting the dust (Figure~\ref{fig:gas_phase_STL}).
    Across both dust and stellar mass scales, the \stl\ ratio exhibits a distinct trend: it initially rises, reaches a peak, and subsequently declines at the highest masses. These three phases correspond to distinct physical regimes: initial grain growth onto seeded small grains, a saturation plateau where growth balances coagulation, and a final coagulation-dominated regime.
    This median trend is similar to observations at $z=0$.

    \item The dust mass function (Figure~\ref{fig:dmf}) shows reasonable agreement within the range of those recovered observationally at $z < 1$. At higher redshifts, the observations cover larger dust masses not probed by the model.
    
    \item The CDMD (Figure~\ref{fig:cdmd_evo}) peaks at $z \sim 1$ with little subsequent evolution, contrasting with most observational trends. While our predictions agree well with high-redshift observations, they exceed the majority of low-redshift ($z<0.7$) data by $\gtrsim 0.3$~dex, overlapping with only a small subset of observations. The normalisation and shape of the CDMD depend on the gas-phase selection (Figure~\ref{fig:gas_phase_cdmd}), with the decline in the relation from $z=1$ to $z=0$ for the densest phase mirroring the observations.
    
    \item At low metallicities and low stellar masses, the dust scaling relations and the dust mass function depend strongly on the simulated resolution; consequently, the cosmic dust mass density is also resolution-dependent.
    This is primarily driven by the ability of higher-resolution simulations to resolve higher gas densities, which enables an earlier onset of accelerated dust grain growth. 
\end{itemize}

The \colibre\ simulation suite represents a significant step forward by modelling the multiphase interplay between dust, gas, stars, and feedback in large cosmological volumes. 
While the \colibre\ dust model successfully reproduces several observed dust scaling relations, some tensions remain between predictions of the model and observations. 
These discrepancies highlight the non-trivial challenge of simultaneously matching multiple observational constraints in cosmological simulations.

Importantly, these discrepancies must be interpreted by considering both the significant uncertainties in observationally derived scaling relations (\S~\ref{sec:obs_data.caveats}) and the sensitivity of simulated predictions to the specific gas-phase selection (\S~\ref{sec:gas_phase}).
This further underscores the need for \textit{apples-to-apples} comparisons between simulations and observations through rigorous forward modelling of simulated galaxies in the future. 
In this regard, the \colibre\ dust model constitutes a substantial leap forward, as the number of tunable free parameters in the forward modelling of dust is minimal compared to previous large-volume cosmological simulations.

By coupling \colibre\ with forward modelling codes like \textsc{skirt} \cite[]{skirt2020} or \textsc{synthesizer} \cite[]{synth2_2025,synth1_2025}, we can self-consistently generate representative synthetic observations that account for the complex star-dust geometry and extinction curve evolution predicted by the model. 
This approach will enable us to determine whether the apparent discrepancies in the high-redshift dust budget arise from missing physics in galaxy formation models or from biases in observationally inferred dust masses. 
Moreover, the detailed tracking of grain sizes and species in \colibre\ provides an ideal laboratory to study the evolution of dust attenuation curves across cosmic time, offering predictions that can be directly tested against the growing library of high-redshift spectroscopy from \jwst.
Finally, forward-modelled predictions of the dust continuum at the redshift frontier will offer testable predictions for the detectability of dust in these extreme high-redshift galaxies.

\section*{Acknowledgements}
We gratefully acknowledge everyone who has contributed to the \colibre\ project for their input. This work used the DiRAC@Durham facility managed by the Institute for Computational Cosmology on behalf of the STFC DiRAC HPC Facility (www.dirac.ac.uk). The equipment was funded by BEIS capital funding via STFC capital grants ST/K00042X/1, ST/P002293/1, ST/R002371/1 and ST/S002502/1, Durham University and STFC operations grant ST/R000832/1. DiRAC is part of the National eInfrastructure.
We thank Minju M. Lee, M\'onica Relano, Robert M. Yates, Stephen M. Wilkins, Steven R. Gillman, for helpful discussions and providing datasets for comparisons done in the paper. 

APV acknowledge support from the Sussex Astronomy Centre STFC Consolidated Grant
(ST/X001040/1). 
ABL acknowledges support by the Italian Ministry for Universities (MUR) program “Dipartimenti di Eccellenza 2023-2027” within the Centro Bicocca di Cosmologia Quantitativa (BiCoQ), and support by UNIMIB’s Fondo Di Ateneo Quota Competitiva (project 2024-ATEQC-0050).
EC acknowledges support from STFC consolidated grant ST/X001075/1.
SP acknowledges support by the Austrian Science Fund (FWF) through grant-DOI: 10.55776/V982. 

We would like to acknowledge the use of the following software in this work: \texttt{SwiftSimIO} \cite[]{Borrow2020}, \href{https://automeris.io}{\texttt{WebPlotDigitizer}}.

\section*{Data Availability}
We will make the dust scaling relations, dust mass functions, and cosmic dust mass density data available on the \colibre\ website (\href{https://colibre.strw.leidenuniv.nl}{https://colibre.strw.leidenuniv.nl/}) on the acceptance of the paper. The scripts to reproduce all the plots in this work will be made available through Github on the acceptance of the paper.
All the \colibre\ simulation data will be eventually made publicly available. 
A public version of the \texttt{SWIFT} code \cite[]{Schaller2024_swift} is available
at \href{http://www.swiftsim.com}{http://www.swiftsim.com}.



\bibliographystyle{mnras}
\bibliography{dust_sim,dust_obs,colibre} 



\appendix

\section{Effect of resolution}\label{sec:app_resolution}
\begin{figure*}
    \centering
    \includegraphics[width=\textwidth]{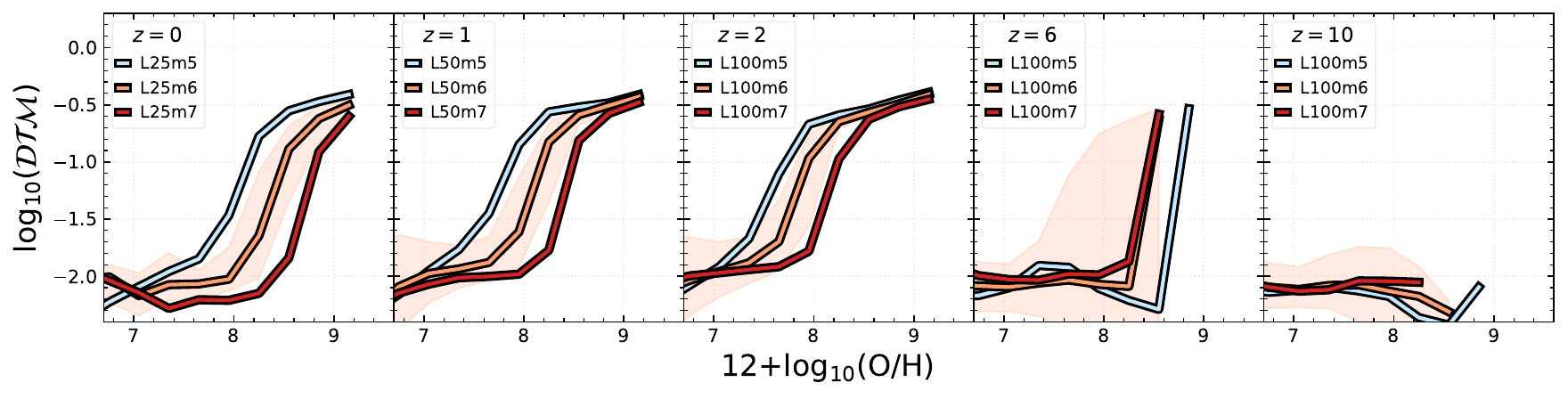}
    \includegraphics[width=\textwidth]{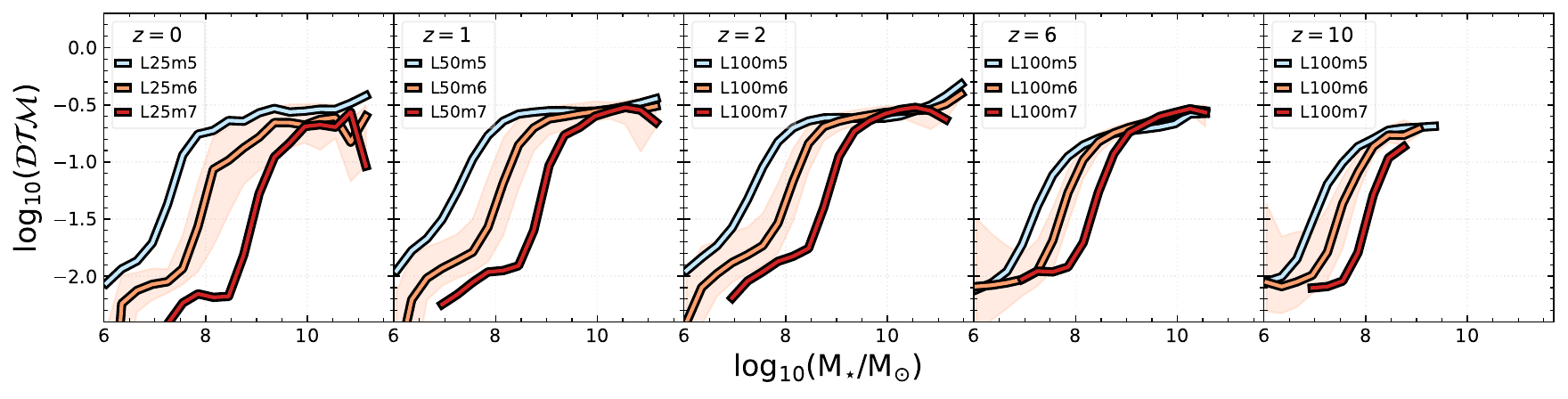}
    \caption{The median dust-to-metal ratio as a function of metallicity (top panel) and stellar mass (bottom panel) respectively for different \colibre\ simulation resolutions (m5, m6, and m7 denoted by light blue, orange and red lines) for $z=0,1,2,6\, {\rm and}\, 10$. The shaded region denotes the $1\sigma$ spread for the m6 simulations. The volume used for the different resolutions is determined by the biggest available volume for the m5 simulation at that redshift.}
    \label{fig:DTM_res_conv}
\end{figure*}
\begin{figure*}
    \centering
    \includegraphics[width=\textwidth]{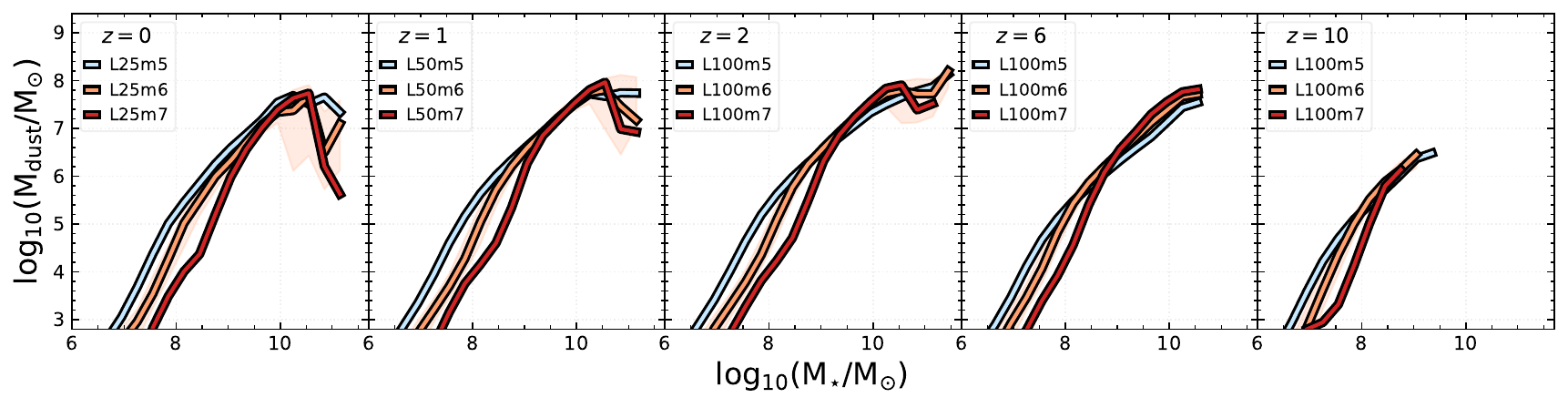}
    \caption{The median dust mass as a function of stellar mass for different \colibre\ simulation resolutions (m5, m6, and m7 denoted by light blue, orange and red lines) for $z=0,1,2,6\, {\rm and}\, 10$. The shaded region denotes the $1\sigma$ spread for the m6 simulations. The volume used for the different resolutions is determined by the biggest available volume for the m5 simulation  at that redshift.}
    \label{fig:mdust_mstar_res_conv}
\end{figure*}
\begin{figure*}
    \centering
    \includegraphics[width=\textwidth]{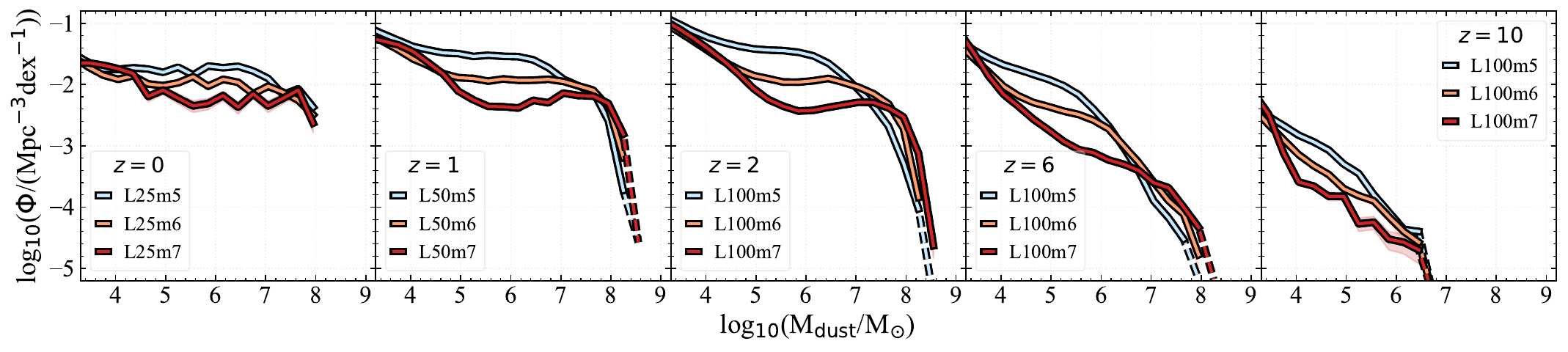}
    \caption{The dust mass function for different \colibre\ simulation resolutions (m5, m6, and m7 denoted by light blue, orange and red lines) for $z=0,1,2,6\, {\rm and}\, 10$. Bins with less than 5 galaxies are denoted by coloured dashed lines. The volume used for the different resolutions is determined by the biggest available volume for the m5 simulation at that redshift.}
    \label{fig:dmf_res_conv}
\end{figure*}
In the \colibre\ model, the free parameters in the subgrid model are  varied between different resolutions \cite[see Table~1 in][]{schaye_colibre2025}. The simulation adopts the `weak convergence' principle for the subgrid parameters from \cite{Schaye2015}, which acknowledges the fact that some observables (e.g. galaxy stellar masses) depend strongly on unresolved feedback physics that cannot be predicted ab initio, so recalibration is unavoidable as one changes the resolution.

We compare galaxy dust properties across the different particle resolutions, m5, m6, and m7, which correspond to gas masses of $2 \times 10^{5}\, \Msun$, $1.8 \times 10^{6}\, \Msun$, and $1.47 \times 10^{7}\, \Msun$, respectively.
We focus specifically on the dust-to-metal (\dtm) ratio, the dust mass - stellar mass relation, and the dust mass function to demonstrate that our conclusions remain robust within the same volume, as other properties show only minor variations from those in the main text. 
The chosen volume corresponds to the largest simulation size available for the m5 resolution. No cut is imposed on the number of stellar particles for the galaxies in this analysis. 
While most results are consistent with the main text, some discrepancies appear at high redshift due to the absence of a particle count cut. Detailed information regarding differences between resolutions can be found in the corresponding sections of the main text; here, we focus primarily on the deviations from those findings.


Figure~\ref{fig:DTM_res_conv} shows the \dtm\ ratio as a function of metallicity (top panel) and stellar mass (bottom panel) across different resolutions for $z=0,1,2,6\, {\rm and}\, 10$. 
The \dtm\ ratio for the different resolutions converges for the same metallicity and stellar mass ranges as discussed in the main text.
At $z=6$ and $z=10$, the absence of a particle count cut alters the relationship with metallicity compared to that seen in \S~\ref{sec:dustscaling.dtm}. 
The \dtm\ ratio appears essentially flat because galaxies with fewer than 10 star particles (visible in the bottom panel) dominate the entire metallicity range in these volumes.
These galaxies likely experienced few enrichment events and thus are primarily tracing the \dtm\ ratio at injection.

Figure~\ref{fig:mdust_mstar_res_conv} shows the dust mass as a function of the galaxy stellar mass for different resolutions  for $z=0,1,2,6\, {\rm and}\, 10$. 
Similar to the above figures, the convergence across resolution occurs at  high stellar mass (M$_{\star}\gtrsim 10^{8}\, \Msun$ for m5 and m6; M$_{\star}\gtrsim 10^{9}\, \Msun$ for m6 and m7).
However, they start to diverge again at the highest stellar masses (M$_{\star}\gtrsim 10^{11}\, \Msun$) for $z<6$.
This is due to a higher quiescent fraction in the lower-resolution simulations in this stellar mass regime \cite[see][]{Chandro-Gomez2025,Chaikin2025_smf_sfrf}. As discussed in \S~\ref{sec:dustscaling.dmass_smass}, we see the lower normalisation of the mass - metallicity relation of higher-resolution simulations, manifesting as higher dust masses at fixed stellar mass in the regime where the simulations are converged at $z>1$.



Figure~\ref{fig:dmf_res_conv} illustrates the dust mass function (DMF) across different \colibre\ simulation resolutions (m5, m6, m7) for $z=0,1,2,6\, {\rm and}\, 10$.
The DMF converges at the low dust mass end, $\approx 10^{4.5}\, \Msun$ for m6 and m7, while it is $\approx 10^{4}\, \Msun$ for the m5 and m6 resolutions. 
This agreement was not observed in Figure~\ref{fig:dmf} because of the imposed stellar particle cut of 10. 
We attribute this low-mass convergence to dust production being primarily sustained through stellar dust injection.
It is evident that these points correspond to galaxies with different stellar mass, with the higher resolution simulations hosting lower stellar mass galaxies at this end (see Figure~\ref{fig:mdust_mstar_res_conv}).
While all resolutions converge at the high-mass end (M$_{\rm dust}>10^7\, \Msun$) at $z=0$, this convergence weakens at higher redshifts, with the normalisation of the DMF agreeing within $\sim 0.6$ dex across progressively higher resolutions. 
As discussed in \S~\ref{sec:dmf}, the lower resolution simulations exhibit higher normalisation in this regime. Given that we are comparing similar volumes, this trend is not a result of the larger volume of the lower-resolution simulations plotted in Figure~\ref{fig:dmf}. Instead, it results from the higher normalisation of the mass -- metallicity relation in lower \colibre\ resolutions, which drives the higher dust masses. As noted, the galaxy metallicity is the primary driver of the \dtm\ ratio in the \colibre\ simulations.

\section{Hybrid AGN feedback model}\label{sec:app_hybrid}
\begin{figure*}
    \centering
    \includegraphics[width=0.8\textwidth]{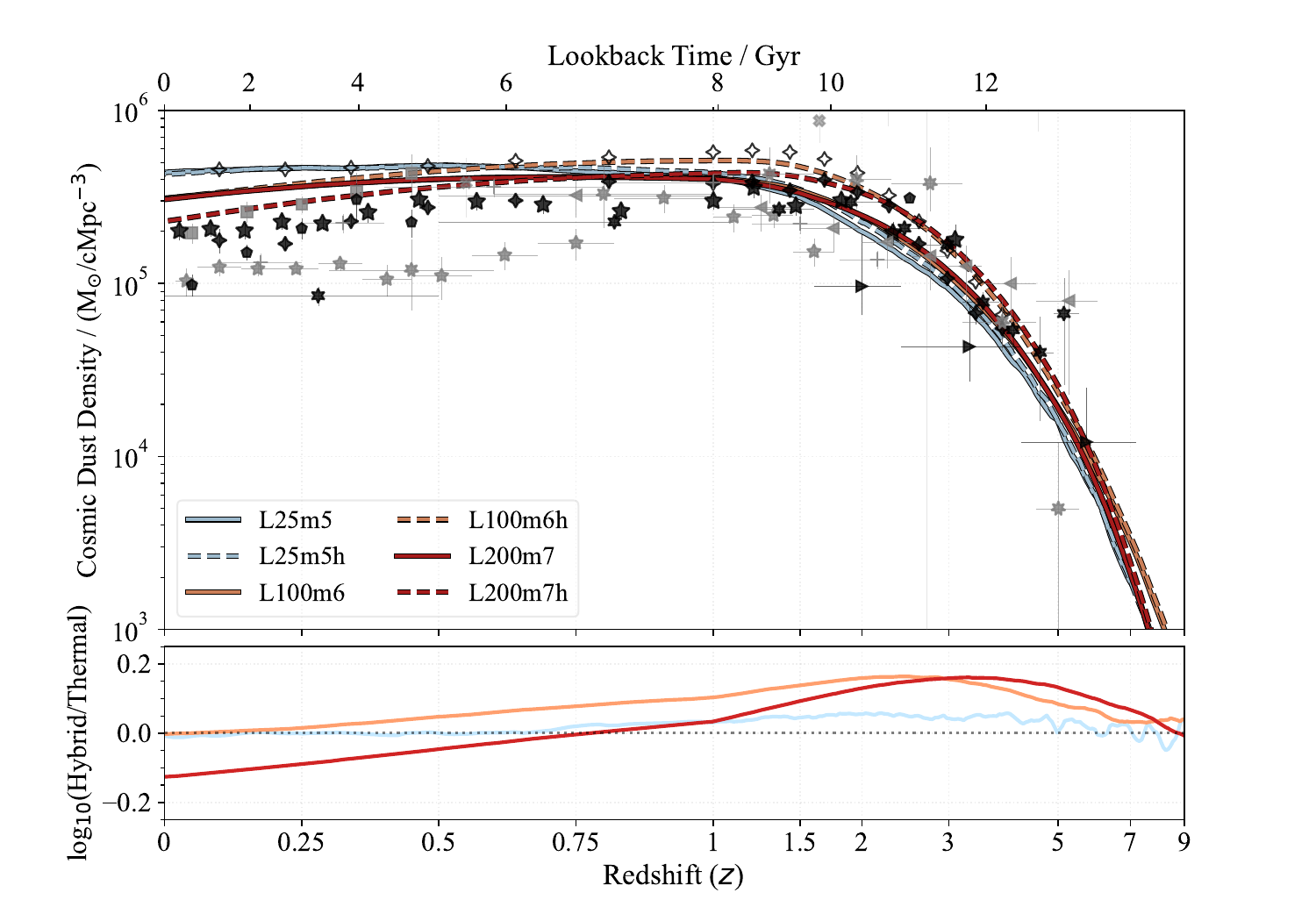}
    \caption{Same as Figure~\ref{fig:cdmd_evo}, but now for the thermal (solid line) and hybrid (dashed line) AGN models for the m5, m6, and m7 resolutions. We also show the relative difference between the hybrid and thermal AGN models in the bottom panel. The observational data are the same ones shown in Figure~\ref{fig:cdmd_evo}.}
    \label{fig:app_cdmd_hybrid_comparison}
\end{figure*}
All the discussions in the main text used the fiducial AGN (`Thermal' AGN model) feedback implementation. Here we compare the fiducial model against the `Hybrid' AGN model as implemented in \cite{Husko2025_hybridAGN}, briefly discussed in \S~\ref{sec:sims.subgrid.smbh}. We compare the two models using the $(25\, {\rm cMpc})^3$, $(100\, {\rm cMpc})^3$, and $(200\, {\rm cMpc})^3$ volumes for the m5, m6, and m7 resolutions, respectively.


Figure~\ref{fig:app_cdmd_hybrid_comparison} compares the redshift evolution of the cosmic dust mass density (CDMD) between the thermal and hybrid AGN model.
The m5 resolution is almost identical for the two models. However, both the m6 and m7 resolution exhibit higher CDMD in the redshift range $0.75 < z \lesssim 7$ compared to the thermal AGN model. 
The hybrid model still show good agreement with the observations in this redshift regime. 
\cite{Husko2025_hybridAGN} showed that the total AGN energy injected in the Hybrid model is $\sim 50\%$ more than the fiducial mode for $z<5$. 
Additionally, \cite{Sharda2026_MZR} find that for m6 resolution, the normalisation of the mass--metallicity relation at $z=1$ and $3$ is higher in the hybrid model than in the thermal model at high stellar mass (M$_{\star}>10^{10}$), whereas it is similar at $z=0$.
This is reflected in the rapid decline in the normalisation of the DMF and the cosmic dust mass density (Figure~\ref{fig:app_cdmd_hybrid_comparison}) from $z \sim 2$ to $0$ compared to the thermal AGN model.
The magnitude of the decline from $z=1\to0$ seen in the cosmic dust mass density from  of the m6 and m7 hybrid AGN model is similar to what is seen for several observations \cite[\eg][]{Dunne2011,Peroux&Howk2020,Magnelli2020,Pozzi2020,Ealse2024}.
This implies that the hybrid AGN model is highly effective at destroying dust, and suppressing its subsequent regrowth in this redshift regime ($z\sim2 \to 0$).
The m6 hybrid AGN simulation shows similar CDMD value to the thermal model below $z=0.25$.
At $z<0.5$, the CDMD of the m7 resolution hybrid model lies below the corresponding thermal model ($\approx 0.1$ by $z=0$), showing better agreement with the observations in this redshift regime.

For most of the other scaling relations explored in this work, the median relations show only negligible differences. 


\section{Effect of aperture on dust mass}\label{sec:app.aperture}
\begin{figure}
    \centering
    \includegraphics[width=\linewidth]{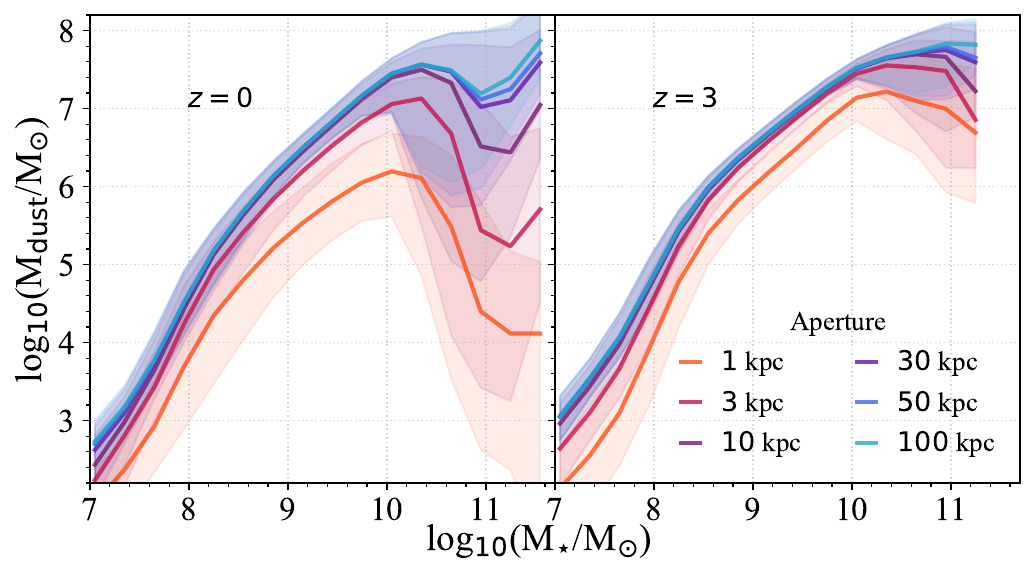}
    \caption{Median galaxy dust mass in different apertures ($1$ kpc, $3$ kpc, $10$ kpc, $30$ kpc, $50$ kpc, and $100$ kpc) as a function of stellar mass for the L200m6 \colibre\ simulation at $z=0\, {\rm and}\, 3$.}
    \label{fig:app.aperture_dust}
\end{figure}
Figure~\ref{fig:app.aperture_dust} shows the median galaxy dust mass in $1$ kpc, $3$ kpc, $10$ kpc, $30$ kpc, $50$ kpc, and $100$ kpc as a function of stellar mass for the L200m6 \colibre\ simulations at $z=0\, {\rm and}\, 3$. All the results discussed in the main text used properties within an aperture of $50$ kpc. Note that the stellar mass is measured within a $50$ kpc aperture.

At $z=3$, there is negligible difference between the dust mass measured in apertures $\ge 3$ kpc, while the $1$ kpc aperture is exhibits lower dust masses by $\approx 0.4$ dex across the plotted stellar mass range. At $z=0$, we see significant differences ($\gtrsim 4$ dex between 1 and 50 kpc) at the highest stellar masses, M$_{\star} > 10^{10}\, \Msun$. This is due to a large fraction of passive galaxies at these stellar masses, with a significant scarcity of dense gas in the central regions.

Essentially, since most of the dust mass resides within $\le 10$~pkpc, they fall well within the limits of typical observational apertures.

\section{Median scaling relation}\label{sec:app.median_scaling}
\begin{figure*}
    \centering
    \includegraphics[width=0.9\textwidth]{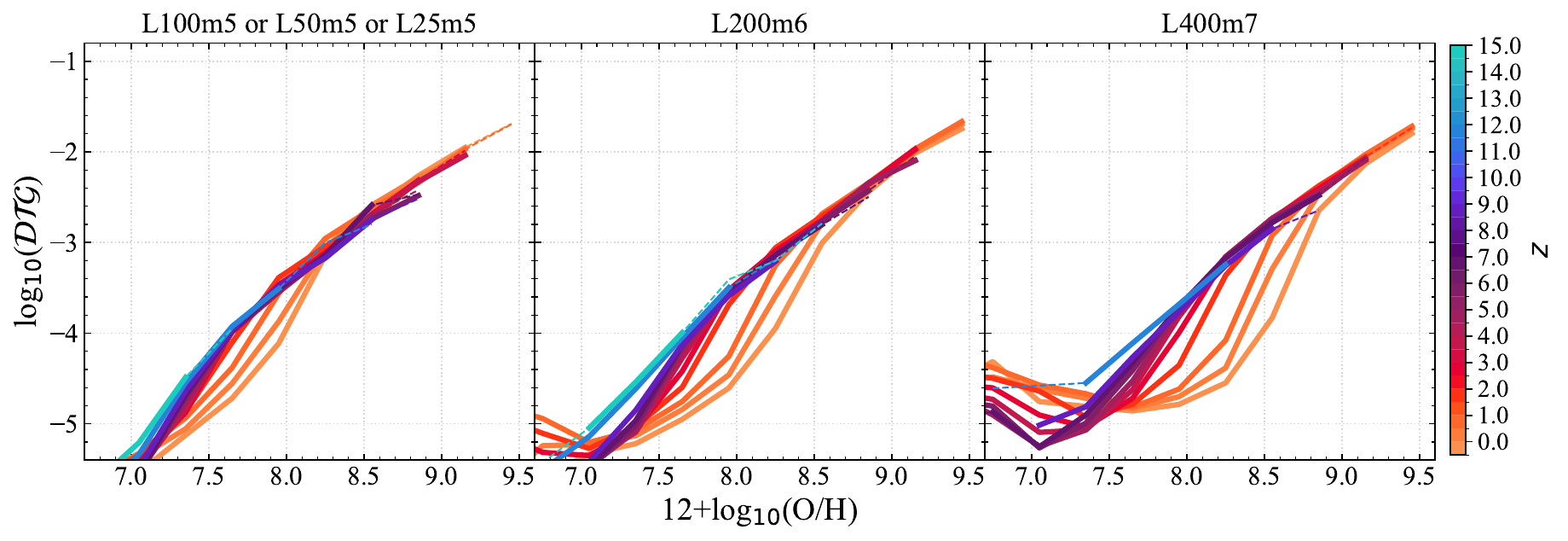}
    \caption{The median dust-to-gas (\dtg) ratio of galaxies in the \colibre\ m5, m6, and m7 simulations as a function of gas phase metallicity for $z \in [0,15]$. The solid coloured lines show the median results; bins with fewer than 5 galaxies are plotted as dashed lines.
    Note that for the m5 resolution, we switch from L100m5 simulation to L50m5 for $z<3$, and to L25m5 for $z<1$. This is similar to Figure~\ref{fig:dgr_Z_z0_9}, but shows the redshift evolution of individual resolutions in a single panel.}
    \label{fig:dtg_met_median}
\end{figure*}
\begin{figure*}
    \centering
    \includegraphics[width=0.9\textwidth]{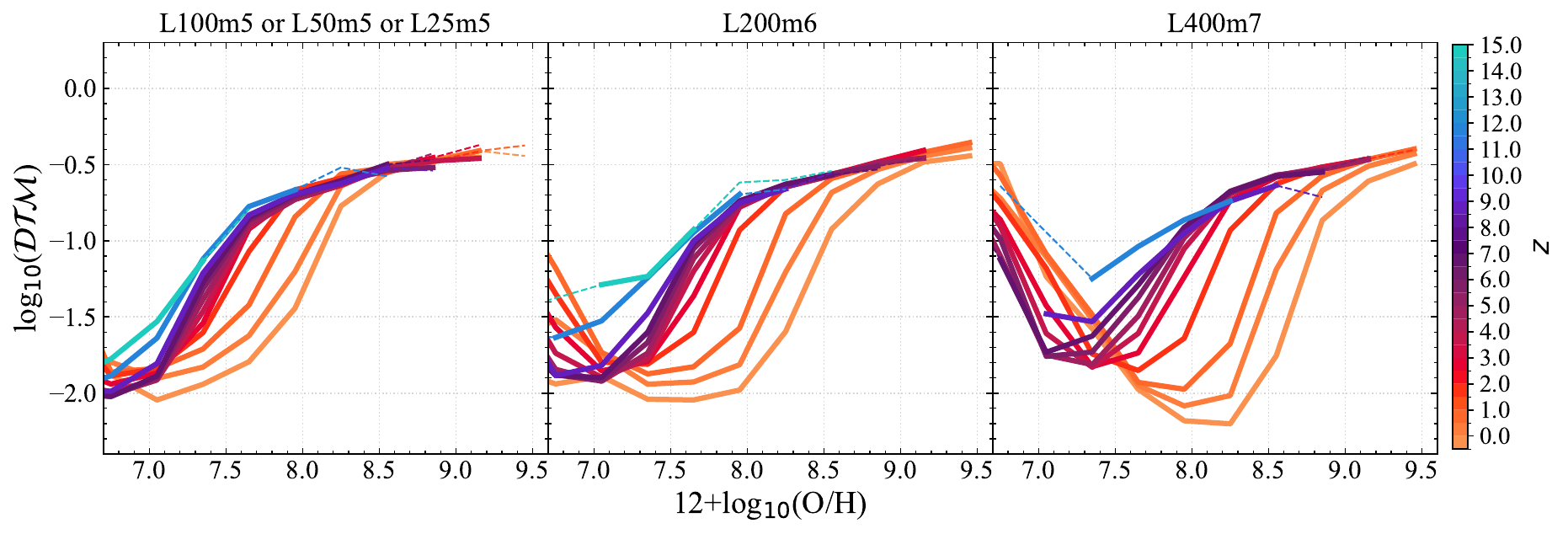}
    \caption{The median dust-to-metal (\dtm) ratio of galaxies in the \colibre\ m5, m6, and m7 simulations as a function of gas phase metallicity for $z \in [0,15]$. The solid coloured lines show the median results; bins with fewer than 5 galaxies are plotted as dashed lines.
    Note that for the m5 resolution, we switch from L100m5 simulation to L50m5 for $z<3$, and to L25m5 for $z<1$. This is similar to Figure~\ref{fig:DTM_vs_Z}, but shows the redshift evolution of individual resolutions in a single panel.}
    \label{fig:dtm_met_median}
\end{figure*}
\begin{figure*}
    \centering
    \includegraphics[width=0.9\textwidth]{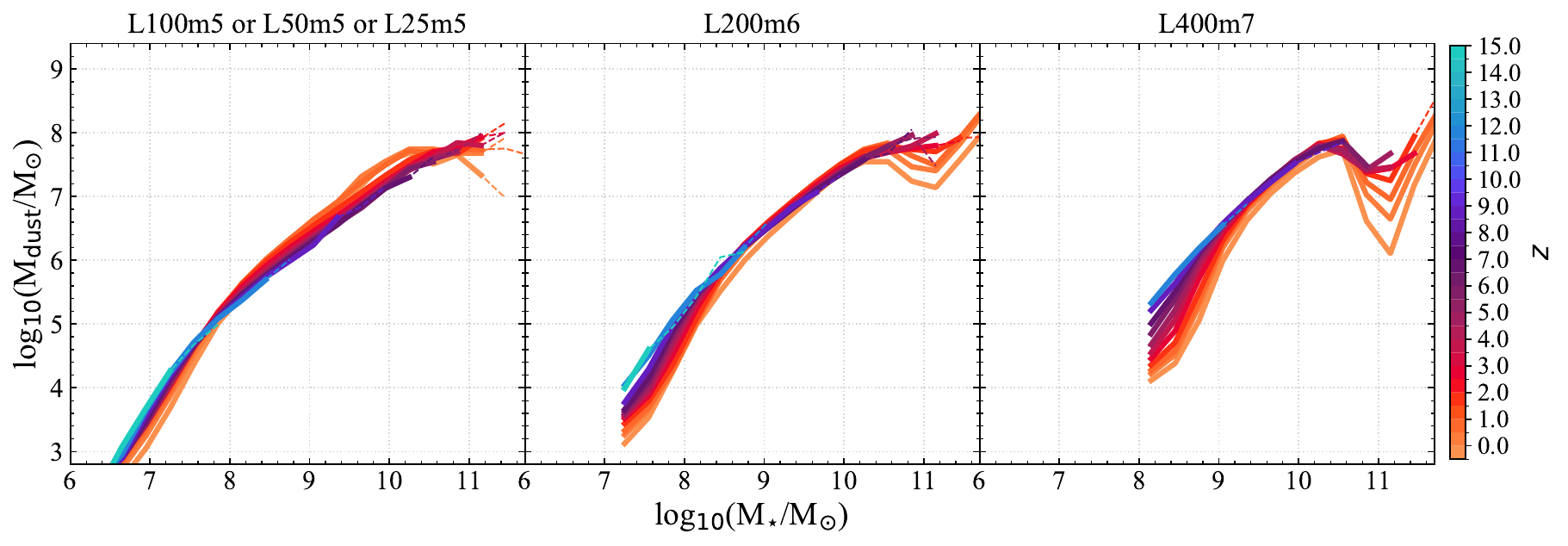}
    \caption{The median galaxy dust mass in the \colibre\ m5, m6, and m7 simulations as a function of their stellar mass for $z \in [0,15]$. The solid coloured lines show the median results; bins with fewer than 5 galaxies are plotted as dashed lines.
    Note that for the m5 resolution, we switch from L100m5 simulation to L50m5 for $z<3$, and to L25m5 for $z<1$. This is similar to Figure~\ref{fig:dmsm_z0_10}, but shows the redshift evolution of individual resolutions in a single panel.}
    \label{fig:mdust_mstar_median}
\end{figure*}
We show the median relation of the \dtg\ ratio with metallicity, the \dtm\ ratio with metallicity, and the dust mass with stellar mass in Figures~\ref{fig:dtg_met_median}, \ref{fig:dtm_met_median}, and \ref{fig:mdust_mstar_median}, respectively. We show the individual resolutions in a single panel for readers to better follow the evolution of the median relation as a companion to the corresponding figures (Figures~\ref{fig:dgr_Z_z0_9}, \ref{fig:DTM_vs_Z}, and \ref{fig:dmsm_z0_10}) in the main text.
\section{Dust mass-to-stellar mass ratio}\label{sec:app.dustscaling.dmass_smass_ratio}
\begin{figure*}
    \centering
    \includegraphics[width=\textwidth]{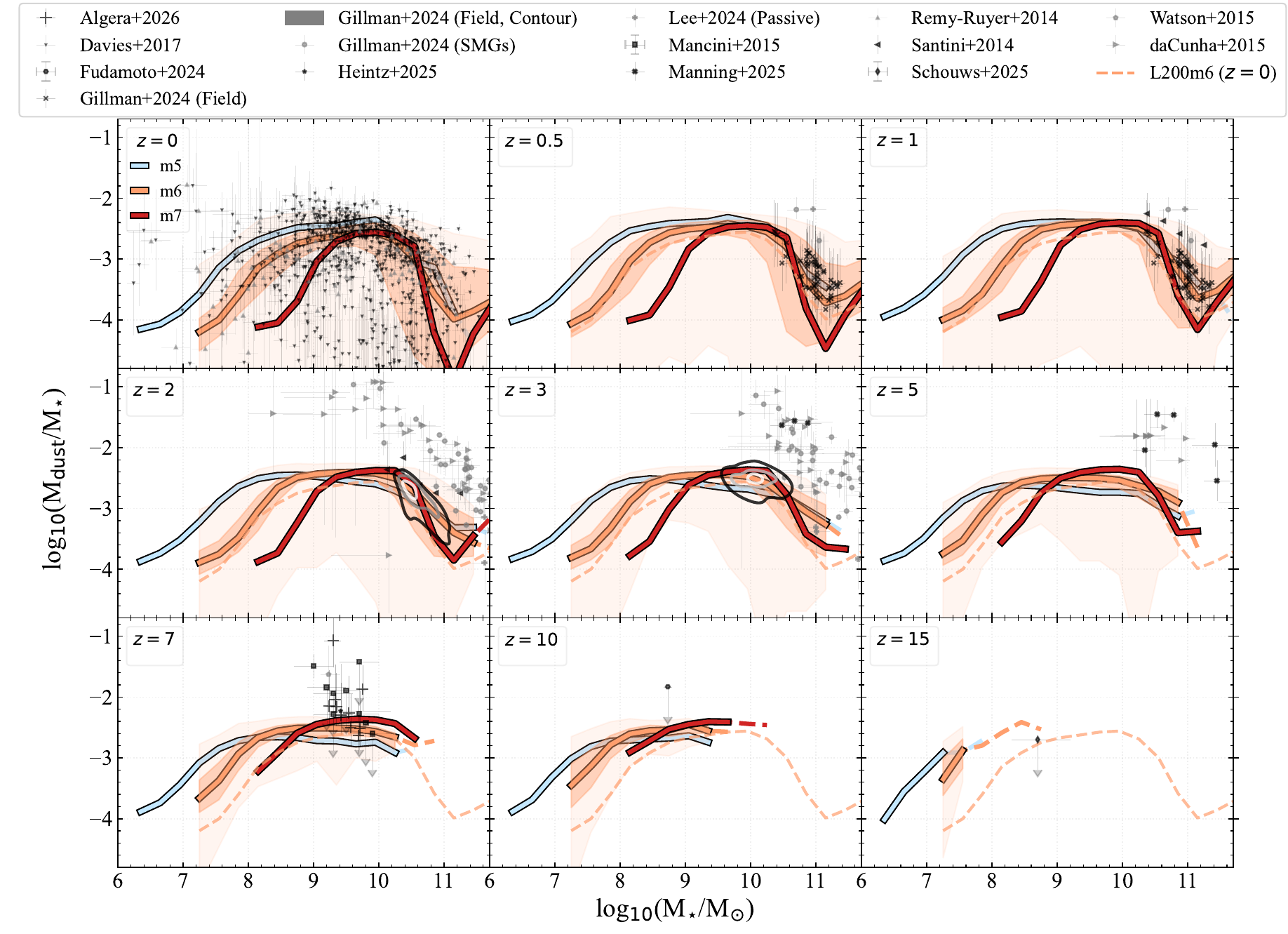}
    \caption{Same as Figure~\protect\ref{fig:dmsm_z0_10}, with the y-axis now showing the ratio of the dust mass to the stellar mass.}
    \label{fig:dmsm_ratio_z0_10}
\end{figure*}
In Figure~\ref{fig:dmsm_ratio_z0_10}, we plot the ratio of the dust mass to stellar mass as a function of stellar mass for galaxies across the redshift range, $z=0\text{--}15$. 
This ratio characterises the dust yield per unit stellar mass of galaxies, and is commonly used to infer the dust production efficiency of SNe in the high-redshift Universe, due to the over-abundance of dust seen in many galaxies. This apparent over-abundance also motivates invoking highly efficient dust grain growth in the dense gas of high-redshift galaxies \cite[\eg][]{Michalowski2010,Vijayan_dust2019}.

At all redshifts, the median relation has an inverted parabolic shape: low values at low stellar masses, a steady rise toward a maximum, a plateau at intermediate stellar masses, and a final decline at the highest stellar masses. 
In the low stellar mass regime, this steady increase is driven by stellar dust production and efficient grain growth in dense gas rapidly building up the dust mass. The exact stellar mass regime at which this transition occurs is resolution dependent, with the threshold increasing as resolution decreases (M$_{\star} \sim 10^{7}\, \Msun$, M$_{\star} \sim 10^{8}\, \Msun$, M$_{\star} \sim 10^{9}\, \Msun$ for m5, m6, and m7 respectively).
This growth reaches a steady state in the intermediate stellar mass range, where dust destruction balances grain growth. 
Beyond this regime, additional mechanisms such as AGN feedback which can heat and eject the surrounding gas while destroying or expelling dust from the galaxies. This disrupts the established balance between dust production and destruction, ultimately enhancing dust destruction and removal.
Moreover, this high-mass regime hosts a significant population of quiescent galaxies with intrinsically low dust content, further suppressing the median dust-to-stellar mass ratio.

As expected from Figure~\ref{fig:dmsm_z0_10}, most observational data lie in good agreement with the simulated trends, but with sub-mm galaxies \cite[from][]{daCunha2015,Gillman2024,Manning2025} continuing to exhibit dust abundances beyond the most extreme objects in the simulation.
Although we previously noted caveats regarding cosmic variance and the difficulties of estimating dust masses for these objects, taken at face value, they represent a population not reproduced by the model.

The median dust-to-stellar mass ratio increases gradually with increasing redshift at a fixed stellar mass, an effect that is more pronounced at low stellar masses. 
This trend can be attributed to higher redshift galaxies in general having higher SFRs and more cold, dense gas at fixed stellar mass, enhancing dust production while suppressing dust destruction.
Observational data showing high dust-to-stellar mass ratios ($> 0.01$) at high-redshift ($z>5$) point toward scenarios involving both high dust yields from stellar sources (with minimal to no dust destruction) and rapid dust grain growth. 
Producing such galaxies in simulations remains difficult when using standard stellar yields and grain growth timescales within such a short cosmic timescale, a challenge widely recognised as the "dust budget crisis" in the high-redshift Universe \cite[]{Gall2018}.
\section{Influence of Gas Phase selection}\label{sec:app_gas_phase}

\begin{figure*}
    \centering
    \includegraphics[width=0.7\textwidth]{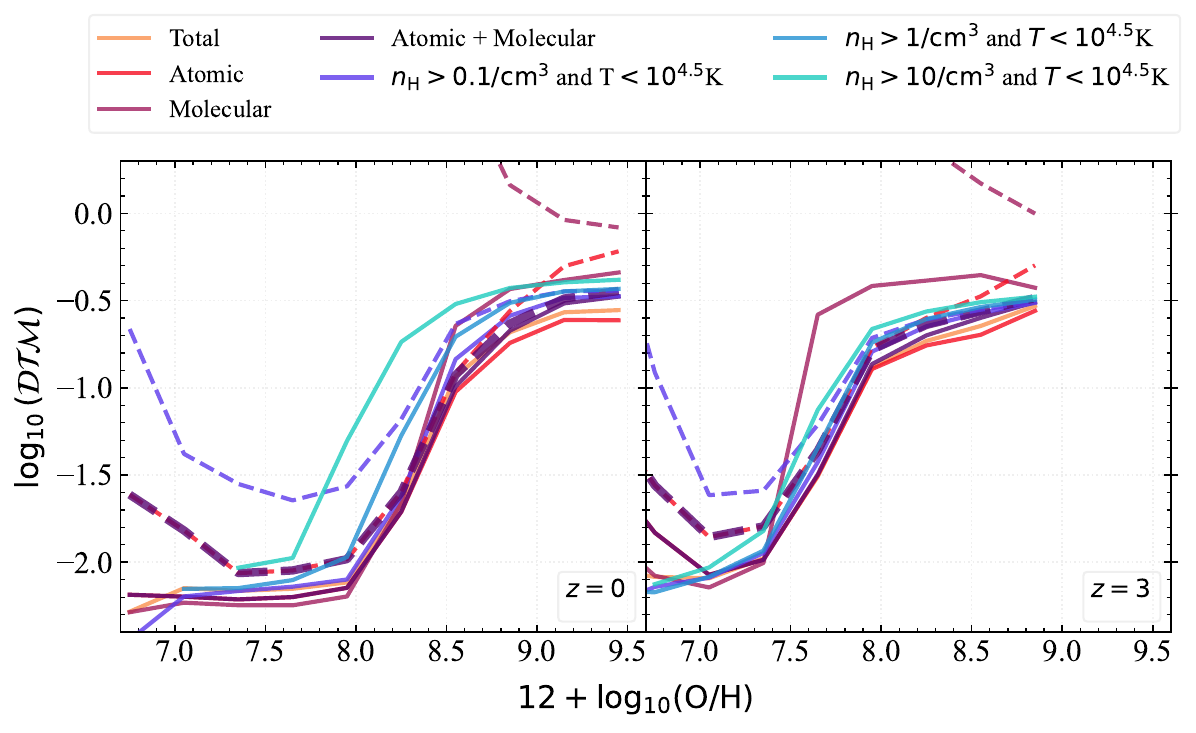}
    \caption{The median dust-to-metal (\dtm) ratio as a function of galaxy gas-phase metallicity for the L100m6 \colibre\ simulation at $z=0\, {\rm and}\, 3$. The different coloured lines show same information as in Figure~\ref{fig:gas_phase_dgr}.  The fiducial model choice is represented by the dashed line corresponding to `Atomic + Molecular'.}
    \label{fig:app_gas_phase_dtm}
\end{figure*}

\begin{figure*}
    \centering
    \includegraphics[width=0.8\textwidth]{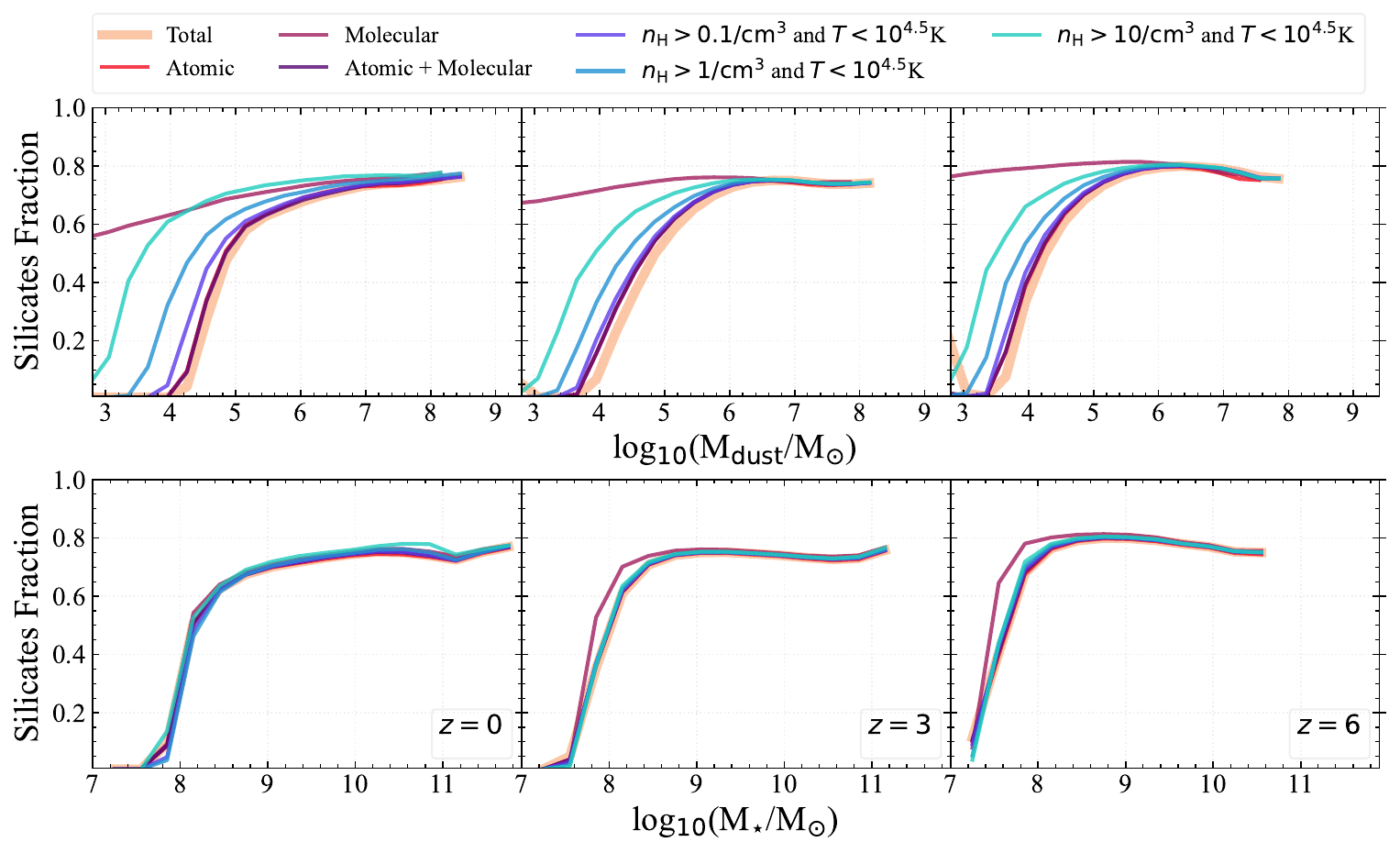}
    \caption{The median silicates fraction as a function of galaxy dust mass (top panel) and stellar mass (bottom panel) for the L100m6 \colibre\ simulation at $z=0, 3,\, {\rm and}\, 6$. The different coloured lines show same information as in Figure~\ref{fig:gas_phase_STL}. The fiducial model choice is represented by the solid line corresponding to `Total'.}
    \label{fig:app_gas_phase_silicates}
\end{figure*}

Figure~\ref{fig:app_gas_phase_dtm} shows the median relation between the \dtm\ ratio and gas-phase metallicity at $z=0$ and $3$. The trends seen here are similar to those exhibited by the relation between \dtg\ and metallicity (Figure~\ref{fig:gas_phase_dgr}).

Figure~\ref{fig:app_gas_phase_silicates} shows the median relation between the silicates fraction as a function of galaxy dust mass (top panel) and stellar mass (bottom panel) for the L100m6 \colibre\ simulation at $z=0, 3,\, {\rm and}\, 6$. 
We find a negligible dependence of the silicate fraction on stellar mass; however, denser gas phases exhibit a higher silicate fraction at low dust masses (M$_{\rm dust} < 10^5\,\Msun$).



\bsp	
\label{lastpage}
\end{document}